\documentclass[prd,superscriptaddress,nofootinbib,preprintnumbers]{revtex4}
\usepackage{amsmath,amsfonts,amssymb}
\usepackage{graphicx}
\usepackage{dcolumn}
\usepackage[mathlines]{lineno}
\usepackage{float}
\usepackage{hyperref}
\usepackage[usenames,dvipsnames]{color}
\usepackage{etex}
\usepackage{booktabs}
\usepackage{adjustbox}
\usepackage{url}

\begin{document}

\title{Radiative Neutrino Mass in a Nonholomorphic $T'$ Modular Invariant Model}

\author{Mohamed Amin Loualidi}
\email{ma.loualidi@uaeu.ac.ae}
\affiliation{Department of physics, United Arab Emirates University, Al-Ain, UAE}

\author{Mohamed Miskaoui}
\email{m.miskaoui@gmail.com}
\affiliation{Department of physics, Faculty of Science, Mohammed V University in Rabat, 10090 Rabat, Morocco}

\author{Salah Nasri}
\email{snasri@uaeu.ac.ae, salah.nasri@cern.ch}
\affiliation{Department of physics, United Arab Emirates University, Al-Ain, UAE}

\begin{abstract}
The \texttt{T4-2-i} topology provides a one-loop realization of Majorana neutrino mass and may be viewed as a radiative extension of the type-II seesaw, with a scalar triplet, two inert scalar doublets, and singlet fermions propagating in the loop. A central difficulty in realizing this topology lies in the simultaneous presence of tree-level type-I and type-II seesaw contributions arising from the same particle content. In addition, the stability of the dark matter candidate typically requires the introduction of an ad hoc discrete symmetry. In this work, we revisit the \texttt{T4-2-i} topology within a nonholomorphic modular invariant framework based on the double cover group $T'$. The presence of both even and odd weight polyharmonic Maa{\ss} forms considerably enlarges the space of allowed modular structures. An exact $Z_2$ parity inherited from the $T'$ modular symmetry stabilizes the lightest odd state for arbitrary values of $\tau$, whereas the vicinity of $\tau=i$ determines the flavor textures. The modular assignments exclude the tree-level type-I operator and determine the flavor structure, while lepton number conservation in the Yukawa sector excludes the tree-level type-II coupling. The modular invariant Majorana mass matrix provides the soft $\Delta L = 2$ breaking required by the radiative mechanism. We confront the model with neutrino oscillation data, charged lepton flavor violating bounds, electroweak precision observables, the Higgs diphoton signal strength, the observed dark matter relic abundance, the cosmological bound on the sum of neutrino masses, and direct detection limits. Focusing on the fermionic dark matter candidate, in which the lightest odd state is the Majorana fermion $N_1$, we find that both normal and inverted neutrino mass orderings remain viable. In the allowed region, the relic abundance is largely controlled by coannihilation with the inert scalar partners, while the Higgs-mediated scalar contribution to the spin-independent direct-detection rate is loop suppressed.
\end{abstract}


\maketitle
\section{Introduction}
\label{introduction} 
The origin of neutrino masses and the identity of dark matter (DM) remain among the clearest indications of physics beyond the Standard Model (SM). Neutrino oscillation experiments have established that neutrinos are massive and mixed \cite{Super-Kamiokande:2001bfk,SNO:2002tuh,SNO:2002hgz}, while a large body of astrophysical and cosmological observations points to the existence of a non-luminous matter component that cannot be accommodated within the SM \cite{Bertone:2004pz,Planck:2018vyg}. These two issues are often treated separately, but radiative neutrino mass models provide an attractive common framework in which the same new fields that generate neutrino masses at loop level may also supply a viable DM candidate. In this way, the smallness of neutrino masses, the structure of lepton mixing, and the stability of DM can be traced back to the same underlying dynamics. One class of radiative models is the topological classification of one-loop realizations of the dimension-five Weinberg operator \cite{Bonnet:2012kz}. Within this classification, the \texttt{T4} family comprises one-loop extensions of the conventional seesaw mechanisms. The \texttt{T4-3-i} topology provides radiative completions of the type-I and type-III seesaws and has recently been realized in two complementary field-theoretic constructions~\cite{Belfkir:2026slp,Bamwidhi:2026vdu}. Here, we focus on \texttt{T4-2-i}, which constitutes the corresponding one-loop extension of the type-II seesaw. Its field content contains a scalar triplet, two inert scalar doublets, and singlet fermions running in the loop. At the same time, this topology is also particularly challenging: the scalar triplet generically permits the usual tree-level type-II seesaw contribution, while the singlet fermions can simultaneously induce a type-I seesaw term, unless the corresponding operators are forbidden by a suitable underlying symmetry. Furthermore, the topology alone does not ensure the stability of the lightest beyond the SM state, and therefore does not by itself guarantee a viable DM candidate. \\ 

The first explicit field-theoretic realization of topology \texttt{T4-2-i} was presented in Ref.~\cite{Loualidi:2020jlj}, where the model was constructed within the traditional flavor-symmetric framework based on $\mathbb{D}_4\times \mathbb{Z}_3\times \mathbb{Z}_5\times \mathbb{Z}_2$ and supplemented by additional flavon fields. In that setup, the enlarged symmetry was used to forbid the unwanted tree-level type-I and type-II seesaw operators and, at the same time, to stabilize a DM candidate upon its breaking. This work established that the \texttt{T4-2-i} topology can indeed be promoted to a viable field-theoretic model, and that the particles running in the loop may also furnish DM candidates. The construction, however, relied on a comparatively large symmetry structure and an enlarged scalar sector, while the DM phenomenology itself was discussed only at a qualitative level. Subsequent studies explored the same topology from different symmetry viewpoints. In Ref.~\cite{Kashav:2026jjg}, this topology was realized by means of non-invertible selection rules, providing a conceptually different mechanism in which generalized fusion relations both remove the unwanted tree-level operators and stabilize the DM candidate; see also Refs.~\cite{Kobayashi:2024cvp,Kobayashi:2025znw,Kobayashi:2025ldi,Nomura:2025yoa,Nomura:2025sod,Jin:2026fkb,Okada:2025kfm,Jangid:2025krp,Okada:2025nap,Nomura:2025tvz,Okada:2025adm,Okada:2026gxl,Kashav:2026jjg,Okada:2026pek} for applications to different lepton models. These works show that \texttt{T4-2-i} is a useful testing ground for new symmetry ideas. An interesting modular realization was considered in \cite{Kashav:2022kpk} with a modular $A_4$ supersymmetric framework where modular weights were employed to suppress the tree-level terms and to stabilize the fields running in the loop, leading to a more economical realization of \texttt{T4-2-i}. \\

On the other hand, a consistent non-supersymmetric modular framework was formulated in terms of nonholomorphic modular flavor symmetry, in which the Yukawa couplings are described by polyharmonic Maa{\ss} forms rather than by the holomorphic modular forms characteristic of supersymmetric constructions~\cite{Qu:2024rns}. Since then, this framework has been explored in several flavor models, including nonholomorphic realizations based on $A_4$, $A_5$, and $S_3$~\cite{Cheshta:2026fls,Abbas:2026siw,Okada:2025nap,Priya:2026ehe,Majhi:2026jdk,Nasri:2026nbf,Gao:2025jlw,Nanda:2025lem,Priya:2025wdm,Zhang:2025dsa,Nomura:2025raf,Nomura:2025ovm,Nomura:2025bph,Nomura:2024atp,Kumar:2024uxn,Nomura:2024vus,Kobayashi:2025hnc,Okada:2025jjo,Li:2024svh,Ding:2024inn,Loualidi:2025tgw}. An important further development was the explicit construction of odd weight polyharmonic Maa{\ss} forms~\cite{Shang:2026qkh,Qu:2025ddz,Zhang:2026kyy,Li:2025kcr,Wang:2020lxk}. For homogeneous finite modular groups, and especially for the double cover group $T'$, these odd weights are not merely a formal extension; they enlarge the available modular multiplets and allow Yukawa textures that are absent in the even weight sector. From the model building point of view, this additional freedom is particularly valuable in radiative neutrino mass models, where one needs enough structure to distinguish the loop fields from the SM sector, control the allowed tree-level interactions. In this sense, the group $T'$ provides a particularly natural framework in which the topology \texttt{T4-2-i} can be revisited in a fully non-supersymmetric modular context. \\

In this paper, our purpose is therefore twofold. First, we construct a nonholomorphic modular invariant realization of the \texttt{T4-2-i} topology based on the double cover group $T'$. The even- and odd-weight polyharmonic Maa{\ss} forms determine the allowed flavor structures and exclude the tree-level Dirac interaction responsible for
the type-I seesaw. The Yukawa sector conserves lepton number, thereby excluding an a tree-level $LL\Delta$ interaction, whereas the modular invariant Majorana mass matrix of the singlet fermions provides the sole soft source of $\Delta L = 2$ breaking. This constitutes the first fully non-supersymmetric realization of \texttt{T4-2-i} built from nonholomorphic $T'$ modular forms with both even and odd weights. Second, we perform a comprehensive phenomenological study of the model. In particular, the parameter space is fitted to the charged lepton masses, the neutrino oscillation observables, the Higgs diphoton signal strength, the electroweak (EW) precision parameters, and the DM relic abundance, while the non-oscillatory neutrino observables, the charged lepton flavor violation (cLFV) bounds, the cosmological limit on $\sum_i m_i$, and the DM direct-detection limits are imposed as additional constraints. The resulting field assignments allow either a fermionic or a scalar DM candidate. Although both possibilities were investigated numerically, the results presented in this work focus on the fermionic Majorana candidate $N_1$. In the viable region, the relic abundance is mainly controlled by coannihilation with the inert scalar partners, while the spin-independent (SI) direct detection rate remains naturally suppressed because it is induced only through a loop-generated Higgs portal. Numerically, the model is explored through a broad parameter scan with point-by-point reconstruction of the physical spectrum and lepton mixing. The surviving points are then selected by the full set of phenomenological requirements and ranked by the total $\chi^2$ function. \\

This paper is organized as follows. In Sec.~II we review the nonholomorphic modular framework, modular invariance, modular forms, and polyharmonic Maa{\ss} forms. In Sec.~III we implement the \texttt{T4-2-i} topology and derive the charged-lepton, heavy neutrino, scalar, and one-loop light-neutrino mass structures implied by the modular assignments. In Sec.~IV we collect the main theoretical and experimental constraints, including perturbativity and vacuum stability, the Higgs diphoton branching ratio, the oblique parameters, cLFV observables, and the fermionic DM framework relevant for the scan. In Sec.~V we present the numerical strategy and the phenomenological results for both neutrino orderings. Finally, Sec.~VI contains our summary and conclusions. Appendix~\ref{app1} summarizes the group theoretical ingredients of the homogeneous finite modular group $T'$ used in the construction. Appendix~\ref{app2} collects the loop functions entering the EW precision and cLFV analyses, and Appendix~\ref{app3} gives the effective loop-induced coupling relevant for direct detection. The scalar-sector perturbative-unitarity analysis and the electroweak input scheme used to define the precision observable diagnostic are presented in Appendices \ref{app4} and \ref{app5}, respectively.
\section{Modular symmetry and its nonholomorphic realization}
\label{Sec2}
Modular symmetry has emerged as a powerful framework for constructing flavor theories. It provides a predictive alternative to conventional non Abelian discrete symmetries, which have long been employed to describe fermion mass hierarchies and mixing structures.
In contrast to traditional approaches requiring multiple flavon fields
and auxiliary symmetries, modular invariance constrains the Yukawa sector
through the transformation properties of fields under the modular group,
thereby reducing the number of free parameters and enhancing theoretical predictivity. The full modular group $\mathrm{SL}(2,\mathbb{Z})$ plays a central role in
modular invariant constructions of particle physics flavor models.
It is defined as the group of $2\times2$ matrices with integer entries and unit determinant,
\begin{equation}
\mathrm{SL}(2,\mathbb{Z})
=
\left\{
\begin{pmatrix}
a & b \\
c & d
\end{pmatrix}
\;\middle|\;
a,b,c,d \in \mathbb{Z}, \quad ad-bc=1
\right\}.
\end{equation}
This group naturally acts on the complex modulus $\tau$ living in the
upper half-plane
$\mathbb{H}=\{\tau\in\mathbb{C}\,|\,\mathrm{Im}\,\tau>0\}$ through the following transformations,
\begin{equation}
\tau \;\rightarrow\; \gamma\tau
= \frac{a\tau+b}{c\tau+d},
\qquad
\gamma \in \mathrm{SL}(2,\mathbb{Z}).
\end{equation}
Because the matrices $\gamma$ and $-\gamma$ induce the same
transformation on $\tau$, the modular
group is described
by the projective modular group
$\mathrm{PSL}(2,\mathbb{Z})=\mathrm{SL}(2,\mathbb{Z})/\{\pm I\}$, where $\pm I$ is the two-dimensional unit element. The modular group is generated by two fundamental elements,
\begin{equation}
S:\ \tau \rightarrow -\frac{1}{\tau},
\qquad
T:\ \tau \rightarrow \tau + 1,
\end{equation}
which satisfy the defining algebraic relations $S^{2} = (ST)^{3} = I$. Finite modular groups are obtained by taking the quotient of the modular group by its principal congruence subgroup of level $N$. The principal congruence subgroup is defined as
\begin{equation}
\Gamma(N)
=
\left\{
\gamma \in \mathrm{SL}(2,\mathbb{Z})
\;\middle|\;
\gamma \equiv I \; (\mathrm{mod}\, N)
\right\},
\end{equation}
which is an infinite normal subgroup of $\mathrm{SL}(2,\mathbb{Z})$. For $N=1,2$, we define
$\overline{\Gamma}(N)=\Gamma(N)/\{\pm I\}$, whereas
for $N>2$, since $-I\notin\Gamma(N)$, we simply have
$\overline{\Gamma}(N)=\Gamma(N)$. The corresponding inhomogeneous finite modular group is defined as
\begin{equation}
\Gamma_{N}
\equiv
\mathrm{PSL}(2,\mathbb{Z}) / \overline{\Gamma}(N),
\end{equation}
This group can be generated by two elements $S$ and $T$ satisfying the relation $S^{2} = (ST)^{3} = T^{N} = I$. For the lowest levels, these groups coincide with the usual discrete groups, $\Gamma_{2}\simeq S_{3}$, $\Gamma_{3}\simeq A_{4}$, $\Gamma_{4}\simeq S_{4}$, and $\Gamma_{5}\simeq A_{5}$~\cite{deAdelhartToorop:2011re}. Such finite groups have been widely used as flavor symmetries in models
of quark and lepton masses and mixing ~\cite{Kobayashi:2023zzc,Cheshta:2026fls,Abbas:2026siw,Okada:2025nap,Priya:2026ehe,Majhi:2026jdk,Nasri:2026nbf,Gao:2025jlw,Nanda:2025lem,Priya:2025wdm,Zhang:2025dsa,Nomura:2025raf,Nomura:2025ovm,Nomura:2025bph,Nomura:2024atp,Kumar:2024uxn,Nomura:2024vus,Kobayashi:2025hnc,Okada:2025jjo,Li:2024svh,Ding:2024inn,Loualidi:2025tgw}.
A different finite structure is obtained by instead considering the quotient $\Gamma'_{N} \equiv \mathrm{SL}(2,\mathbb{Z}) / \Gamma(N)$,
which defines the homogeneous finite modular group ~\cite{Liu:2019khw} .
In contrast to $\Gamma_{N}$, this construction is performed at the level
of $\mathrm{SL}(2,\mathbb{Z})$ rather than its projective quotient, thereby providing a double covering of the inhomogeneous group.
The homogeneous group is generated by $S$, $T$, and an additional
central element $R$ related to $-I\in \mathrm{SL}(2,\mathbb{Z})$. The element $R$ lies in the center of the group and
commutes with all other elements of $ \mathrm{SL}(2,\mathbb{Z})$ group. These generators satisfy the relations
\begin{equation}
S^{2} = R,
\qquad
(ST)^{3} = 1,
\qquad
T^{N} = 1,
\qquad
R^{2} = 1,
\qquad
RT = TR .
\end{equation}
Because of this extended algebraic structure, $\Gamma'_{N}$ naturally
accommodates modular forms of both even and odd weights, which is
essential for realistic model building. Several phenomenologically viable flavor models based on double cover modular symmetries have been constructed by exploiting this property ~\cite{Shang:2026qkh,Qu:2025ddz,Zhang:2026kyy,Li:2025kcr,Wang:2020lxk}. In the phenomenologically important case $N=3$, the homogeneous group
$\Gamma'_{3}$ is isomorphic to the binary tetrahedral group $T'$,
namely the double covering of $A_{4}$.
This enlarged symmetry contains additional irreducible
representations—particularly doublets—that are absent in $A_{4}$ and
thus provides greater flexibility for constructing viable lepton flavor
models ~\cite{Qu:2025ddz}.
A modular form of weight $k$ and level $N$ is defined as a holomorphic
function $f(\tau)$ on $\mathbb{H}$ transforming according to
\begin{equation}
f\!\left(\frac{a\tau+b}{c\tau+d}\right)
= (c\tau+d)^{k} f(\tau),
\qquad
\gamma \in \Gamma(N),
\end{equation}
and remaining regular at the cusps.
For $N=1$ and $N=2$, the element $-I$ belongs to $\Gamma(N)$, implying that nonvanishing modular forms must carry even modular weight. For $N>2$, odd weight modular forms are allowed, which is a crucial feature in constructions based on homogeneous finite modular groups. In the following sections we specialize to the case $N=3$, where the homogeneous finite modular group $\Gamma'_3 \simeq T'$ enables the systematic construction of modular multiplets and their application to realistic models of lepton masses and mixing.
In the supersymmetric framework, chiral supermultiplets transform nontrivially under finite modular symmetries derived from $\Gamma_N$ or its double cover $\Gamma_N'$. A chiral superfield $\phi_i$, assigned to an irreducible representation $\rho_i$ and modular weight $-k_i$, transforms as
\begin{equation}
\phi_i \rightarrow (c\tau+d)^{-k_i}\,\rho_i(\gamma)\,\phi_i.
\end{equation}
Consider a generic trilinear contribution to the superpotential, ${\cal W}(\tau,\phi)
\supset
Y_{ijk}(\tau)\phi_i\phi_j\phi_k$. Modular invariance of the superpotential requires the total modular weight to vanish and the product of representations to contain the trivial singlet. Consequently, the Yukawa couplings cannot be arbitrary constants, but must themselves
transform nontrivially under the modular group. Their modular weight is fixed by those of the matter superfields, i.e., $k_Y=k_i+k_j+k_k$, whereas the representation $\rho_Y$ of the Yukawa multiplet must satisfy $\rho_Y\otimes\rho_i\otimes\rho_j\otimes\rho_k
\supset
\mathbf{1}$.
Owing to the holomorphic nature of the superpotential, these Yukawa couplings are identified with holomorphic modular forms. A modular form of weight $k_Y$ and level $N$ transforms according to
\begin{equation}
Y_{ijk}(\tau)
\rightarrow
Y_{ijk}(\gamma\tau)=(c\tau+d)^{k_Y}
\rho_Y(\gamma)
Y_{ijk}(\tau),
\end{equation}
In supersymmetric realizations of modular flavor symmetry, the combined requirements of modular invariance and holomorphicity strongly constrain the Yukawa sector, thereby reducing the number of independent parameters. However, in the absence of experimental evidence for supersymmetry, it is well motivated to explore non-supersymmetric implementations of modular invariance while retaining the predictive features of the modular framework.
A non-supersymmetric realization of modular flavor symmetry was developed in Ref.~\cite{Qu:2024rns}, where the modular framework was successfully applied to the lepton sector without invoking SUSY. The basic idea was originally proposed in Ref.~\cite{Feruglio:2022cgv}, where automorphic forms were suggested as a natural extension of ordinary modular forms in non-supersymmetric theories. In this approach, the holomorphicity condition is replaced by a Laplace-type equation defined on the upper-half complex plane. For a single modulus $\tau=x+iy$, the relevant objects are harmonic and polyharmonic Maa{\ss} forms, which constitute nonholomorphic generalizations of conventional modular forms.
A harmonic Maa{\ss} form satisfies $\Delta_kF(\tau)=0$,  where $\Delta_k$ is the weight $k$ hyperbolic
Laplacian operator defined as
\begin{equation}
\Delta_k
=
-y^2
\left(
\frac{\partial^2}{\partial x^2}
+
\frac{\partial^2}{\partial y^2}
\right)
+
iky
\left(
\frac{\partial}{\partial x}
+
i\frac{\partial}{\partial y}
\right)
=
-4y^2
\frac{\partial}{\partial\tau}
\frac{\partial}{\partial\bar{\tau}}
+
2iky
\frac{\partial}{\partial\bar{\tau}}
\end{equation}

Polyharmonic Maa{\ss} forms obey the generalized condition $\Delta_k^mF(\tau)=0$ for some positive integer $m$.  Although these functions
transform with the same automorphy factor as ordinary modular forms, they
depend explicitly on both $\tau$ and $\bar{\tau}$ while remaining modular covariant. In non-supersymmetric modular frameworks, the Yukawa couplings are promoted to polyharmonic Maa{\ss} forms of level $N$, which replace ordinary holomorphic modular forms as the fundamental objects controlling the flavor structure Ref.~\cite{Qu:2024rns}. These functions satisfy the same automorphy condition under modular transformations as conventional modular forms, with the holomorphicity condition substituted by a Laplace-type constraint
\begin{equation}
Y(\gamma\tau)=(c\tau+d)^{k}
Y(\tau),
\qquad 
\Delta_k^mY(\tau)=0
\end{equation}
In addition to modular covariance, polyharmonic Maa{\ss} forms are required to satisfy a moderate-growth condition, namely $Y(\tau)=\mathcal{O}(y^\alpha)$ with $\alpha\in\mathbb{R}$ as ${\rm Im}(\tau)\to\infty$~\cite{Borel1,Borel2}, together with the analogous condition at all cusps. In the context of flavor model building, the resulting framework closely parallels that based on ordinary modular forms. At fixed modular weight and level, polyharmonic Maa{\ss} forms span finite-dimensional vector spaces, thereby imposing nontrivial constraints on the flavor structure of the theory. In non-supersymmetric modular theories, the Yukawa interactions generically take the form
\begin{equation}
-\mathcal{L}_Y
=
Y_{ij}(\tau)\,
\bar{\Psi}_{L_i}\Phi\Psi_{R_j}
+\text{h.c.},
\end{equation}
where $\Psi_{L,R}$ denote fermion multiplets and $\Phi$ is the Higgs field.
Under modular transformations, the fields transform according to
\begin{align}
\bar\Psi_{L_i}
\rightarrow
(c\tau+d)^{-k_{L_i}}
\rho_{L_i}(\gamma)\,
\Psi_{L_i},
\qquad
\Psi_{R_j}
\rightarrow
(c\tau+d)^{-k_{R_j}}
\rho_{R_j}(\gamma)\,
\Psi_{R_j},
\qquad
\Phi
\rightarrow
(c\tau+d)^{-k_\Phi}
\rho_\Phi(\gamma)\,
\Phi.
\end{align}
where $k_{\bar{\Psi}}$, $k_{\Psi}$, and $k_{\Phi}$ denote the integer modular weights, while $\rho_{\bar{\Psi}}$, $\rho_{\Psi}$, and $\rho_{\Phi}$ correspond to irreducible representations of $\Gamma_N$. Modular invariance requires the Yukawa coupling $Y_{ij}(\tau)$ to transform as a multiplet of level-$N$ polyharmonic Maa{\ss} forms with even integer modular weight $k_Y$, belonging to a representation $\rho_Y$ of $\Gamma_N$, namely
\begin{equation}
Y_{ij}(\tau)
\rightarrow
(c\tau+d)^{k_Y}
\rho_Y(\gamma)\,
Y_{ij}(\tau),
\end{equation}
with the modular weight constraint $k_Y=
k_{L_i}+k_{R_j}+k_\Phi$. At the same time, the tensor product of representations must contain the trivial singlet, $\rho_Y
\otimes \rho_{L_i}\otimes\rho_\Phi \otimes \rho_{R_j} \supset \mathbf{1}$.
In the case of inhomogeneous modular symmetry $\Gamma_N$, the associated polyharmonic Maa{\ss} forms carry even modular weights, whereas for the homogeneous groups $\Gamma_N^\prime$ the modular weights can take arbitrary integer values. At fixed weight and level, these forms span finite-dimensional vector spaces, allowing modular invariance to retain its predictive power even in non-supersymmetric frameworks.
An important feature of polyharmonic Maa{\ss} forms is that their Fourier expansion naturally separates into holomorphic and nonholomorphic contributions. For the level-$3$ modular symmetry considered in this work, the Yukawa multiplets admit the expansion ~\cite{Qu:2024rns}
\begin{equation}
Y(\tau)
=
\sum_{n\in\frac1N\mathbb{Z},\,n\ge0}
c^+(n)q^n
+
c^-(0)y^{1-k}
+
\sum_{n\in\frac1N\mathbb{Z},\,n<0}
c^-(n)
\Gamma(1-k,-4\pi ny)\,
q^n,
\qquad
q=e^{2\pi i\tau},
\end{equation}
where the coefficients $c^{\pm}(n)$ and $c^{\pm}(0)$ are constants, and $\Gamma(a,z)=\int_z^\infty e^{-t}t^{a-1}dt$
denotes the incomplete gamma function. The second term corresponds to the genuinely nonholomorphic contribution, which distinguishes polyharmonic Maa{\ss} forms from the ordinary holomorphic modular forms commonly used in modular flavor models. We emphasize that $\tau$ is treated as a vacuum expectation value, not as a propagating field, it enters the Lagrangian only through the modular forms evaluated at $\langle\tau\rangle$, so it has no direct dynamical effect on the observables computed in this work\footnote{Treating $\langle\tau\rangle$ as a free model parameter is common in bottom-up, non-supersymmetric modular model building. Following the non-holomorphic modular framework of Ref.~\cite{Qu:2025ddz}, we scan $\langle\tau\rangle$ over the fundamental domain. Promoting the modulus to a dynamical field and determining its vacuum expectation value would require specifying the modulus-sector dynamics, including an appropriate kinetic term and stabilization potential. Since these ingredients are model dependent, a detailed analysis of modulus stabilization lies beyond the scope of the present effective framework.}.
In this work, we consider a nonholomorphic modular framework based on the homogeneous finite modular group $\Gamma'_3\simeq T'$. The Yukawa sector is built from polyharmonic Maa{\ss} multiplets transforming under irreducible representations of $T'$, providing a nonholomorphic realization of modular flavor symmetry.
\section{Implementing the nonholomorphic $T'$ group in topology T4-2-i}
Unlike modular $A_4$ models where modular forms are restricted to even weights due to the single-valued representation structure, the double covering group $T'$ allows for modular forms of both even and odd weights. This enriches the model-building possibilities and provides a natural framework for incorporating discrete symmetries such as a residual $\mathbb{Z}_2$. A key feature is that the stability of DM is not the result of an ad hoc symmetry, but a consequence of modular invariance under the residual $\mathbb{Z}_2 \subset T'$ at $\tau \simeq i$. This origin makes the model minimal and UV-motivated.
\subsection{Field Content and Assignments}
\begin{figure}[b!]
    \centering
    \includegraphics[width=0.35\linewidth]{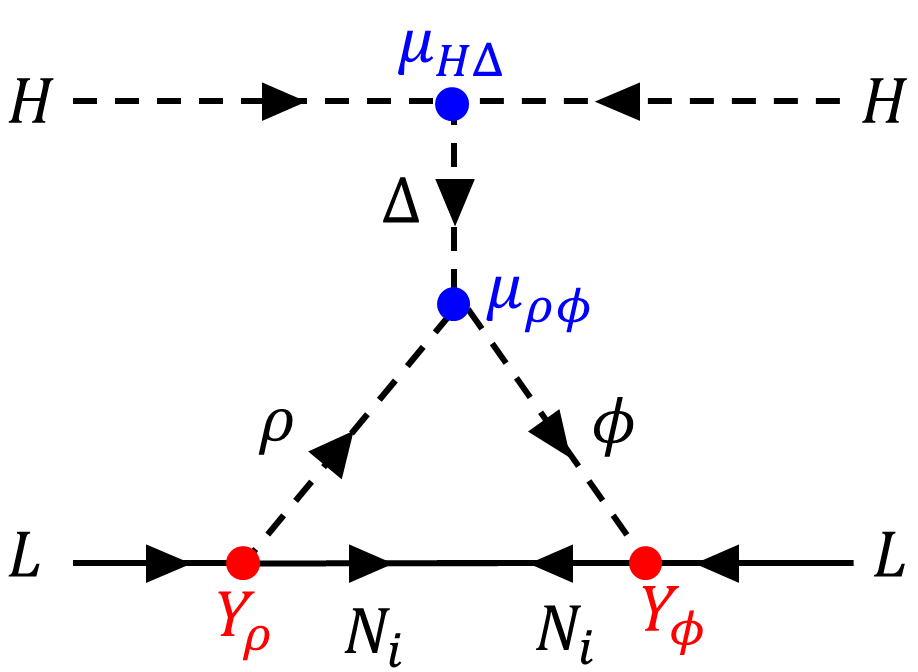}
    \caption{One-loop \texttt{T4-2-i} realization of neutrino mass generation, mediated by the inert doublets $\rho$ and $\phi$, the scalar triplet $\Delta$, and the heavy Majorana neutrinos $N_i$.}
    \label{fig1}
\end{figure}

Among the various one-loop realizations of radiative neutrino mass generation, the \texttt{T4-2-i} topology has received comparatively little attention in the literature. This topology involves a scalar triplet $\Delta$ with hypercharge $Y=2$ and two inert scalar doublets, $\phi$ and $\rho$, running in the loop together with right-handed Majorana neutrinos $N_i$. The corresponding one-loop diagram is shown in Fig.~\ref{fig1}. 
In this setup, neutrino masses are generated radiatively at the one-loop level through the exchange of the scaler fields $\Delta$, $\phi$, $\rho$, and the right-handed neutrinos $N_i$, while the modular weight and lepton number field assignments forbid the corresponding tree-level seesaw contribution.

Although the \texttt{T4-2-i} topology contains a richer scalar sector than minimal scotogenic realizations \cite{Ma:2006km, Avila:2025qsc}, it also introduces two important model building challenges. On the one hand, the simultaneous presence of the scalar triplet and right-handed neutrinos generically permits tree-level type-II and type-I seesaw contributions unless additional selection rules are imposed. On the other hand, the topology alone does not guarantee DM stability, since the lightest beyond the SM state may in principle decay into SM particles. The present construction
combines the modular $T'$ selection rules with a lepton number conserving Yukawa sector. These ingredients determine the flavor structure, remove the independent tree-level neutrino mass operators, and retain the residual symmetry structure required by the dark sector.

The modular symmetry and lepton number have complementary functions. The $T'$ representations and modular weights determine the allowed flavor structures and exclude $\bar{L}\tilde{H}N$. Lepton number is conserved by the Yukawa and scalar interactions and is broken softly only by the modular invariant Majorana mass matrix of the singlet fermions. Furthermore, in contrast to conventional radiative neutrino mass models, which typically require the introduction of ad hoc discrete symmetries to forbid unwanted interactions and stabilize the DM candidate, the present construction exploits the residual symmetry structure that emerges in the vicinity of the fixed point $\tau=i$. This residual symmetry simultaneously suppresses undesirable operators and provides a natural mechanism for DM stability within the modular framework. More specifically, the SM fields are assigned even modular weights, whereas the dark sector fields $\rho$, $\phi$, and $N_1$ carry odd modular weights. Consequently, modular invariance forbids operators involving an odd number of odd weight fields, thereby preventing the decay of the dark sector states into purely SM particles. As a result, the lightest state among ${\rho,\phi,N_1}$ is stable and can serve as a viable DM candidate. In this way, the modular weight assignments effectively reproduce the role commonly played by an auxiliary dark parity, without the need to impose an additional discrete symmetry by hand. Furthermore, modular invariance strongly constrains the Yukawa sector through the structure of modular forms, significantly reducing the number of independent parameters and leading to predictive textures in both the charged lepton and neutrino sectors. Under $U(1)_L$, the lepton doublets, charged-lepton singlets, and right-handed neutrino
singlets $N_i$ are assigned $L = 1$, whereas all scalar multiplets carry $L = 0$. The particle content and transformation properties under $SU(2)_L\times U(1)_Y$, $T'$, and the modular weights $k_I$ are summarized in Table~\ref{tab:fields}. At the fixed point $\langle\tau\rangle=i$, the vacuum preserves the residual modular symmetry generated by $S$; this residual symmetry is broken when $\langle\tau\rangle$ departs from $i$. DM stability, however, does not rely on this fixed-point symmetry. Instead, it follows from the central element $R=S^{2}=-\mathbb{I}$ of the homogeneous modular group $T'$, which acts trivially on $\tau$ and therefore remains unbroken for an arbitrary value of the modulus~\cite{Liu:2019khw}. For a field $\Phi_I$ of modular weight $k_I$, transforming in the representation $r_I$, the action of $R$ defines the exact parity
\begin{equation}
\mathcal{P}_I=(-1)^{k_I}\rho_{\mathbf{r}_I}(R).
\label{eq:modular-parity}
\end{equation}
With the assignments given in Table~\ref{tab:fields}, the SM fields, $H$, and $\Delta$ are $\mathcal{P}$-even, whereas $N_i$, $\rho$, and $\phi$ are $\mathcal{P}$-odd. The lightest $\mathcal{P}$-odd state is therefore stable. Thus, the proximity of $\langle\tau\rangle$ to i affects the flavor structure but is not required for DM stability.
\begin{table}[H]
\centering
\renewcommand{\arraystretch}{1.2}
\setlength{\tabcolsep}{6pt}
\begin{tabular}{|c|c|c|c|c|c|c|c|c|c|c|}
\hline
\text{Fields} & $\bar{L}$ & $e_R$ & $\mu_R$ & $\tau_R$ & $N_1$ & $N_{2,3}$ & $H$ & $\Delta$ & $\rho$ & $\tilde{\phi}$ \\ \hline
$SU(2)_L \times U(1)_Y$ 
& $(\mathbf{2},+1)$ & $(\mathbf{1},-2)$ & $(\mathbf{1},-2)$ & $(\mathbf{1},-2)$
& $(\mathbf{1},0)$ & $(\mathbf{1},0)$ & $(\mathbf{2},1)$ & $(\mathbf{3},2)$ & $(\mathbf{2},-1)$ & $(\mathbf{2},-1)$ \\ \hline
$T'$ 
& $\mathbf{3}$ & $\mathbf{1}$ & $\mathbf{1''}$ & $\mathbf{1'}$ 
& $\mathbf{1}$ & $\mathbf{2}$ & $\mathbf{1}$ & $\mathbf{1}$ & $\mathbf{1}$ & $\mathbf{1}$ \\ \hline
$k_I$ 
& $-4$ & $4$ & $4$ & $4$ & $1$ & $2$ & $0$ & $0$ & $-1$ & $1$ \\ \hline
\end{tabular}
\caption{Field assignments under $SU(2)_L \times U(1)_Y$, $T'$ and the modular weight $k_I$.}
\label{tab:fields}
\end{table}
In this framework, the Yukawa couplings transform nontrivially under the modular symmetry and are identified with polyharmonic Maa{\ss} forms associated with the double cover modular group $T'$. The modular multiplets appearing in the model consist of triplet forms $Y^{(k)}_{3} \equiv \big(Y^{(k)}_{3,1},\, Y^{(k)}_{3,2},\, Y^{(k)}_{3,3}\big)^{T}$ with modular weights $k=0,2,4,-4$, doublet forms $Y^{(k)}_{2} \equiv \big(Y^{(k)}_{2,1},\, Y^{(k)}_{2,2}\big)^{T}$ with weights $k=1,3$, and doublet forms $Y^{(k)}_{2''} \equiv \big(Y^{(k)}_{2'',1},\, Y^{(k)}_{2'',2}\big)^{T}$ with weights $k=1,3,-3$, transforming in the $\mathbf{3}$, $\mathbf{2}$, and $\mathbf{2''}$ representations of $T'$, respectively. In addition, $Y^{(-2)}_{1}$ corresponds to the $T'$ singlet modular form. For completeness, the explicit $q$-expansions of these modular forms are provided in ~\cite{Qu:2025ddz}.
In our model, the assignment of modular weights is determined by the requirement of reproducing the observed lepton mass matrices while simultaneously forbidding tree-level seesaw contributions and ensuring DM stability. This restricts the modular forms \footnote{We emphasize that the use of polyharmonic Maa\ss\ forms in this work should not be understood as the most general consequence of modular covariance in a non-supersymmetric theory. Rather, it is a selection principle of the framework, where the holomorphy condition of supersymmetric modular models is replaced by a Laplace/polyharmonic condition together with suitable growth assumptions. If one allows completely general nonholomorphic modular covariant functions, additional independent modular structures and invariants may appear, which can reduce predictivity. This point was recently emphasized in Ref. \cite{Tavartkiladze:2025oiq}, where broad classes of nonholomorphic modular forms and the importance of residual symmetries near fixed points were discussed.} entering the construction to weights satisfying $|k|\leq 4$, as summarized in Table~\ref{tab:ModularForms}.
\begin{table}[h!]
\centering
\renewcommand{\arraystretch}{1.2}
\setlength{\tabcolsep}{6pt}
\begin{tabular}{|c|c|c|c|c|c|c|c|c|c|c|}
\hline
\text{Polyharmonic Maa{\ss} forms} & $Y_{3}^{(0)}$ & $Y_{1}^{(-2)}$ & $Y_{2''}^{(-3)}$ & $Y_{3}^{(-4)}$ & $Y_{3}^{(4)}$ & $Y_{2}^{(3)}$ & $Y_{2''}^{(3)}$ & $Y_{3}^{(2)}$ & $Y_{2}^{(1)}$ & $Y_{2''}^{(1)}$ \\ \hline
$T'$ & $\mathbf{3}$ & $\mathbf{1}$ & $\mathbf{2}^{\prime\prime}$ & $\mathbf{3}$ & $\mathbf{3}$ & $\mathbf{2}$ & $\mathbf{2}^{\prime\prime}$ & $\mathbf{3}$ & $\mathbf{2}$ & $\mathbf{2}^{\prime\prime}$ \\ \hline
$-k_I$ & $0$ & $-2$ & $-3$ & $-4$ & $4$ & $3$ & $3$ & $2$ & $1$ & $1$ \\ \hline
\end{tabular}
\caption{Polyharmonic Maa{\ss} form representations under $T'$ and their modular weights $k_I$.}
\label{tab:ModularForms}
\end{table}
Within this framework, every interaction must satisfy modular-weight conservation and $T'$ invariance; however, for the field assignments in Table~\ref{tab:fields} and the modular multiplets listed in Table~\ref{tab:ModularForms}, no modular form has simultaneously the weight and $T'$ representation required to contract $\overline{L}\widetilde{H}N$ into an invariant singlet. The corresponding Dirac Yukawa coupling is therefore forbidden, thereby excluding a tree-level type-I seesaw contribution. By contrast, although the candidate contractions $LL\Delta Y_{\mathbf1}^{(8)}$ and $LL\Delta Y_{\mathbf3}^{(8)}$ can satisfy the modular-symmetry constraints, they are forbidden as independent tree-level Yukawa interactions by the imposed $U(1)_L$ symmetry, since both $\Delta$ and the polyharmonic Maa{\ss} forms carry vanishing lepton number, while $LL\Delta$ carries $\Delta L=2$. The Majorana terms $\Lambda c_iY^{(k)}N_iN_j$, which likewise carry $\Delta L=2$, thus constitute the sole source of soft lepton-number breaking, which is transmitted radiatively to the visible lepton sector and generates the effective $LL\Delta$ interaction through the finite one-loop diagram of topology \texttt{T4-2-i}.
\subsection{Lepton mass matrices}
The gauge- and modular-invariant lepton sector Lagrangian contains lepton number conserving Yukawa interactions and soft $\Delta L = 2$ Majorana mass terms,
\begin{eqnarray}
-\mathcal{L}_{\text{leptons}} &=&
a_1\left(\bar{L} \otimes Y^{(0)}_3\right)_{\mathbf{1}}e_R H
+ a_2 \left(\bar{L} \otimes Y^{(0)}_3\right)_{\mathbf{1'}}\mu_R H
+ a_3 \left(\bar{L} \otimes Y^{(0)}_3\right)_{\mathbf{1''}}\tau_R H \nonumber \\
&& + b_1 \bar{L} \left(N_{\mathbf{2,3}} \otimes Y^{(3)}_{\mathbf{2}}\right)_{\mathbf{3}} \rho
+ b_2 \bar{L} \left(N_{\mathbf{2,3}} \otimes Y^{(3)}_{\mathbf{2''}}\right)_{\mathbf{3}} \rho
+ b_3 \left(\bar{L} \otimes Y^{(4)}_{\mathbf{3}}\right)_{\mathbf{1}} N_{\mathbf{1}} \rho \nonumber \\
&& + b_4 \bar{L} \left(N_{\mathbf{2,3}} \otimes Y^{(1)}_{\mathbf{2}}\right)_{\mathbf{3}} \tilde{\phi}
+ b_5 \bar{L} \left(N_{\mathbf{2,3}} \otimes Y^{(1)}_{\mathbf{2''}}\right)_{\mathbf{3}} \tilde{\phi} 
+ b_6 \left(\bar{L} \otimes Y^{(2)}_{\mathbf{3}}\right)_{\mathbf{1}} N_{\mathbf{1}} \tilde{\phi} \nonumber \\
&& + c_1\,\Lambda \left(N_{\mathbf{2,3}} \otimes N_{\mathbf{2,3}}\right)_{\mathbf{3}} Y^{(-4)}_{\mathbf{3}} 
+ c_2\,\Lambda N_{\mathbf{1}} N_{\mathbf{1}} Y^{(-2)}_{\mathbf{1}} 
+ c_3\,\Lambda N_{\mathbf{1}} \left(N_{\mathbf{2,3}} \otimes Y^{(-3)}_{\mathbf{2''}}\right)_{\mathbf{1}}
+ \mathrm{h.c.}
\label{lagrangian}
\end{eqnarray}
Here, $a_{1,2,3}$, $b_{1,\dots,6}$, and $c_{1,2,3}$ are complex couplings, while $\Lambda$ denotes the cutoff scale. The notation $(A \otimes B)_{\mathbf{R}}$ indicates contraction into the irreducible representation $\mathbf{R}$. After EWSB, the Yukawa interactions in Eq.~(\ref{lagrangian}) generate the mass and interaction matrices of the lepton sector. Using the $T'$ tensor product rules (see Appendix~\ref{app1}), one obtains the charged lepton Yukawa matrix $Y_e$, the heavy Majorana neutrino mass matrix $M_R$, as well as the Dirac Yukawa matrices associated with the inert scalar doublets $\rho$ and $\phi$:
\begin{eqnarray}
Y_e &=& \begin{pmatrix}
a_1 Y_{3,1}^{(0)} & a_2 Y_{3,2}^{(0)} & a_3 Y_{3,3}^{(0)} \\
a_1 Y_{3,3}^{(0)} & a_2 Y_{3,1}^{(0)} & a_3 Y_{3,2}^{(0)} \\
a_1 Y_{3,2}^{(0)} & a_2 Y_{3,3}^{(0)} & a_3 Y_{3,1}^{(0)}
\end{pmatrix}, \quad
M_R = \Lambda \begin{pmatrix}
c_2 Y_{1}^{(-2)} & c_3 Y_{2'',2}^{(-3)} & -c_3 Y_{2'',1}^{(-3)} \\
c_3 Y_{2'',2}^{(-3)} & -c_1 Y_{3,2}^{(-4)} & \frac{c_1}{\sqrt{2}} Y_{3,3}^{(-4)} \\
- c_3 Y_{2'',1}^{(-3)} & \frac{c_1}{\sqrt{2}} Y_{3,3}^{(-4)} & c_1 Y_{3,1}^{(-4)}
\end{pmatrix}, \\
Y_{\rho} &=& \begin{pmatrix}
b_3 Y_{3,1}^{(4)} & \tfrac{b_2}{\sqrt{2}} Y_{2'',2}^{(3)} & b_1 Y_{2,2}^{(3)} + \tfrac{b_2}{\sqrt{2}} Y_{2'',1}^{(3)} \\
b_3 Y_{3,3}^{(4)} & -b_1 Y_{2,1}^{(3)} & b_2 Y_{2'',2}^{(3)} \\
b_3 Y_{3,2}^{(4)} & \tfrac{b_1}{\sqrt{2}} Y_{2,2}^{(3)} - b_2 Y_{2'',1}^{(3)} & \tfrac{b_1}{\sqrt{2}} Y_{2,1}^{(3)}
\end{pmatrix}, \quad
Y_{\phi} = \begin{pmatrix}
b_6 Y_{3,1}^{(2)} & \tfrac{b_5}{\sqrt{2}} Y_{2'',2}^{(1)} & b_4 Y_{2,2}^{(1)} + \tfrac{b_5}{\sqrt{2}} Y_{2'',1}^{(1)} \\
b_6 Y_{3,3}^{(2)} & -b_4 Y_{2,1}^{(1)} & b_5 Y_{2'',2}^{(1)} \\
b_6 Y_{3,2}^{(2)} & \tfrac{b_4}{\sqrt{2}} Y_{2,2}^{(1)} - b_5 Y_{2'',1}^{(1)} & \tfrac{b_4}{\sqrt{2}} Y_{2,1}^{(1)}
\end{pmatrix}. \nonumber
\label{allYukawas}
\end{eqnarray}
The scalar doublets $\rho$ and $\phi$ are inert and therefore do not acquire vacuum expectation values, implying that $Y_{\rho}$ and $Y_{\phi}$ do not contribute to fermion masses at tree level but instead mediate radiative effects. To proceed, it is convenient to work in the basis where the heavy Majorana mass matrix is diagonal. This is achieved via a unitary transformation $U_R$ satisfying $U_R^T M_R U_R = \mathrm{diag}(M_1, M_2, M_3)$. In this basis, the Dirac Yukawa matrices become
\begin{equation}
\hat{Y}_{\rho} = Y_{\rho} U_R, \qquad \hat{Y}_{\phi} = Y_{\phi} U_R,
\end{equation}
which directly enter the loop-induced neutrino mass. Neutrino masses are then generated radiatively at one loop through the exchange of the inert scalars and heavy fermions. The resulting effective mass matrix is given by
\begin{equation}
M_\nu = - \frac{\mu_{\rho\phi}\mu_{H\Delta} v^2}{m_\Delta^2}
\sum_{i=1}^{3}
\left[
\hat y_{\rho i}\hat y_{\phi i}^{T}
+
\hat y_{\phi i}\hat y_{\rho i}^{T}
\right]
M_i\, F(M_i^2,m_\rho^2,m_\phi^2),
\label{numass}
\end{equation}
where $\hat y_{\rho i}$ and $\hat y_{\phi i}$ denote the $i$-th columns of $\hat{Y}_{\rho}$ and $\hat{Y}_{\phi}$, respectively. Although both Yukawa vertices conserve lepton number, the mass-proportional component of the internal Majorana propagator, $\langle N_iN_i \rangle\propto M_i$, provides the required $\Delta L=2$ insertion. The explicit factor of $M_i$ in Eq.~\ref{numass} therefore identifies $M_R$ as the lepton-number-violating spurion responsible for generating the effective $LL\Delta$ interaction at one loop and ensures that the light-neutrino mass vanishes in the symmetry-restoring limit $M_R\to 0$. The loop function $F(M_i^2, m_\rho^2, m_\phi^2)$ encodes the dependence on the internal masses and is given by
\begin{equation}
F(M_i^2, m_\rho^2, m_\phi^2) =
-\frac{1}{(4\pi)^2}
\left[
\frac{m_\rho^2}{m_\rho^2 - M_i^2} \ln\!\left(\frac{m_\rho^2}{M_i^2}\right)
-
\frac{m_\phi^2}{m_\phi^2 - M_i^2} \ln\!\left(\frac{m_\phi^2}{M_i^2}\right)
\right].
\end{equation}
For numerical analysis, it is convenient to factor out the overall scale by writing $M_\nu = \kappa\, \widetilde{M}_\nu$, where $\kappa = \frac{\mu_{\rho\phi}\mu_{H\Delta}}{m_\Delta^2}$ controls the absolute neutrino mass scale.
\subsection{Scalar potential}
The most general renormalizable gauge and modular invariant scalar potential involving the $SU(2)$ doublets $H, \rho, \phi$ and the triplet $\Delta=\Delta^a\sigma^a/\sqrt{2}$ is given by
\footnote{Under a modular transformation $\gamma$, a scalar
field $\phi^{(I)}$ with modular weight $k_I$ transforms as
$\phi^{(I)}\to (c\tau+d)^{-k_I}\,
\rho^{(I)}(\gamma)\,\phi^{(I)}$. Consequently, the combination
$|\phi^{(I)}|^2$ is not modular invariant by itself. The corresponding
modular invariant quadratic structure is $\frac{|\phi^{(I)}|^2} {(-i\tau+i\bar{\tau})^{k_I}} = |\phi^{(I)}|^2 (2\operatorname{Im}\tau)^{-k_I}.$ For example, for $\rho^\dagger\rho$, with $k_\rho=-1$, the invariant structure is $(-i\tau+i\bar{\tau})^{-k_\rho}(\rho^\dagger\rho) = (-i\tau+i\bar{\tau})(\rho^\dagger\rho).$ Once the modulus is fixed at its vacuum expectation value $\langle\tau\rangle$, these modulus-dependent factors become numerical constants and can be absorbed into redefinitions of the corresponding mass parameters and scalar couplings. Thus, the parameters $m_\rho^2$, $m_{\tilde{\phi}}^2$, $\lambda_\rho$, $\lambda_1'$, etc., appearing in the scalar potential should be understood as effective parameters defined after this rescaling. This convention is consistent with the treatment of scalar kinetic and quadratic terms in Refs.~\cite{Nomura:2024vzw}.}
\begin{align}
\mathcal{V} &= -\mu_H^2 H^\dagger H + m_\rho^2 \rho^\dagger \rho
    + m_\phi^2 \phi^\dagger \phi + m_\Delta^2 Tr(\Delta^\dagger \Delta) + \lambda_H (H^\dagger H)^2 + \lambda_\rho (\rho^\dagger \rho)^2 + \lambda_\phi (\phi^\dagger \phi)^2 + \lambda_\Delta Tr(\Delta^\dagger \Delta)^2 \nonumber \\
    &+ \lambda_\Delta^{\prime} [Tr(\Delta^\dagger \Delta)]^2 + \lambda_1 (H^\dagger H) (\rho^\dagger \rho) + \lambda_1^{\prime}|\rho^\dagger H|^2 + \lambda_2 (H^\dagger H) (\phi^\dagger \phi) + \lambda_2^{\prime}|\phi^\dagger H|^2 + \lambda_3|\phi^\dagger \rho|^2 \nonumber \\
    &+ \lambda_4 (H^\dagger H) Tr(\Delta^\dagger \Delta) + \lambda_5 (\rho^\dagger \rho) (\phi^\dagger \phi) + \lambda_6 (\rho^\dagger \rho) Tr(\Delta^\dagger \Delta) + \lambda_7 (\phi^\dagger \phi) Tr(\Delta^\dagger \Delta) + \lambda_8|\Delta^\dagger H|^2 \nonumber \\ &+ \lambda_9|\Delta^\dagger \rho|^2 + \lambda_{10}|\Delta^\dagger \phi|^2 + \mu_{H \Delta} (H^T i\sigma_2 \Delta^\dagger H + h.c.) + \mu_{\rho \phi} (\rho^T i\sigma_2 \Delta \tilde{\phi} + h.c.) 
\label{potential}
\end{align}
Upon electroweak symmetry breaking (EWSB), the SM Higgs doublet and the new states can be expanded in their component form as
\begin{eqnarray}
    H = \begin{pmatrix}
        \eta^+ \\ \frac{1}{\sqrt{2}}(\upsilon_H + h_d + i \omega_0) 
    \end{pmatrix}, \quad \rho = \begin{pmatrix}
        \frac{H_1 + i \omega_1}{\sqrt{2}} \\ \rho^-
    \end{pmatrix}, \quad \phi = \begin{pmatrix}
        \phi^+ \\ \frac{H_2 + i \omega_2}{\sqrt{2}}
    \end{pmatrix}, \quad \Delta = \begin{pmatrix}
        \frac{1}{\sqrt{2}} \delta^+ && \delta^{++} \\
        \frac{\upsilon_\Delta + h_t + i \omega_3}{\sqrt{2}}  && -\frac{1}{\sqrt{2}} \delta^+
    \end{pmatrix}.
    \label{scalars}
\end{eqnarray}
The mass eigenstates in the active sector --- namely the {\it CP}-even, {\it CP}-odd, and charged scalar states --- are obtained through the mixing angles $\theta_h$, $\theta_b$, and $\theta_{h^\pm}$, respectively. Similarly, the mass eigenstates in the inert sector, including the {\it CP}-even and {\it CP}-odd neutral states, are determined by the mixing angles $\theta_n$ and $\theta_a$. These mixings arise from the $\mu_{H\Delta}$, $\lambda_{4,8}$, and $\mu_{\rho\phi}$ terms in Eq.~\ref{potential}. The corresponding field rotations are given by
\begin{eqnarray}
\left(
\begin{array}{c}
h \\
H
\end{array}
\right)
&=&
\left(
\begin{array}{cc}
c_{\theta_h} & s_{\theta_h} \\
-s_{\theta_h} & c_{\theta_h}
\end{array}
\right)
\left(
\begin{array}{c}
h_d \\
h_t
\end{array}
\right),
\quad
\left(
\begin{array}{c}
G^0 \\
A^0
\end{array}
\right)
=
\left(
\begin{array}{cc}
c_{\theta_b} & s_{\theta_b} \\
-s_{\theta_b} & c_{\theta_b}
\end{array}
\right)
\left(
\begin{array}{c}
\omega_0 \\
\omega_3
\end{array}
\right),
\quad
\left(
\begin{array}{c}
G^\pm \\
H^\pm
\end{array}
\right)
=
\left(
\begin{array}{cc}
c_{\theta_{h^\pm}} & s_{\theta_{h^\pm}} \\
-s_{\theta_{h^\pm}} & c_{\theta_{h^\pm}}
\end{array}
\right)
\left(
\begin{array}{c}
\eta^\pm \\
\delta^\pm
\end{array}
\right),
\nonumber \\
\left(
\begin{array}{c}
S_1^0 \\
S_2^0
\end{array}
\right)
&=&
\left(
\begin{array}{cc}
c_{\theta_n} & s_{\theta_n} \\
-s_{\theta_n} & c_{\theta_n}
\end{array}
\right)
\left(
\begin{array}{c}
H_1 \\
H_2
\end{array}
\right),
\quad
\left(
\begin{array}{c}
A_1^0 \\
A_2^0
\end{array}
\right)
=
\left(
\begin{array}{cc}
c_{\theta_a} & s_{\theta_a} \\
-s_{\theta_a} & c_{\theta_a}
\end{array}
\right)
\left(
\begin{array}{c}
\omega_1 \\
\omega_2
\end{array}
\right).
\label{sca_mix}
\end{eqnarray}
Here, we have introduced the shorthand notation $c_{\theta_x} \equiv \cos\theta_x$ and $s_{\theta_x} \equiv \sin\theta_x$. After EWSB, the doublet and triplet Higgs fields acquire VEVs $\langle H \rangle = \upsilon_H / \sqrt{2}$ and $\langle \Delta \rangle = \upsilon_\Delta / \sqrt{2}$ such that $\upsilon = \sqrt{\upsilon_H^2 + 2 \upsilon_\Delta^2} = 246~\text{GeV}$, while the scalar doublets $\rho$ and $\phi$ remain inert with vanishing vacuum expectation values. The conditions for minimizing the potential, which determine these VEVs as functions of the potential parameters, are given by
\begin{equation}
    \mu_H^2 = \lambda_H \upsilon_H^2 + \frac{\lambda_4 + \lambda_8}{2} \upsilon_\Delta^2 - \sqrt{2} \mu_{H \Delta} \upsilon_\Delta, \quad
    m_\Delta^2 = \frac{\mu_{H \Delta} \upsilon_H^2}{\sqrt{2} \upsilon_\Delta} - \frac{\lambda_4 + \lambda_8}{2} \upsilon_H^2 - ( \lambda_\Delta + \lambda_\Delta^\prime ) \upsilon_\Delta^2
\end{equation}
After diagonalizing the mass matrices of the charged and neutral scalar fields, three states ($G^0$ and $G^\pm$) are identified as Goldstone bosons. These are absorbed through the Higgs mechanism and become the longitudinal components of the $W^\pm$ and $Z$ gauge bosons. The remaining states correspond to physical scalar particles. In the active sector (as in the usual Higgs triplet model \cite{Konetschny:1977bn,Mohapatra:1979ia,Magg:1980ut,Lazarides:1980nt,Schechter:1980gr,Cheng:1980qt}), the physical spectrum consists of two neutral {\it CP}-even scalars ($h$ and $H$), one neutral {\it CP}-odd scalar ($A$), a pair of singly charged scalars ($H^\pm$), and a pair of doubly charged scalars ($\delta^{\pm\pm}$). The mass-squared matrices for the the {\it CP}-even, {\it CP}-odd and singly charged scalars are given by
\begin{equation}
    M_{h^0,H^0}^2 = \begin{pmatrix}
        A & B \\ B & C
    \end{pmatrix}, \quad M_{G^0,A^0}^2 = \begin{pmatrix}
        D & E \\ E & F
    \end{pmatrix}, \quad M_{G^\pm,H^\pm}^2 = \begin{pmatrix}
        I & J \\ J & K
    \end{pmatrix},
    \label{squa_mat}
\end{equation}
where the matrix elements are explicitly expressed as
\begin{eqnarray}
    A &=& 2 \lambda_H \upsilon_H^2, \quad B = (\lambda_4 + \lambda_8)\upsilon_H \upsilon_\Delta - \sqrt{2} \mu_{H \Delta} \upsilon_H, \quad C = 2(\lambda_\Delta + \lambda_\Delta^\prime) \upsilon_\Delta^2 + \frac{\mu_{H\Delta} \upsilon_H^2}{\sqrt{2}\upsilon_\Delta} \nonumber \\
    D &=& 2\sqrt{2} \mu_{H \Delta} \upsilon_\Delta, \quad E = -\sqrt{2} \mu_{H \Delta} \upsilon_H, \quad F = \frac{\mu_{H\Delta} \upsilon_H^2}{\sqrt{2} \upsilon_\Delta} \\
    I &=& \sqrt{2} \mu_{H \Delta} \upsilon_\Delta - \frac{\lambda_8 \upsilon_\Delta^2}{2}, \quad J = \frac{\lambda_8}{2 \sqrt{2}} \upsilon_H \upsilon_\Delta - \mu_{H \Delta} \upsilon_H, \quad K = \frac{\mu_{H\Delta }\upsilon_H^2}{\sqrt{2} \upsilon_\Delta} - \frac{\lambda_8 \upsilon_H^2}{4}. \nonumber
    \label{rot_diag}
\end{eqnarray}
These symmetric matrices are diagonalized by the orthogonal transformations defined in the first line of Eq.~\ref{sca_mix}. The resulting mass eigenvalues for the physical scalar states are
\begin{eqnarray}
    m_{h^0, H^0}^2 &=& \frac{1}{2} (A+C~\mp \sqrt{(A-C)^2~+ 4~B^2}), \quad m_{A^0}^2 = \frac{\mu_{H \Delta}}{\sqrt{2} \upsilon_\Delta} ( \upsilon_H^2 + 4 \upsilon_\Delta^2), \nonumber \\ m_{H^\pm}^2 &=& \frac{\mu_{H\Delta}}{\sqrt{2} \upsilon_\Delta} (\upsilon_H^2 + 2 \upsilon_\Delta^2) - \lambda_8 (\frac{\upsilon_H^2}{4} + \frac{\upsilon_\Delta^2}{2}), \quad m_{\delta^{\pm\pm}}^2 = \frac{\mu_{H \Delta} \upsilon_H^2}{\sqrt{2} \upsilon_\Delta} - \lambda_\Delta \upsilon_\Delta^2 - \frac{\lambda_8 \upsilon_H^2}{2}
    \label{phys_mass}
\end{eqnarray}
where the mass of the doubly charged scalar is obtained directly from the scalar potential by identifying the coefficient of the bilinear term $\delta^{++}\delta^{--}$. All mixing angles satisfy
\begin{equation}
    \tan2\theta_X = \frac{2 [M_{m,n}^2]_{12}}{[M_{m,n}^2]_{22} - [M_{m,n}^2]_{11}},
\end{equation}
for $[\theta_h, M_{h^0, H^0}^2]$, $[\theta_b, M_{G^0, A^0}^2]$, and $[\theta_{h^\pm}, M_{G^\pm, H^\pm}^2]$. In this study, we identify $h^0$ as the SM-like Higgs with $m_{h^0} = 125~\text{GeV}$. After algebraic manipulations of Eqs. \ref{squa_mat}–\ref{phys_mass}, we require $\mu_{H\Delta} > 0$ to ensure that $A^0$ is non-tachyonic. In addition, the mixing angles in the {\it CP}-odd and charged scalar sectors take the simplified forms $\tan \theta_b = 2 \upsilon_\Delta / \upsilon_H$, and $\tan \theta_{h^\pm} = \sqrt{2} \upsilon_\Delta / \upsilon_H$.

The inert sector consists of two neutral {\it CP}-even scalars ($S_1^0$, $S_2^0$), two neutral {\it CP}-odd scalars ($A_1^0$, $A_2^0$), and two pairs of singly charged scalars ($\rho^\pm$, $\phi^\pm$). The mass-squared matrices for the neutral {\it CP}-even and {\it CP}-odd states are identical and can be written as
\begin{equation} M_{S_{1,2}^0,A_{1,2}^0}^2 = \begin{pmatrix} P & Q \\ Q & S \end{pmatrix} \text{ with } P = m_\rho^2 + \frac{\lambda_1 \upsilon_H^2}{2} + \frac{\lambda_6 \upsilon_\Delta^2}{2}, \quad Q = \mu_{\rho \phi} \upsilon_\Delta / \sqrt{2}, \quad S = m_\phi^2 + \frac{\lambda_2 + \lambda_2^\prime}{2} \upsilon_H^2 + \frac{\lambda_7 + \lambda_{10}}{2} \upsilon_\Delta^2. \label{squa_mat2} 
\end{equation}
Since the {\it CP}-even and {\it CP}-odd mass matrices are the same, they are diagonalized by the same orthogonal transformation, implying $\theta_n = \theta_a$. Consequently, each {\it CP}-even state is degenerate in mass with its {\it CP}-odd partner, and the mass eigenvalues are given by
\begin{equation}
m_{S_1^0}^2 = m_{A_1^0}^2 = \frac{1}{2}\left(P + S - \sqrt{(P - S)^2 + 4 Q^2}\right), \quad m_{S_2^0}^2 = m_{A_2^0}^2 = \frac{1}{2}\left(P + S + \sqrt{(P - S)^2 + 4 Q^2}\right).
\end{equation}
This degeneracy follows from the fact that the scalar potential is {\it CP}-conserving and does not contain any operators that differentiate between the real and imaginary parts of the neutral inert fields. As a result, the quadratic terms for the {\it CP}-even and {\it CP}-odd components are identical, leading to equal mass matrices and hence degenerate eigenvalues.\newline
In contrast to the neutral sector, there is no mixing in the inert charged scalar sector. Therefore, $\rho^\pm$ and $\phi^\pm$ are already mass eigenstates. Their masses are obtained directly from the scalar potential by extracting the coefficients of the bilinear terms $\rho^+ \rho^-$ and $\phi^+ \phi^-$, respectively
\begin{equation}
m_{\rho^\pm}^2 = m_\rho^2 + \frac{\lambda_1 + \lambda_1^\prime}{2}\upsilon_H^2 + \frac{\lambda_6 + \lambda_9}{2}\upsilon_\Delta^2,
\quad
m_{\phi^\pm}^2 = m_\phi^2 + \frac{\lambda_2}{2} \upsilon_H^2 + \frac{\lambda_7}{2} \upsilon_\Delta^2.
\end{equation}
The independent parameters of the scalar sector are chosen as follows
\begin{equation}
    \{ m_{h^0,H^0},~m_{A^0},~m_{H^\pm},~m_{\delta^{\pm\pm}},~m_{S_{1,2}^0},~m_{A_{1,2}^0},~m_{\rho^\pm},~m_{\phi^\pm},~\lambda_{1,2,3,5,6,7},~\lambda_{1,2}^\prime,~\upsilon_\Delta,~\theta_{h,a,b,n,h^\pm} \}
\end{equation}
The remaining parameters are expressed as 
\begin{eqnarray}
    \lambda_H &=& \frac{1}{2 \upsilon_H^2} (c_h^2 m_h^2 + s_h^2 m_H^2), \quad \mu_{H \Delta} = \frac{\sqrt{2}\upsilon_\Delta}{\upsilon_H^2 + 4 \upsilon_\Delta^2} m_{A^0}^2, \quad \mu_{\rho \phi}=\frac{\sqrt{2}}{\upsilon_\Delta} c_n s_n (m_{S_2^0}^2 - m_{S_1^0}^2) \nonumber \\
    \lambda_4 &=& \frac{1}{\upsilon_H \upsilon_\Delta} \left[s_h c_h (m_H^2 - m_h^2) + \sqrt{2} \mu_{H\Delta} \upsilon_H \right], \quad \lambda_8 = \frac{\frac{\mu_{H\Delta}}{\sqrt{2} \upsilon_\Delta} (\upsilon_H^2 + 2 \upsilon_\Delta^2) - m_{H^\pm}^2}{\frac{\upsilon_H^2}{4} + \frac{\upsilon_\Delta^2}{2}} \nonumber \\
    \lambda_\Delta &=& \frac{1}{\upsilon_\Delta^2} \left[ \frac{\mu_{H \Delta} \upsilon_H^2}{\sqrt{2} \upsilon_\Delta} - \frac{\lambda_8 \upsilon_H^2}{2} - m_{\delta^{\pm\pm}}^2 \right], \quad \lambda_\Delta^\prime = \frac{1}{2 \upsilon_\Delta^2} \left[ s_h^2 m_h^2 + c_h^2 m_H^2 - \frac{\mu_{H \Delta} \upsilon_H^2}{\sqrt{2} \upsilon_\Delta} \right] - \lambda_\Delta \\
    \lambda_9 &=& \frac{1}{\upsilon_\Delta^2} \left[ 2(m_{\rho^\pm}^2 - c_n^2 m_{S_1^0}^2 - s_n^2 m_{S_2^0}^2) - \lambda_1^\prime \upsilon_H^2 \right], \quad \lambda_{10} = \frac{1}{\upsilon_\Delta^2} \left[ 2(s_n^2 m_{S_1^0}^2 + c_n^2 m_{S_2^0}^2 - m_{\phi^\pm}^2) - \lambda_2^\prime \upsilon_H^2 \right] \nonumber \\
    m_\rho^2 &=& c_n^2 m_{S_1^0}^2 + s_n^2 m_{S_2^0}^2 - \frac{\lambda_1 \upsilon_H^2}{2} - \frac{\lambda_6 \upsilon_\Delta^2}{2}, \quad m_\phi^2 = m_{\phi^\pm}^2 - \frac{\lambda_2 \upsilon_H^2}{2} - \frac{\lambda_7 \upsilon_\Delta^2}{2}.  \nonumber
\end{eqnarray}
The expressions associated with the active sector are consistent with those given in Ref.~\cite{Arhrib:2011uy}.
\section{Theoretical and experimental constraints}
\subsection{Perturbativity and vacuum stability}
To ensure that the theory remains within the perturbative regime, we impose bounds on the quartic couplings appearing in the scalar potential. Perturbativity require the quartic couplings to be less than $4\pi$, i.e, $|\lambda_i| < 4\pi$. In addition, the scalar potential must be bounded from below in order to be physically viable. This means that it must not approach negative infinity in any direction of field space. At large field values, the asymptotic behavior of the potential is dominated by the quartic terms, while cubic contributions become negligible. We therefore apply co-positivity conditions to the quartic part of the potential to guarantee stability in all field directions~\cite{Kannike:2012pe,Kannike:2016fmd}.
\newline
For field configurations in which only a single multiplet acquires a nonzero value, the bounded-from-below conditions reduce to
\begin{eqnarray}
    \lambda_H > 0, \quad \lambda_\rho > 0, \quad 
    \lambda_\phi > 0, \quad \lambda_\Delta^{\text{eff}} > 0,~~ \text{where}~~    \lambda_\Delta^{\text{eff}} = \min\!\left(\lambda_\Delta + \lambda_\Delta^\prime,\; \frac{\lambda_\Delta}{2} + \lambda_\Delta^\prime \right).
    \label{bfb1}
\end{eqnarray}
For directions in which two scalar multiplets are simultaneously nonzero, additional constraints arise. When two doublets are active, the conditions are
\begin{eqnarray}
    \tilde{\lambda}_{H\rho} + 2\sqrt{\lambda_H \lambda_\rho} > 0, \quad 
    \tilde{\lambda}_{H\phi} + 2\sqrt{\lambda_H \lambda_\phi} > 0, \quad 
    \tilde{\lambda}_{\rho\phi} + 2\sqrt{\lambda_\rho \lambda_\phi} > 0,
\end{eqnarray}
with
\begin{eqnarray}
    \tilde{\lambda}_{H\rho} = \lambda_1 + \min(0,\lambda_1^\prime), \quad
    \tilde{\lambda}_{H\phi} = \lambda_2 + \min(0,\lambda_2^\prime), \quad
    \tilde{\lambda}_{\rho\phi} = \lambda_5 + \min(0,\lambda_3).
\end{eqnarray}
When one doublet and the triplet $\Delta$ are nonzero, the stability conditions become
\begin{eqnarray}
    \tilde{\lambda}_{H\Delta} + 2\sqrt{\lambda_H \lambda_\Delta^{\text{eff}}} > 0, \quad 
    \tilde{\lambda}_{\rho\Delta} + 2\sqrt{\lambda_\rho \lambda_\Delta^{\text{eff}}} > 0, \quad 
    \tilde{\lambda}_{\phi\Delta} + 2\sqrt{\lambda_\phi \lambda_\Delta^{\text{eff}}} > 0,
\end{eqnarray}
where
\begin{eqnarray}
    \tilde{\lambda}_{H\Delta} = \lambda_4 + \min(0,\lambda_8), \quad
    \tilde{\lambda}_{\rho\Delta} = \lambda_6 + \min(0,\lambda_9), \quad
    \tilde{\lambda}_{\phi\Delta} = \lambda_7 + \min(0,\lambda_{10}).
    \label{bfb2}
\end{eqnarray}
Collectively, Eqs. \ref{bfb1}–\ref{bfb2} provide necessary bounded-from-below conditions by ensuring positivity of the quartic potential along the one- and two-multiplet field directions considered above. These pairwise conditions do not, by themselves, constitute a sufficient bounded-from-below criterion for completely general field configurations in which three or more multiplets are simultaneously nonzero, nor do they establish that the electroweak stationary point is the global minimum of the full scalar potential.

The bounded-from-below conditions are supplemented by tree-level partial-wave unitarity constraints. In the high-energy limit, scalar masses and dimensionful trilinear couplings are negligible in $2\to2$ scalar scattering, so the amplitudes are determined entirely by the quartic interactions. For each eigenvalue $e_a$ of the coupled-channel scattering matrix, the $s$-wave condition
$\left|\operatorname{Re}a_{0,a}\right|<\frac12$, with  $a_{0,a}=e_a/16\pi$, implies
\begin{equation}
    |e_a| < 8\pi.
    \label{eigenval}
\end{equation} 
The complete set of scattering channels and the corresponding eigenvalues is presented in Appendix D. Applying Eq.~\ref{eigenval} point by point confirms that all parameter points retained in our numerical scans, for both neutrino-mass orderings, satisfy tree-level perturbative unitarity.
\subsection{The decay $h \rightarrow \gamma\gamma$ in the \texttt{T4-2-i} model}
One important phenomenological consequence of the scalar sector in our model is its effect on the Higgs diphoton decay channel, $h\to\gamma\gamma$. In the SM, this loop-induced process is dominated by the $W^\pm$ boson contribution, with an additional subleading contribution from the top quark loop \cite{Ellis:1975ap,Shifman:1979eb,Okun:1982ap,Gavela:1981ri}. In the present model, these amplitudes are modified by the extra charged scalars. We therefore consider the ratio of the diphoton decay width to its SM prediction
\begin{equation}
    R_{\gamma\gamma}
    =
    \frac{\Gamma(h \rightarrow \gamma \gamma)^{\texttt{T4-2-i}}}
    {\Gamma(h \rightarrow \gamma \gamma)^{\rm SM}} .
\end{equation}
This observable has been measured at the LHC. In particular, the ATLAS Collaboration reports \cite{ATLAS:2022vkf}, $R_{\gamma\gamma} = 1.088^{+0.095}_{-0.089}$. In the present model, the partial decay width for $h \rightarrow \gamma \gamma$ is given by
\begin{equation}
    \Gamma(h \rightarrow \gamma \gamma)^{\texttt{T4-2-i}}
    =
    \frac{\alpha^2 G_F m_h^3}{128\sqrt{2}\pi^3}
    \left|
    \kappa_W A_1(\tau_W)
    +
    \frac{4}{3}\kappa_t A_{1/2}(\tau_t)
    +
    \sum_S Q_S^2 \kappa_S A_0(\tau_S)
    \right|^2 ,
    \label{hgaga}
\end{equation}
where $\tau_i = m_h^2/(4m_i^2)$, and $Q_S$ denotes the electric charge of the charged scalar $S$. The dimensionless loop functions are
\begin{eqnarray}
    A_1(\tau_W) &=& - \left[ 2\tau_W^3 + 3\tau_W + 3(2\tau_W - 1) f(\tau_W) \right]\tau_W^{-2}, \quad A_{1/2}(\tau_t) = 2 \left[ \tau_t + (\tau_t - 1) f(\tau_t)
    \right]\tau_t^{-2},
    \nonumber \\
    A_0(\tau_S) &=& - \left[ \tau_S - f(\tau_S) \right]\tau_S^{-2}.
\end{eqnarray}
Since $h$ is not a purely SM-like Higgs boson, the first two terms in Eq.~\ref{hgaga}, representing the SM loop contributions, differ from their SM values due to the modified couplings of $h$ to the $W^\pm$ bosons and the top quark. The relevant reduced couplings (relative to the SM ones) are given by
\begin{equation}
    \kappa_W = \frac{g_{h W^\pm W^\mp}^{\texttt{T4-2-i}}}{g_{h W^\pm W^\mp}^{SM}} = \frac{c_{\theta_h} \upsilon_H + 2 s_{\theta_h} \upsilon_\Delta}{\upsilon}, \quad \kappa_t = \frac{g_{ht\bar{t}}^{\texttt{T4-2-i}}}{g_{ht\bar{t}}^{SM}} = \frac{c_{\theta_h}}{c_{\theta_h^\pm}}
\end{equation}
The summation in the last term of Eq. \ref{hgaga} is being performed over all the physical charged scalars of the model, i.e., $H^\pm,~\delta^{\pm\pm},~\rho^\pm, \phi^\pm$. Explicitly, we have
\begin{eqnarray}
    \sum_S Q_S^2\kappa_S A_0(\tau_S) &=& \frac{m_W}{g_2 m_{H^\pm}^2} g_{h H^\pm H^\mp} A_0(\tau_{H^\pm})~+~4 \frac{m_W}{g_2 m_{\delta^{\pm \pm}}^2} g_{h \delta^{\pm\pm} \delta^{\mp \mp}} A_0(\tau_{\delta^{\pm \pm}}) \nonumber \\ &+& \frac{m_W}{g_2 m_{\rho^\pm}^2} g_{h \rho^\pm \rho^\mp} A_0(\tau_{\rho^\pm})~+~\frac{m_W}{g_2 m_{\phi^\pm}^2} g_{h \phi^\pm \phi^\mp} A_0(\tau_{\phi^\pm}),
\end{eqnarray}
For later use in the numerical discussion, it is convenient to absorb the electric-charge factor appearing in Eq.~\ref{hgaga} into scalar-specific reduced coefficients. Comparing the previous expression term by term, we define
\begin{align}
\kappa_{H^\pm}
&\equiv
\frac{m_W}{g_2m_{H^\pm}^2}\,
g_{hH^\pm H^\mp},
&
\kappa_{\delta^{\pm\pm}}
&\equiv
4\,\frac{m_W}{g_2m_{\delta^{\pm\pm}}^2}\,
g_{h\delta^{\pm\pm}\delta^{\mp\mp}},
\nonumber\\
\kappa_{\rho^\pm}
&\equiv
\frac{m_W}{g_2m_{\rho^\pm}^2}\,
g_{h\rho^\pm\rho^\mp},
&
\kappa_{\phi^\pm}
&\equiv
\frac{m_W}{g_2m_{\phi^\pm}^2}\,
g_{h\phi^\pm\phi^\mp}.
\label{eq:scalar_kappas_hgg}
\end{align}
These quantities are not additional free parameters; they are only a compact way of displaying the charged-scalar contributions to the diphoton amplitude. In particular, the factor of $4$ in $\kappa_{\delta^{\pm\pm}}$ comes from the doubly charged state, $Q_{\delta^{\pm\pm}}^2=4$. These are the same coefficients used as the color palettes in the $R_{\gamma\gamma}$ plots below. The trilinear couplings entering the definitions are
\begin{eqnarray}
    g_{h~H^\pm~H^\mp} &=& \left[ -2 \mu_{H\Delta } c_{\theta_h^\pm} s_{\theta_h} - \lambda_4 \upsilon_H  c_{\theta_h^\pm}^2 - \frac{\lambda_8 \upsilon_H}{\sqrt{2}} c_{\theta_h^\pm}^2 - 2 \lambda_H \upsilon_H s_{\theta_h^\pm}^2 - \frac{\lambda_8 \upsilon_\Delta}{\sqrt{2}} c_{\theta_h^\pm} s_{\theta_h^\pm} \right]  c_{\theta_h} \nonumber \\
    &-& \left[ 2(\lambda_\Delta + \lambda_\Delta^\prime) \upsilon_\Delta c_{\theta_h^\pm}^2 + \lambda_4 \upsilon_\Delta  s_{\theta_h^\pm}^2 - \frac{\lambda_8 \upsilon_H}{\sqrt{2}} c_{\theta_h^\pm} s_{\theta_h^\pm}  \right] s_{\theta_h} \nonumber \\
    g_{h~\delta^{\pm\pm}~\delta^{\mp\mp}} &=& - \lambda_4 \upsilon_H c_{\theta_h} - 2 (\lambda_\Delta + \lambda_\Delta^\prime) \upsilon_\Delta s_{\theta_h} , \quad \quad g_{h~\phi^\pm~\phi^\mp} = - \lambda_2 \upsilon_H c_{\theta_h} - \lambda_7 \upsilon_\Delta s_{\theta_h} \\
    g_{h~\rho^\pm~\rho^\mp} &=& -(\lambda_1 + \lambda_1^\prime) \upsilon_H c_{\theta_h} -(\lambda_6 + \lambda_9) \upsilon_\Delta s_{\theta_h} \nonumber
\end{eqnarray}
\subsection{Oblique parameters}
\label{EWPO}
New-physics corrections to the gauge-boson self-energies are quantified by the oblique shifts $\Delta S$, $\Delta T$, and $\Delta U$ relative to a specified SM reference point~\cite{Peskin:1991sw}. The shifts $\Delta S$ and $\Delta T$ are particularly sensitive to mass splittings among the scalar states. For $\Delta U=0$, the global electroweak fit is centered at $(\Delta S_{\rm fit},\Delta T_{\rm fit})=(-0.05,0.00)$, with standard deviations $\sigma_S=0.07$ and $\sigma_T=0.06$~\cite{ParticleDataGroup:2024cfk}.
Both tree-level and loop-level effects contribute to modifications of these parameters. At tree level, changes in the VEV affect the masses of the $W$ and $Z$ bosons, which in turn modify the electroweak $\rho$ parameter, defined as the ratio of the $W$ and $Z$ boson masses. In our framework, the modification of the electroweak gauge boson masses follows the same structure as in the SM extended by an $SU(2)_L$ scalar triplet $\Delta$ (the type-II seesaw scenario). The relevant terms arise from the scalar kinetic Lagrangian
\begin{equation}
\mathcal{L}_{\text{kin}} = (D_\mu H)^\dagger (D^\mu H) +
\text{Tr}\left[(D_\mu \Delta)^\dagger (D^\mu \Delta)\right],
\end{equation}
where the covariant derivatives are defined as
\begin{equation}
D_\mu H = \left(\partial_\mu - i g_2 \tau^a W_\mu^a - i g_1 Y_H B_\mu\right) H,
\qquad
D_\mu \Delta = \partial_\mu \Delta - i g_2 [\tau^a W_\mu^a, \Delta] - i g_1 Y_\Delta B_\mu \Delta.
\end{equation}
After EWSB, the VEVs $v_H$ and $v_\Delta$ generate the gauge boson masses
\begin{equation}
m_W^2 = \frac{g_2^2}{4}\left(v_H^2 + 2 v_\Delta^2\right),
\qquad
m_Z^2 = \frac{g_2^2}{4 c_W^2}\left(v_H^2 + 4 v_\Delta^2\right),
\end{equation}
with $c_W = \cos\theta_W$. The electroweak scale is therefore fixed by $v^2 \equiv v_H^2 + 2 v_\Delta^2 \simeq (246\ \mathrm{GeV})^2$. The tree-level $\rho$ parameter is given by
\begin{equation}
\rho \equiv \frac{m_W^2}{m_Z^2 c_W^2}
= \frac{v_H^2 + 2 v_\Delta^2}{v_H^2 + 4 v_\Delta^2}.
\end{equation}
In the phenomenologically relevant regime $v_\Delta^2 \ll v_H^2$, this expression reduces to $\rho \simeq 1 - 2v_\Delta^2 / v_H^2$. Hence, unlike in the SM, a non-vanishing triplet VEV induces a deviation of $\rho$ from unity already at tree level. Defining $\delta\rho \equiv \rho - 1 \simeq -2v_\Delta^2 / v^2$, one finds that precision electroweak measurements impose a stringent upper bound on $v_\Delta$. The most recent global electroweak fit reported by the Particle Data Group yields $\rho_0 = 1.00031 \pm 0.00019$ \cite{ParticleDataGroup:2024cfk}, where $\rho_0 = 1$ corresponds to the SM prediction. Since the present framework predicts $\rho \le 1$, compatibility with the experimental result at the 95\% confidence level implies $v_\Delta \lesssim 1.45~\text{GeV}$ for $v_H \simeq 246\ \mathrm{GeV}$. The tree-level custodial-breaking measure may equivalently be written as $\delta\rho/\alpha \simeq -2\upsilon_\Delta^2/(\alpha\upsilon_H^2)$, where $\alpha$ is the electromagnetic fine-structure constant.

Since $\rho^{\rm tree}\neq1$, the electroweak sector requires four independent input parameters, with the fourth input providing an additional renormalization condition at one loop. A convenient on-shell choice is  $\{\alpha_{\rm em},G_F,m_Z,\hat s_W^2\}$, where $\hat s_W^2$ is fixed by the effective leptonic weak mixing angle at the $Z$ pole~\cite{Blank:1997qa,Kanemura:2012rs}. At tree level these inputs determine the triplet vacuum expectation value through
\begin{equation}
 v_\Delta^2=
 \frac{\hat s_W^2\hat c_W^2}{2\pi\alpha_{\rm em}}m_Z^2
 -\frac{1}{2\sqrt{2}G_F}.
 \label{eq:vdelta-four-input}
\end{equation}
At one loop the counterterm $\delta\hat s_W^2$, or equivalently $\delta v_\Delta$, is fixed by the renormalization condition associated with the fourth input, while physical predictions such as $m_W$ and $\rho_{\rm EW}$ are obtained from the complete renormalized gauge-boson self-energies together with the corresponding vertex and box corrections~\cite{Blank:1997qa,Kanemura:2012rs}. Because custodial symmetry is already broken at tree level, the conventional active doublet–triplet contribution to $\Delta T$ is not a separately ultraviolet-finite prediction~\cite{Drauksas:2023isd}; a complete next-to-leading-order electroweak analysis must instead be formulated directly in terms of physical observables within a specified four-input renormalization scheme.

The numerical study adopts a reduced electroweak diagnostic. The exact tree-level expression for $\rho_{\rm EW}$ is imposed as a separate constraint. Loop effects from the active doublet--triplet sector are expanded about the $v_\Delta\to0$ custodial reference theory; only the finite leading contribution to $\Delta S$ is retained,
\begin{equation}
    \Delta S_{H\Delta}^{(0)} = -\frac{1}{3\pi} \ln\frac{m_{\delta^{++}}^2}{m_{H^0}^2} - \frac{2}{\pi} \left\{ (1 - 2 s_W^2)^2 \xi(x_{\delta^{++}}, x_{\delta^{++}}) + s_W^4 \xi(x_{H^+}, x_{H^+}) + \xi(x_{H^0}, x_{H^0}) \right\},
\end{equation}
where $x_X=m_X^2/m_Z^2$ and the loop function $\xi(X,Y)$ is given in Appendix~\ref{app2}~\cite{Lavoura:1993nq}. A consistent active-sector contribution to $\Delta T$ belongs to the complete one-loop calculation in the four-input renormalization scheme described above.

For the inert sector, whose doublets have vanishing vacuum expectation values, the one-loop contributions to $T$ and $S$ are finite. The one-loop inert contribution is given by
\begin{align}
\Delta T_{\text{inert}} &= \frac{1}{16 \pi s_W^2 c_W^2}
\Bigg\{ c_n^2 F(m_{\rho^\pm}^2, m_{S_1^0}^2) + s_n^2 F(m_{\rho^\pm}^2, m_{S_2^0}^2)
+ s_n^2 F(m_{\phi^\pm}^2, m_{S_1^0}^2) + c_n^2 F(m_{\phi^\pm}^2, m_{S_2^0}^2)
\nonumber \\
&\quad + c_a^2 F(m_{\rho^\pm}^2, m_{A_1^0}^2) + s_a^2 F(m_{\rho^\pm}^2, m_{A_2^0}^2)
+ s_a^2 F(m_{\phi^\pm}^2, m_{A_1^0}^2) + c_a^2 F(m_{\phi^\pm}^2, m_{A_2^0}^2)
\nonumber \\
&\quad - c_n^2 s_a^2 F(m_{S_1^0}^2, m_{A_2^0}^2) - s_n^2 c_a^2 F(m_{S_2^0}^2, m_{A_1^0}^2)
\Bigg\}.
\label{eq:T-inert-degenerate}
\end{align}
where we used the fact that the exact mass degeneracies $m_{A_1^0}=m_{S_1^0}$ and $m_{A_2^0}=m_{S_2^0}$ imply $F(m_{S_1^0}^2,m_{A_1^0}^2)=F(m_{S_2^0}^2,m_{A_2^0}^2)=0$. The inert contribution to $\Delta S$ is
\begin{align}
\Delta S_{\text{inert}} &= \frac{1}{24 \pi}
\Bigg\{
(2 s_W^2 - 1)^2 \Big[ G(m_{\rho^\pm}^2, m_{\rho^\pm}^2, m_Z^2) + G(m_{\phi^\pm}^2, m_{\phi^\pm}^2, m_Z^2) + G(m_{\rho^\pm}^2, m_{\phi^\pm}^2, m_Z^2)
\Big]
\nonumber \\
&\quad + c_n^2 c_a^2 G(m_{S_1^0}^2, m_{A_1^0}^2, m_Z^2) + c_n^2 s_a^2 G(m_{S_1^0}^2, m_{A_2^0}^2, m_Z^2) + s_n^2 c_a^2 G(m_{S_2^0}^2, m_{A_1^0}^2, m_Z^2) + s_n^2 s_a^2 G(m_{S_2^0}^2, m_{A_2^0}^2, m_Z^2)
\nonumber \\
&\quad + c_n^2 \ln\!\left(\frac{m_{S_1^0}^2}{m_{\rho^\pm}^2}\right)
+ s_n^2 \ln\!\left(\frac{m_{S_2^0}^2}{m_{\phi^\pm}^2}\right)
+ c_a^2 \ln\!\left(\frac{m_{A_1^0}^2}{m_{\rho^\pm}^2}\right)
+ s_a^2 \ln\!\left(\frac{m_{A_2^0}^2}{m_{\phi^\pm}^2}\right)
\Bigg\}
\end{align}
where the loop functions $F(X,Y)$ and $G(X,Y,Z)$ are given in Appendix~\ref{app2}.

For the numerical scan, the quantities
\begin{equation}
    \Delta S=\Delta S_{H\Delta}^{(0)}+\Delta S_{\rm inert},
    \qquad
    \Delta T=\Delta T_{\rm inert}
    \label{reduced-oblique}
\end{equation}
define a reduced oblique diagnostic. It retains the finite inert-doublet contributions and the leading custodial-reference contribution $\Delta S_{H\Delta}^{(0)}$, while the active-sector $T$ term is left outside this two-parameter quantity. This diagnostic assumes $\Delta U=0$, negligible non-oblique vertex and box corrections, and scalar masses sufficiently above the electroweak scale for the oblique expansion to be applicable. In our numerical study, the correlated likelihood is constructed from Eq.~\eqref{reduced-oblique}, using $\Delta S_{\rm fit}=-0.05\pm0.07$ and $\Delta T_{\rm fit}=0.00\pm0.06$~\cite{ParticleDataGroup:2024cfk}. It is used together with the exact tree-level $\rho_{\rm EW}$ constraint and is not identified with a complete one-loop prediction in the four-input scheme. Appendix~\ref{app5} summarizes the resulting allowed parameter ranges.
\subsection{Charged lepton flavor violation processes}
\label{cLFV}
The only source of cLFV in our model is the pair of Yukawa interactions $LN\rho$ and $LN\tilde{\phi}$. Consequently, all cLFV observables arise at one loop from the same particles that generate neutrino mass, namely the heavy Majorana neutrinos $N_i$ and the charged components of the inert doublets $\rho$ and $\phi$. By contrast, the triplet $\Delta$ enters neutrino masses only through the scalar sector and has no direct tree-level couplings to charged leptons, so its direct contribution to cLFV amplitudes is absent at leading order.
\newline
In components, the charged lepton interactions with the charged scalars are derived from Eq. \ref{lagrangian}
\begin{equation}
-\mathcal{L}_{\text{int}} \supset  \overline{\ell_\alpha} P_R N_i
\left[ (\hat Y_\rho)_{\alpha i}\,\rho^- - (\hat Y_{\phi})_{\alpha i} \phi^-  \right] + \text{h.c.},
\qquad \alpha=e,\mu,\tau,
\label{yrpint}
\end{equation}
so that the radiative decay $\ell_\alpha \to \ell_\beta \gamma$ is generated at one loop by the exchange of $N_i$ together with the charged inert scalars $\rho^\pm$ and $\phi^\pm$.
The corresponding effective dipole operator can be written as
\begin{equation}
\mathcal{L}_{\rm eff}
=
e\,m_{\ell_\alpha}\,
\overline{\ell_\beta}\,
\sigma^{\mu\nu}
\left(
A_L^{\beta\alpha} P_L + A_R^{\beta\alpha} P_R
\right)
\ell_\alpha\,
F_{\mu\nu}
+ \text{h.c.}
\end{equation}
For the chirality structure of Eq.~\ref{yrpint}, only one dipole coefficient is generated at leading order, namely $A_L^{\beta\alpha} \simeq 0$ and
\begin{equation}
A_R^{\beta\alpha}
= 
\frac{1}{32\pi^2}
\sum_{i=1}^3
\left[
\frac{(\hat Y_\rho)_{\beta i} (\hat Y_\rho)_{\alpha i}^\ast}{m_{\rho^\pm}^2}
F_2\!\left(\frac{M_i^2}{m_{\rho^\pm}^2}\right)
+
\frac{(\hat Y_{\phi})_{\beta i} (\hat Y_{\phi})_{\alpha i}^\ast}{m_{\phi^\pm}^2}
F_2\!\left(\frac{M_i^2}{m_{\phi^\pm}^2}\right)
\right],
\label{eq:ARclfv}
\end{equation}
where the loop function is listed in Appendix \ref{app2} together with the off-shell photon and box functions used below. Hence, the branching ratio for the radiative decay $l_\alpha \rightarrow l_\beta \gamma$ is given by
\begin{align}
{\rm Br}(\ell_\alpha \to \ell_\beta \gamma)
=
\frac{3\alpha_{\rm em}}{64\pi G_F^2}
\Bigg|
\sum_{i=1}^3
\Bigg[
\frac{(\hat Y_\rho)_{\beta i} (\hat Y_\rho)_{\alpha i}^\ast}{m_{\rho^\pm}^2}
F_2\!\left(\frac{M_i^2}{m_{\rho^\pm}^2}\right)
+
\frac{(\hat Y_{\phi})_{\beta i} (\hat Y_{\phi})_{\alpha i}^\ast}{m_{\phi^\pm}^2}
F_2\!\left(\frac{M_i^2}{m_{\phi^\pm}^2}\right)
\Bigg]
\Bigg|^2
{\rm Br}(\ell_\alpha \to \ell_\beta \nu_\alpha \overline{\nu_\beta}).
\label{eq:BRclfv}
\end{align}
In our numerical analysis, we impose the current experimental upper bounds on the cLFV decays (see Table \ref{tab:clfv}) which provide some of the strongest limits on the Yukawa structures $\widehat{Y}_\rho$ and $\widehat{Y}_\phi$ that enters both the cLFV branching ratio formulas and the neutrino mass matrix. In relation to this, since the radiative neutrino mass depends on the mixing product $\widehat Y_\rho {\rm diag} (M_iF_i) \widehat Y_{\phi}^T$ whereas cLFV amplitudes depends on the Hermitian combinations $\widehat Y_\rho \widehat Y_\rho^\dagger$ and $\widehat Y_\phi \widehat Y_\phi^\dagger$, neutrino masses can remain small even when the cLFV rates are phenomenologically relevant. From Eq.~\ref{numass}, the couplings in the heavy-neutrino mass basis are $\hat Y_\rho = Y_\rho U_R$ and $\hat Y_{\phi}=Y_{\phi}U_R$, so that all flavor violation is encoded in the columns of these two matrices. In particular, the first column is associated with the $T'$-odd singlet $N_1$, whereas the second and third columns originate from the $T'$ doublet $(N_2,N_3)$. Therefore, once the modulus $\tau$ is fixed, the flavor pattern of each column is no longer arbitrary but is sharply constrained by the modular symmetry.
\begin{table}[t]
\centering
\renewcommand{\arraystretch}{1.15}
\begin{tabular}{lllll}
\hline
\textbf{Process} & \textbf{Current bound} & \textbf{Reference} & \textbf{Future sensitivity} & \textbf{Reference} \\
\hline
$\mathrm{Br}(\mu\to e\gamma)$ & $1.5\times10^{-13}$ & \cite{MEGII:2025muegamma} & $6.0\times10^{-14}$ & \cite{Baldini:2018uhj} \\
$\mathrm{Br}(\tau\to e\gamma)$ & $3.3\times10^{-8}$ & \cite{BaBar:2010taulgamma} & $3.0\times10^{-9}$ & \cite{Belle-II:2018jsg,Aoki:2025clfv} \\
$\mathrm{Br}(\tau\to \mu\gamma)$ & $4.2\times10^{-8}$ & \cite{Belle:2021taugamma} & $1.0\times10^{-9}$ & \cite{Belle-II:2018jsg,Aoki:2025clfv} \\
$\mathrm{Br}(\mu\to 3e)$ & $1.0\times10^{-12}$ & \cite{SINDRUM:1988mu3e} & $2\times10^{-15}$ (Phase I), $10^{-16}$ (Phase II) & \cite{Blondel:2021fji,Aoki:2025clfv} \\
$\mathrm{Br}(\tau\to 3e)$ & $2.7\times10^{-8}$ & \cite{Belle:2010tau3l} & $5.0\times10^{-10}$ & \cite{Belle-II:2018jsg} \\
$\mathrm{Br}(\tau\to 3\mu)$ & $2.1\times10^{-8}$ & \cite{Belle:2010tau3l} & $4.0\times10^{-10}$ & \cite{Belle-II:2018jsg} \\
${\rm CR}(\mu\!-\!e,\mathrm{Ti})$ & $4.3\times10^{-12}$ & \cite{ParticleDataGroup:2024cfk} & $10^{-18}$ (PRISM/PRIME benchmark) & \cite{Aoki:2025clfv} \\
${\rm CR}(\mu\!-\!e,\mathrm{Au})$ & $7.0\times10^{-13}$ & \cite{ParticleDataGroup:2024cfk} & -- & -- \\
${\rm CR}(\mu\!-\!e,\mathrm{Pb})$ & $4.6\times10^{-11}$ & \cite{ParticleDataGroup:2024cfk} & -- & -- \\
\hline
\end{tabular}
\caption{Current bounds and future sensitivities for the charged-lepton-flavor-violating observables discussed in this work. All current bounds are quoted at $90\%$ C.L. The future entries refer to the standard design or benchmark sensitivities most commonly adopted in current phenomenological studies.}
\label{tab:clfv}
\end{table}

The same Yukawa interactions that induce the radiative decays also generate the three-body modes $\mu\to 3e$, $\tau\to 3e$, and $\tau\to 3\mu$, as well as the $\mu$--$e$ conversion rate in a target of atomic nuclei. Unlike the radiative modes, these observables are sensitive not only to the on-shell dipole form factor, but also to off-shell photon penguin, Z-penguin, and box contributions. The relevant present limits and future sensitivities are collected in Table \ref{tab:clfv}. In the present model the charged lepton interactions with the inert charged scalars are of the form $\overline{\ell_\alpha}P_R N_i\, s^-$ with $s=\rho,\phi$, so the dominant lepton flavor violating form factors are right-handed. The three-body channels then depend on the set of form factors $\{A_D,A_{ND},F_{RR},F_{RL},B\}$, where $A_{ND}$ denotes the off-shell photon coefficient, $F_{RR}$ and $F_{RL}$ the $Z$-penguin coefficients, and $B$ the box contribution. Each of these quantities receives separate $\rho^\pm$ and $\phi^\pm$ contributions, while the box amplitude also contains mixed $(\rho,\phi)$ terms. Using the standard notation of Refs.~\cite{Toma:2013zsa,Kitano:2002mt}, the branching ratio for $\ell_\alpha\to \ell_\beta\bar\ell_\beta\ell_\beta$ can be written as
\begin{eqnarray}
\mathrm{Br}(\ell_\alpha \to \ell_\beta \bar{\ell}_\beta \ell_\beta)
 &=&
\frac{3(4\pi)^2\alpha_{\rm em}^2}{8G_F^2}
\Bigg[ |A_{ND}|^2 + |A_D|^2 \left( \frac{16}{3}\ln\frac{m_{\ell_\alpha}}{m_{\ell_\beta}} -\frac{22}{3} \right)
 + \frac{1}{3}\left(2|F_{RR}|^2+|F_{RL}|^2\right) + \frac{1}{6}|B|^2
\nonumber \\
 &+& \left( - 2A_{ND}A_D^\ast + \frac{1}{3}A_{ND}B^\ast - \frac{2}{3}A_DB^\ast + \mathrm{h.c.}  \right) \Bigg]
\times \mathrm{Br}(\ell_\alpha\to \ell_\beta\nu_\alpha\bar{\nu}_\beta),
\label{eq:3bodyfull}
\end{eqnarray}
Here $A_D$ is the same dipole coefficient already introduced in Eq.~(\ref{eq:ARclfv}), $A_{ND}$ is the off-shell photon form factor proportional to the loop function $G_2$, and $B$ denotes the sum of the pure and mixed box terms built from the loop functions $D_1$ and $D_2$. The explicit loop functions are collected in Appendix~\ref{app2}. In practice the $Z$-penguin contributions are numerically smaller than the dipole and box pieces because of the additional weak-coupling and charged-lepton-mass suppressions, but they are retained for completeness. For coherent $\mu$--$e$ conversion, the full rate is target dependent and involves matching the effective dipole, vector, scalar, and box operators onto nucleon matrix elements, followed by the corresponding nuclear overlap integrals~\cite{Kitano:2002mt}. Since the phenomenological role of conversion in the present work is to provide an additional cLFV consistency test, we adopt the standard dipole-dominant gold benchmark
\begin{equation}
{\rm CR}(\mu\text{-}e,{\rm Au})\simeq 8\times 10^{-3}\,{\rm Br}(\mu\to e\gamma),
\label{eq:mueconvdipole}
\end{equation}
which reproduces the usual order-of-magnitude relation between dipole-dominated conversion in gold and $\mu\to e\gamma$. We then impose the SINDRUM~II gold limit searched for $\mu \rightarrow e$ conversion on gold and given in Table \ref{tab:clfv} \cite{ParticleDataGroup:2024cfk}. A more refined treatment should replace Eq.~(\ref{eq:mueconvdipole}) by the full target-dependent conversion formula.
\subsection{Dark matter in \texttt{T4-2-i} model}
\begin{figure}[H]
\centering
\includegraphics[width=0.49\linewidth]{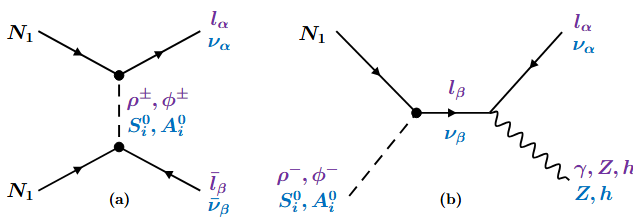}\hfill
\includegraphics[width=0.49\linewidth]{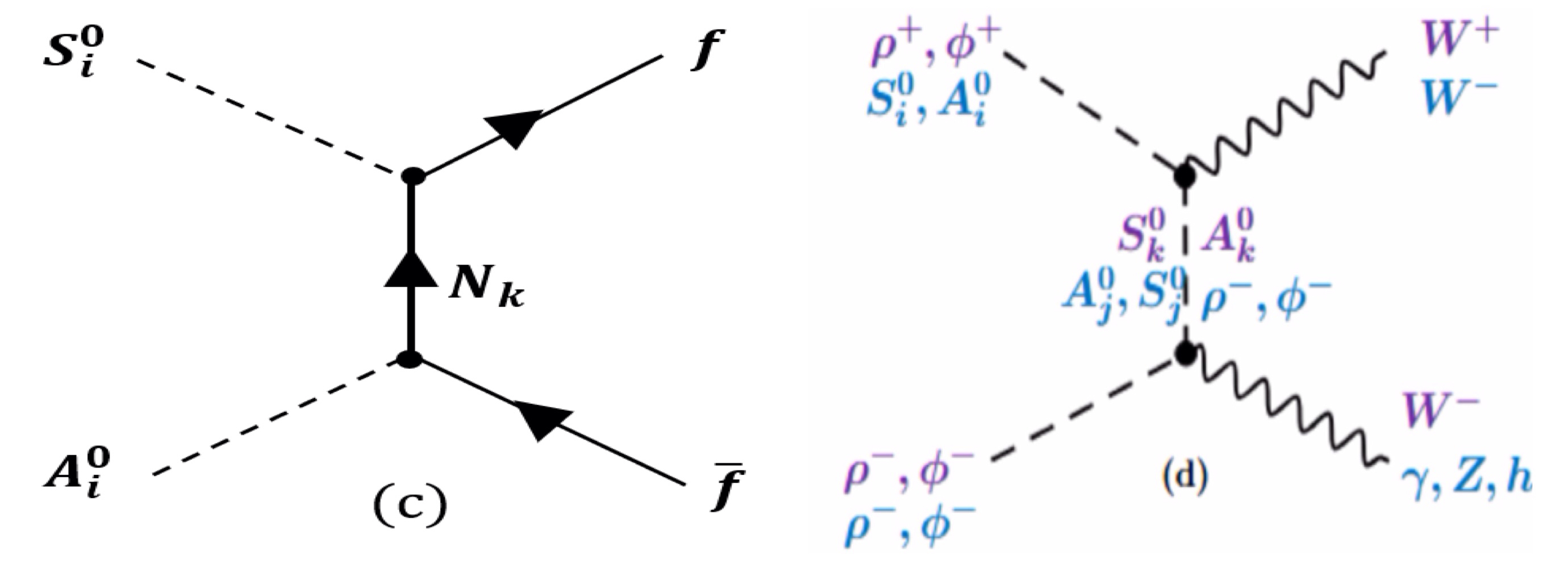}
\caption{Annihilation and coannihilation topologies contributing to $\langle \sigma_{\rm eff} v\rangle$. Panels (a) and (b) represent pure $N_1N_1$ annihilation and mixed $N_1$--inert coannihilation, respectively, while panels (c) and (d) shows scalar--scalar coannihilation into fermionic and bosonic final states, respectively.}
\label{fig2}
\end{figure}

Consider a set of particles $\chi_i$ ($i=1,\ldots,p$) that are odd under the exact modular parity $\mathcal{P}$ defined in Eq.~\eqref{eq:modular-parity}. Since this parity remains unbroken throughout the scanned region of $\tau$, the lightest odd particle is stable. The model therefore admits two DM possibilities. If $N_1$ is the lightest odd state, the DM is a Majorana fermion. If instead the lightest odd state lies in the neutral inert sector, the DM is the exactly degenerate complex scalar associated with the lighter inert pair $\Phi_1^0=\frac{1}{\sqrt2}\left(S_1^0+iA_1^0\right)$. In the present study, we focus on the fermionic DM by taking $N_1$ to be the lightest odd state. The relic-density and direct-detection formulas then follow the standard WIMP and coannihilation framework~\cite{Jungman:1995df,Arcadi:2017kky}. For a dark sector in thermal equilibrium with the SM bath, the total odd-sector number density obeys the usual Boltzmann equation \cite{Lee:1977ua,Kolb:1990vq,Gondolo:1990dk,Griest:1990kh}
\begin{equation}
\frac{dn}{dt}+3Hn=
-\langle\sigma_{\rm eff} \upsilon\rangle
\left(n^2-n_{\rm eq}^2\right),
\label{eq:boltzmann_dm}
\end{equation}
where $H$ is the Hubble parameter, $n_{\rm eq}$ is the equilibrium total number density, and $\langle\sigma_{\rm eff}v_{\rm rel}\rangle$ includes all annihilation and coannihilation channels involving $N_1$ and the odd states that remain thermally populated at freeze-out. In the standard freeze-out approximation one may write \cite{Lee:1977ua}
\begin{equation}
\Omega_{\rm DM}h^2
\simeq
\frac{3\times10^{-26}\ {\rm cm^3\,s^{-1}}}
{\langle\sigma_{\rm eff} \upsilon \rangle},
\label{eq:omega_approx_dm}
\end{equation}
so an excessively large annihilation rate yields an underabundant relic density, while a too-small rate leads to overabundance. When the inert partners are close in mass to $N_1$, the relevant quantity is the coannihilation-weighted effective cross section \cite{Griest:1990kh}
\begin{equation}
\langle\sigma_{\rm eff} \upsilon \rangle
=
\sum_{i,j}
\langle\sigma_{ij}v_{ij}\rangle r_i r_j,
\quad
r_i=
\frac{g_i(1+\Delta_i)^{3/2}e^{-x_f\Delta_i}}
{\sum_k g_k(1+\Delta_k)^{3/2}e^{-x_f\Delta_k}}, \quad \text{where} \quad 
\Delta_i=\frac{m_i-M_{N_1}}{M_{N_1}},
~~ \text{and} ~~
x_f=\frac{M_{N_1}}{T_f},
\label{sigmaeff}
\end{equation}
Here $T_f$ denotes the freeze-out temperature and $g_i$ is the number of internal degrees of freedom of the coannihilating species $\chi_i$. The Boltzmann weights $r_i$ show explicitly that states with $\Delta_i\ll1$ remain thermally relevant at freeze-out. The effective cross section in Eq.~\ref{sigmaeff} is not determined by $N_1N_1$ annihilation alone. Once the inert neutral and charged scalars remain sufficiently close in mass to $N_1$, the freeze-out dynamics receives important contributions from mixed fermion--scalar coannihilation and from scalar--scalar coannihilation among the thermally populated odd states. In the present model these channels are induced by the same Yukawa and electroweak interactions that connect the odd sector to leptons, gauge bosons, and the Higgs field. The corresponding process classes are summarized in Fig.~\ref{fig2}.

Direct detection probes the elastic scattering of $N_1$ on nuclei. Since $N_1$ is an electroweak-singlet Majorana fermion, it has no tree-level $Z$ or Higgs interaction with quarks. The leading spin-independent amplitude is instead generated by a loop-induced Higgs portal, which may be parameterized as
\begin{equation}
{\cal L}_{\rm eff}\supset
-\frac12\,g_{hN_1N_1}\,h\,\overline{N_1^c}N_1,
\end{equation}
with charged and neutral inert scalars running in the loop together with SM leptons. At the nucleon level one may write schematically
\begin{equation}
\sigma_{\rm SI}^{N_1}
\simeq
\frac{\mu_N^2}{\pi}
\left[
\frac{f_Nm_N}{v\,m_h^2}\,
g_{hN_1N_1}
\right]^2,
\label{si_fermion}
\end{equation}
where $\mu_N=m_NM_{N_1}/(m_N+M_{N_1})$ is the $N_1$--nucleon reduced mass and $f_N$ is the scalar nucleon form factor. The complete one-loop expression for $g_{hN_1N_1}$ in the present model is collected in Appendix~\ref{app3}.
\begin{figure}[H]
\centering
\includegraphics[width=0.9\linewidth]{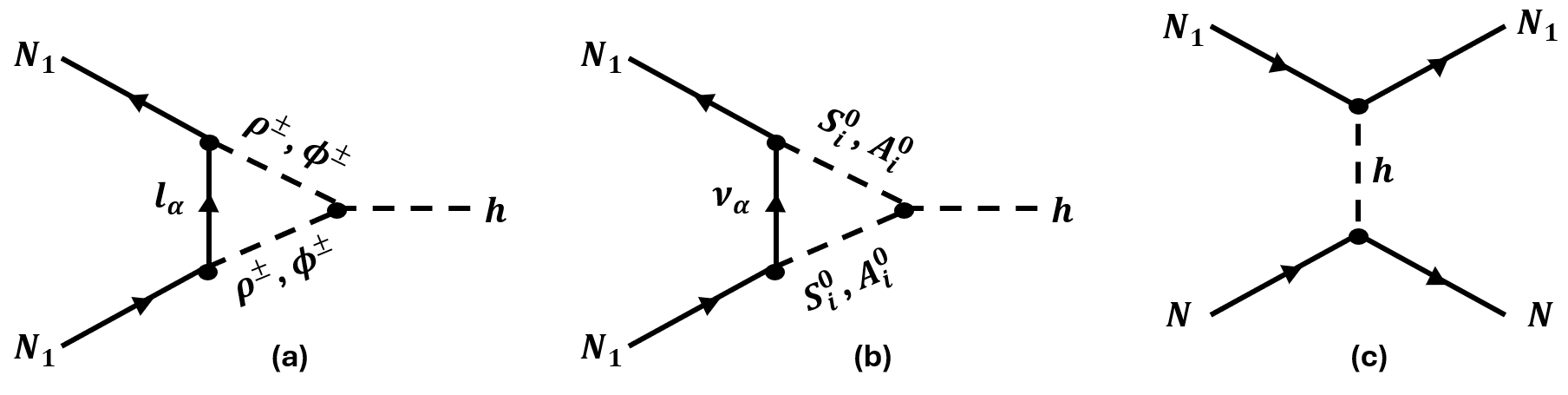}
\caption{Representative one-loop origin of the effective $hN_1N_1$ coupling and the resulting Higgs-mediated SI scattering on a nucleon.}
\label{fig3}
\end{figure}
Figure \ref{fig3} displays the scalar part of the elastic-scattering amplitude. The first loop topology contains charged inert scalars and charged leptons, whereas the second contains neutral inert scalars and neutrinos. Both generate $g_{hN_1N_1}^{\rm eff}$, followed by Higgs exchange with the nucleon. Eq. \eqref{si_fermion} therefore gives the SI cross section associated with the scalar nonrelativistic operator $\mathcal O_1$ and does not include the recoil responses induced by photon or $Z$-boson penguins. The relevant nonrelativistic operators are
\begin{equation}
 \mathcal O_1=\mathbf{1}_\chi\mathbf{1}_N,
 \qquad
 \mathcal O_8=\mathbf S_\chi\cdot\mathbf v^\perp,
 \qquad
 \mathcal O_9=i\,\mathbf S_\chi\cdot
 \left(\mathbf S_N\times\frac{\mathbf q}{m_N}\right),
 \label{NR-operators}
\end{equation}
Here $\chi\equiv N_1$, $N$ denotes a nucleon, $\mathbf S_\chi$ and $\mathbf S_N$ are the corresponding spin operators, $\mathbf q$ is the momentum transfer, and $\mathbf v^\perp$ is the transverse relative velocity satisfying $\mathbf v^\perp\cdot\mathbf q=0$. The vertices relevant to these amplitudes follow directly from Eqs.~\eqref{lagrangian}, \eqref{scalars}, and \eqref{sca_mix}. In the heavy-neutrino mass basis, the lepton--odd-sector interactions are
\begin{equation}
-\mathcal L_Y \supset
 \overline{\ell_\alpha}P_RN_i
 \left[(\widehat Y_\rho)_{\alpha i}\rho^-
       -(\widehat Y_\phi)_{\alpha i}\phi^-\right]
 +\overline{\nu_\alpha}P_RN_i
 \left[y^S_{\alpha i j}S_j^0+y^A_{\alpha i j}A_j^0\right]
 +\mathrm{h.c.},
 \label{dd-yukawa-vertices}
\end{equation}
where
\begin{eqnarray}
y^S_{\alpha i1} =& \frac{c_n(\widehat Y_\rho)_{\alpha i}+s_n(\widehat Y_\phi)_{\alpha i}}{\sqrt2}, \qquad
y^S_{\alpha i2} =& \frac{-s_n(\widehat Y_\rho)_{\alpha i}+c_n(\widehat Y_\phi)_{\alpha i}}{\sqrt2},\nonumber\\
y^A_{\alpha i1} =& \frac{i[c_a(\widehat Y_\rho)_{\alpha i}-s_a(\widehat Y_\phi)_{\alpha i}]}{\sqrt2}, \qquad
y^A_{\alpha i2} =& -\frac{i[s_a(\widehat Y_\rho)_{\alpha i}+c_a(\widehat Y_\phi)_{\alpha i}]}{\sqrt2}.
\end{eqnarray}
Through these interactions, the photon couples to the charged lepton and to $\rho^\pm,\phi^\pm$, while the $Z$ boson couples to charged leptons, neutrinos, and charged inert scalars, and also to the derivative neutral-scalar current $Z_\mu S_i^0\overleftrightarrow{\partial^\mu}A_j^0$. Up to an overall sign fixed by the covariant-derivative convention, its mixing coefficient is
\begin{equation}
C_{ij}=(R_n)_{i1}(R_a)_{j1}-(R_n)_{i2}(R_a)_{j2}.
\label{ZSA-coefficient}
\end{equation}
Because the $N_i$ are electroweak singlets and do not mix with the active neutrinos, no tree-level $Z N_i N_j$ interaction is present; nevertheless, the vertices above generate photon- and $Z$-penguin amplitudes in addition to the Higgs penguin~\cite{Schmidt:2012yg,Ibarra:2016dlb,Herrero-Garcia:2018koq}.
For an on-shell Majorana field,
\begin{equation}
 \overline{N_1}\gamma^\mu N_1=0, \qquad \overline{N_1}\sigma^{\mu\nu}N_1=0,
 \qquad
 \overline{N_1}\sigma^{\mu\nu}\gamma_5N_1=0,
 \label{majorana-bilinears}
\end{equation}
so the diagonal vector current and the electric and magnetic dipole form factors vanish identically. The only allowed diagonal electromagnetic form factor is the anapole interaction~\cite{Herrero-Garcia:2018koq},
\begin{equation}
 {\cal L}_{\rm anapole} = {\cal A}_{N_1} \overline{N_1}\gamma^\mu\gamma_5N_1 \partial^\nu F_{\mu\nu}.
 \label{anapole}
\end{equation}
After matching to the nuclear electromagnetic current, Eq.~\ref{anapole} induces the non-relativistic operators $\mathcal O_8$ and $\mathcal O_9$ defined in Eq.~\ref{NR-operators}. The operator $\mathcal O_8$ is proportional to the transverse relative velocity but can receive coherent enhancement through its coupling to the nuclear charge, whereas $\mathcal O_9$ depends on the momentum transfer and nucleon spin. Neither operator is included in the Higgs-mediated scalar cross section given in Eq.~\ref{si_fermion}. The $Z$-penguin amplitude also contains the axial Majorana current $\bar N_1\gamma^\mu\gamma^5N_1$, whose contraction with the quark vector current matches onto $\mathcal O_8$ and $\mathcal O_9$, while its contraction with the quark axial current generates the leading spin–spin structure $\mathbf S_\chi\cdot\mathbf S_N$. The relative sizes of the Higgs-, photon-, and $Z$-mediated contributions depend on the Yukawa matrices, scalar mixing angles, and mass splittings and must therefore be determined numerically. Accordingly, the curves based on Eq.~\ref{si_fermion} represent only the Higgs-mediated scalar contribution, while a combined recoil-spectrum analysis incorporating the anapole and $Z$-mediated interactions lies beyond the scope of the present study.
\section{Numerical results}
\label{sec4}
\subsection{Scan setup}
Our numerical analysis focuses on the fermionic DM candidate $N_1$. At each scan point, the polyharmonic Maa{\ss} forms are evaluated at the corresponding value of $\tau$ and used to construct the charged-lepton Yukawa matrix $Y_e$, the leptonic Yukawa structures $Y_\rho$ and $Y_\phi$ associated with the $\bar{L}N\rho$ and $\bar{L}N\tilde{\phi}$ interactions, as well as the heavy Majorana mass matrix $M_R$. The scan is performed in the vicinity of $\tau=i$, where a residual $\mathbb{Z}_2$ subgroup of $T'$ is preserved. This choice ensures the stability of the DM candidate while retaining a predictive flavor framework. In this region, the model successfully reproduces the observed neutrino mass spectrum and lepton mixing pattern. As a result, neutrino and dark sectors are governed by a reduced set of modular parameters, giving rise to strong correlations among low energy observables. Our numerical analysis focuses on the fermionic DM candidate $N_1$. For each scan point we evaluate the polyharmonic Maa\ss\ forms at the chosen $\tau$ and use them to construct the charged lepton Yukawa matrix $Y_e$, the leptonic Yukawa matrices $Y_\rho$ and $Y_\phi$ associated with the $\bar L N\rho$ and $\bar L N\tilde\phi$ interactions, and the heavy Majorana neutrino mass matrix $M_R$. The charged lepton matrix is diagonalized by singular value decomposition, while $M_R$ and the loop induced light neutrino mass matrix $M_\nu$ are diagonalized by Takagi factorization. The scalar spectrum is obtained independently from the scalar potential after EWSB. The model is implemented in {\tt FeynRules}~\cite{Alloul:2013bka}, which produces the corresponding {\tt CalcHEP} model
files~\cite{Belyaev:2012qa}, and then passed to {\tt micrOMEGAs}~6.2.3~\cite{Belanger:2013oya,Alguero:2023zol} to compute the relic abundance, the direct-detection likelihood, and the annihilation/coannihilation rates from the same Lagrangian input. For each neutrino mass ordering we generate an initial Latin-hypercube sample~\cite{mckay2000comparison} with the SciPy quasi-Monte Carlo utilities~\cite{virtanen2020scipy} and subsequently perform three local rounds around the best-ranked points, giving $10660$ evaluated points per ordering.

The scan retained for the final analysis is parametrized in terms of a reduced set of physical inputs. The odd neutral sector is described by the lightest inert mass $M_{S_1}$ and the inert splitting $ \Delta M_S^{\rm inert}\equiv M_{S_2}-M_{S_1}=M_{A_2}-M_{A_1}$.
The relations $M_{A_1}=M_{S_1}$, $M_{A_2}=M_{S_2}$, and $\theta_a=\theta_n$ are imposed exactly, so that the scanned mass and mixing ranges are
\begin{eqnarray}
100 &\le& M_{S_1}\le 1200~{\rm GeV}, \qquad
300\le M_{A^0}\le 3000~{\rm GeV}, \qquad
100\le \Lambda\le 10^4~{\rm GeV}, \nonumber \\
2 &\le& \Delta M_S^{\rm inert}\le 9~{\rm GeV}\ ({\rm NO}), \qquad
5 \le \Delta M_S^{\rm inert}\le 9~{\rm GeV}\ \ ({\rm IO}), \qquad 0.24 \le \upsilon_\Delta \le 0.40~{\rm GeV}.
\end{eqnarray}
The ranges for $M_{S_1}$ and $M_{A^0}$ are chosen to cover both the sub-TeV region and the multi-TeV decoupling regime, whereas the restricted interval for $\Delta M_S^{\rm inert}$ reflects the coannihilation pattern required by fermionic DM. The moderate range for $v_\Delta$ preserves the triplet interpretation of the active scalar sector and helps maintain electroweak-precision consistency. All remaining dimensionless couplings are scanned in perturbative domains; in particular, the quartics controlling the triplet and inert sectors are chosen so that the reconstructed spectrum remains flexible while the scalar potential stays perturbative. Rather than scanning all masses independently, the charged inert states $m_{\rho^\pm}$ and $m_{\phi^\pm}$, together with the scalar masses $M_{H^0}$, $M_{H^\pm}$, and $m_{\delta^{\pm\pm}}$, are reconstructed from the scanned inputs and the target quartics entering the scalar potential. The broad consistency windows
\begin{equation}
50\le M_{S_{1,2}},\,m_{\rho^\pm},\,m_{\phi^\pm}\le 1500~{\rm GeV},\qquad
100\le M_{H^0},\,M_{H^\pm},\,m_{\delta^{\pm\pm}},\,M_{A^0}\le 5000~{\rm GeV},
\end{equation}
are imposed together with $1\le M_{N_i}\le 10^4$~GeV. To keep $N_1$ as the dark matter candidate and at the same time remain in the coannihilation regime, we require the gap to the lightest odd scalar to satisfy $M_{\rm odd}^{\rm lightest} - M_{N_1} \in [6 - 9]\rm GeV$. This strategy keeps the active sector broad, avoids scanning highly correlated masses independently, and preserves the compressed odd spectra that are phenomenologically relevant in the fermionic DM sector.

At each sampled point the six complex coefficients $b_i$ entering the one-loop light neutrino mass matrix are readjusted so that the model fits the oscillation observables $\{\Delta m_{21}^2,\Delta m_{3\ell}^2,\sin^2\theta_{12},\sin^2\theta_{23},\sin^2\theta_{13},\delta_{\rm CP}\}$. For $\Delta m_{3l}^2$, $l=1$ for NO and $l=2$ for IO. The fit is performed with {\tt scipy.optimize.least\_squares}~\cite{virtanen2020scipy}, using the trust-region reflective algorithm~\cite{Branch:1999}. This parametrization is numerically convenient because the radiative neutrino sector naturally favors hierarchical couplings spanning several orders of magnitude. To compare the viable points in a uniform way, we define the total chi-square function
\begin{equation}
\chi^2_{\rm tot}
=
\chi^2_{\Omega}
+\chi^2_{\nu}
+\chi^2_{ST}
+\chi^2_{\gamma\gamma}
+\chi^2_{\ell},
\label{eq:chi2tot}
\end{equation}
This quantity measures how close a given point is to the preferred experimental values of the observables that admit a pull-type treatment. It includes the relic density, the neutrino oscillation observables, the oblique parameters $\Delta S$ and $\Delta T$, the Higgs diphoton branching ratio, and the charged-lepton masses. A smaller value of $\chi_{\rm tot}^2$ therefore indicates a better overall agreement with these data. By contrast, observables that are currently constrained mainly by upper bounds are not added to Eq. \ref{eq:chi2tot}; they are imposed separately as hard cuts. In this way $\chi_{\rm tot}^2$ is used to order
the accepted points, while the full selection still respects all present exclusion limits. The different contributions are defined in the usual way. The relic-density term is taken as a Gaussian pull around the Planck target value, while $\chi_{\ell}^2$ measures the quality of the charged-lepton fit,
\begin{equation}
\chi^2_{\Omega}
=
\left(
\frac{\Omega h^2-0.120}{0.001}
\right)^2,
\qquad
\chi^2_{\ell}
=
\sum_{\alpha=e,\mu,\tau}
\frac{\big[\ln(m_\alpha^{\rm fit}/m_\alpha^{\rm exp})\big]^2}{10^{-8}},
\end{equation}
The oscillation term $\chi_\nu^2$ is the sum of the six pulls associated with $\Delta m_{21}^2$, $\Delta m_{3\ell}^2$, $\sin^2\theta_{12}$, $\sin^2\theta_{23}$, $\sin^2\theta_{13}$, and $\delta_{\rm CP}$. Each of these is computed with respect to the NuFIT 6.0 best-fit value. Since the quoted allowed ranges are asymmetric, we use different upper and lower $1\sigma$ widths extracted from the published $3\sigma$ intervals \cite{Esteban:2024eli}. The diphoton term $\chi_{\gamma\gamma}^2$ is defined in the same way from the ATLAS best-fit value of $R_{\gamma\gamma}$ and its asymmetric uncertainty. For the oblique shifts we use the correlated form
\begin{equation}
\chi^2_{ST} = \Delta X^{\rm T}C_{ST}^{-1}\Delta X, \qquad
\Delta X=(\Delta S-\Delta S_{\rm fit},\Delta T-\Delta T_{\rm fit})^{\rm T},
\end{equation}
Here $(\Delta S_{\rm fit},\Delta T_{\rm fit})=(-0.05,0.00)$ is the experimental best-fit point, $\sigma_S=0.07$ and $\sigma_T=0.06$ are the corresponding standard deviations, and $C_{ST}$ is the covariance matrix containing these uncertainties and their correlation. In this way, $\chi^2_{\rm tot}$ combines the observables that admit a meaningful pull interpretation, while quantities constrained only by one-sided bounds are treated separately. In practice, a point is kept only if it satisfies the relic-density window, the NuFIT $3\sigma$ ranges, the correlated oblique bound, the allowed ATLAS range for $R_{\gamma\gamma}$, the cosmological condition $\sum_i m_i<0.12$~eV, the direct-detection likelihood returned by {\tt micrOMEGAs}, and the present cLFV bounds summarized in Table~\ref{tab:clfv}.

The oscillation observables entering the fit are extracted from the Pontecorvo-Maki-Nakagawa-Sakata (PMNS) matrix $U$ according to $\sin^2\theta_{13}=|U_{e3}|^2$, $\sin^2\theta_{12}=|U_{e2}|^2 / (1-|U_{e3}|^2)$, and $\sin^2\theta_{23}=|U_{\mu3}|^2 /( 1-|U_{e3}|^2)$, while the Dirac phase is read in the standard PMNS parametrization~\cite{ParticleDataGroup:2024cfk}. After this fit, we also evaluate the derived neutrino observables that do not enter $\chi^2_{\rm tot}$ but are important for the phenomenological discussion, namely the effective Majorana mass $m_{\beta\beta}$ in neutrinoless double beta decay experiments, the effective neutrino mass $m_\beta$ in beta decay experiments and the sum of three neutrino masses $\sum m_i$ from cosmological observations defined respectively as $m_{\beta\beta} = |\sum_{i=1}^3 U_{ei}^2~m_i|$, $m_\beta = (\sum_{i=1}^3 |U_{ei}|^2m_i^2)^{1/2}$, and $\sum_i m_i=m_1+m_2+m_3$.
\subsection{Results}
Here, we summarize the results obtained from the scan and present scatter plots for the main parameters and observables discussed in the previous sections. In all plots, the blue points represent scan points that fail at least one of the imposed phenomenological requirements, whereas the red points correspond to points that satisfy the full set of constraints. The star marks the point with the smallest value of $\chi^2_{\rm tot}$ in each neutrino mass ordering. In the fermionic-DM regime considered here, the scan points that reproduce the observed relic abundance are found predominantly in a region where $N_1$ and the lightest odd scalar are nearly degenerate. The resulting compressed odd-sector spectrum enhances coannihilation processes, which control the relic abundance over most of the viable parameter space. After imposing the neutrino-oscillation constraints, the reduced electroweak precision diagnostic, the Higgs diphoton constraint, the cosmological limit on $\sum m_\nu$, direct-detection limits, and the cLFV bounds, only a restricted subset of the scanned points remains viable. This reduction reflects the correlations between the relic-density requirement, the compressed spectrum, and the other phenomenological constraints imposed on the model.
\begin{figure}[!]
\centering
\includegraphics[width=0.32\linewidth]{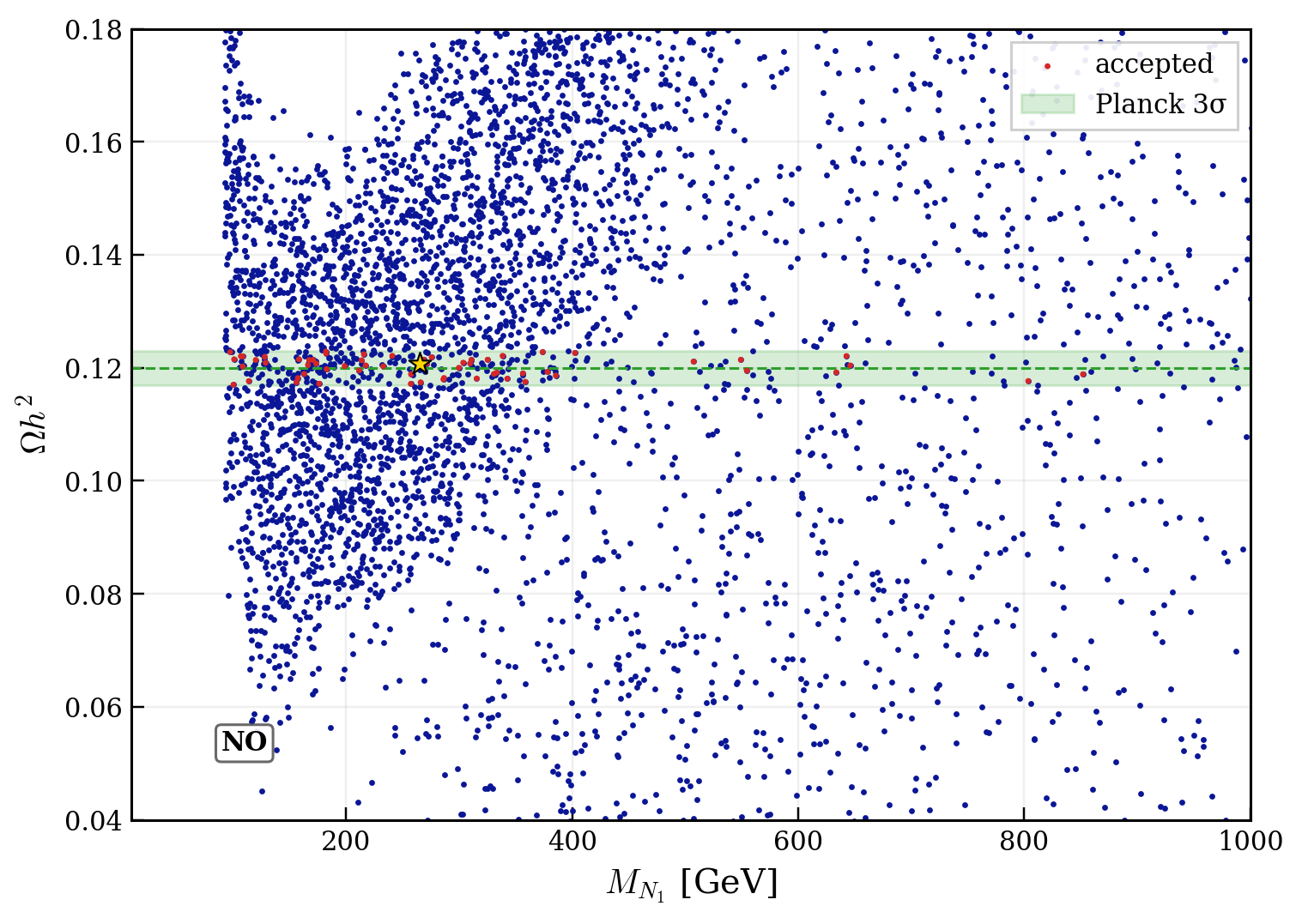}
\includegraphics[width=0.3\linewidth]{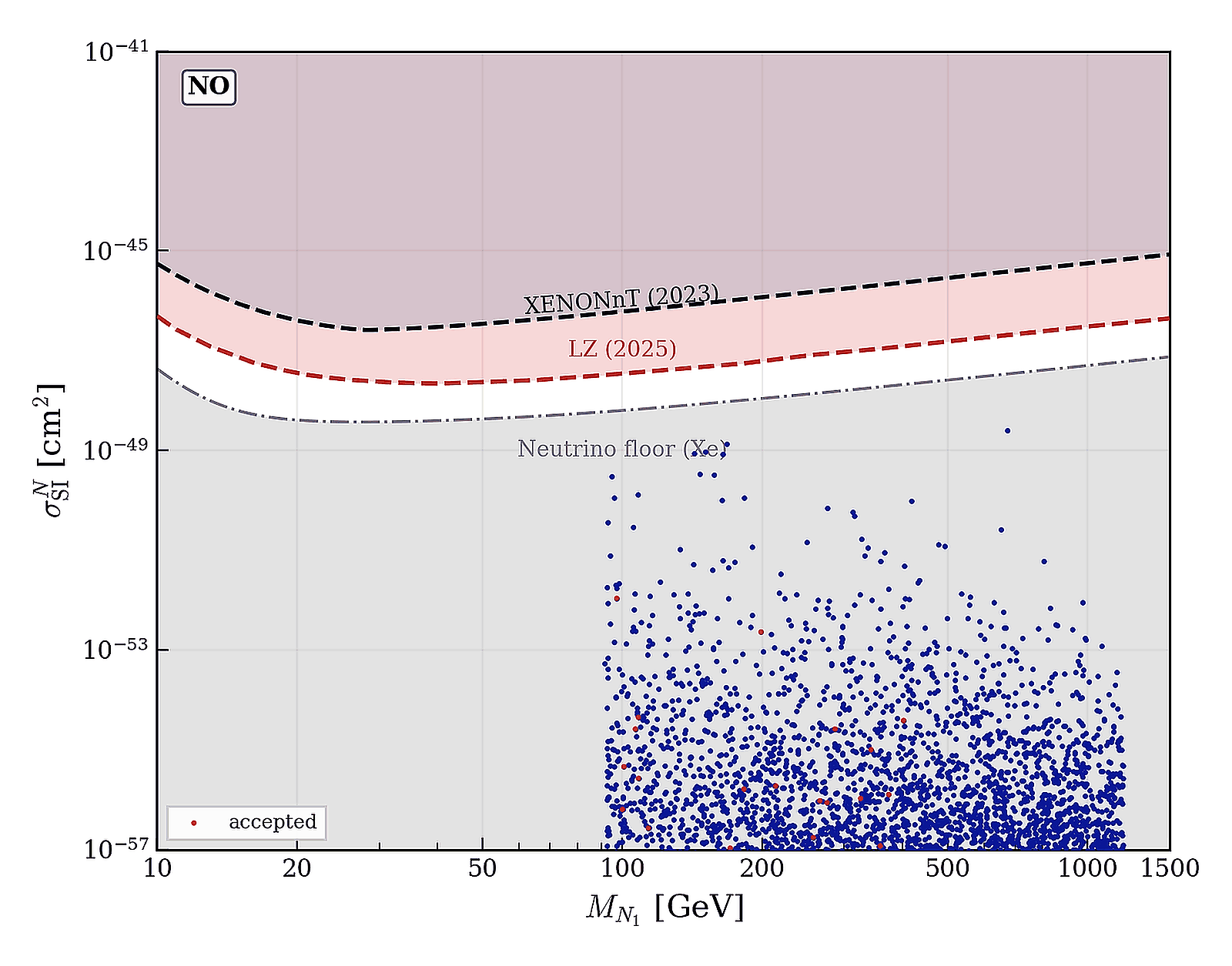}
\includegraphics[width=0.295\linewidth]{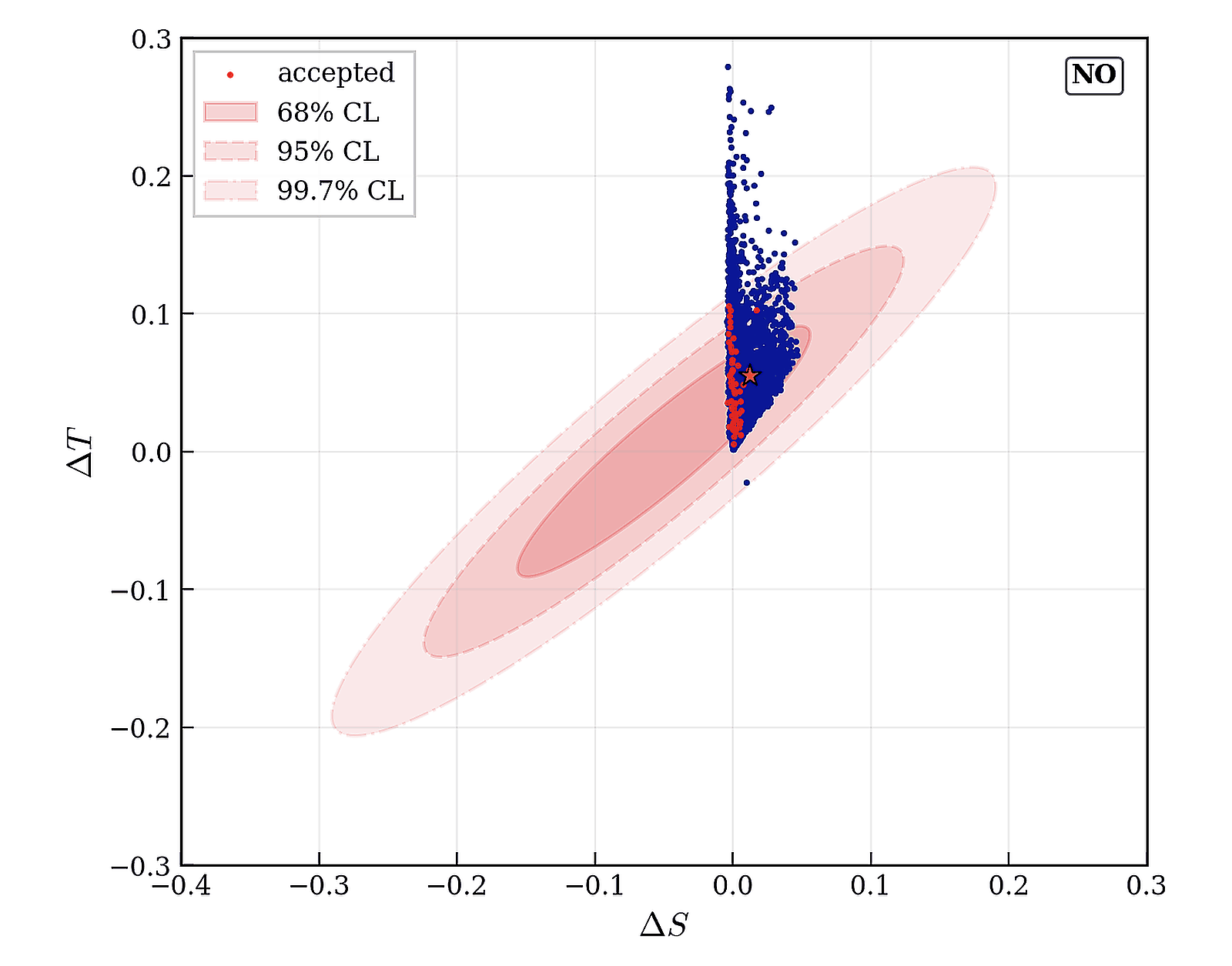}
\caption{From left to right, the panels show the relic density versus $M_{N_1}$, the spin-independent nucleon cross section versus $M_{N_1}$, and the $(\Delta S,\Delta T)$ plane. The horizontal band and the direct-detection curves correspond to Planck~\cite{Planck:2018vyg}, XENONnT~\cite{XENON:2023cxc}, and LZ~\cite{LZ:2024zvo}, while the ellipse is the preferred electroweak-fit region~\cite{ParticleDataGroup:2024cfk}.}
\label{fig4}
\end{figure}
\begin{figure}[!]
\centering
\includegraphics[width=0.245\textwidth]{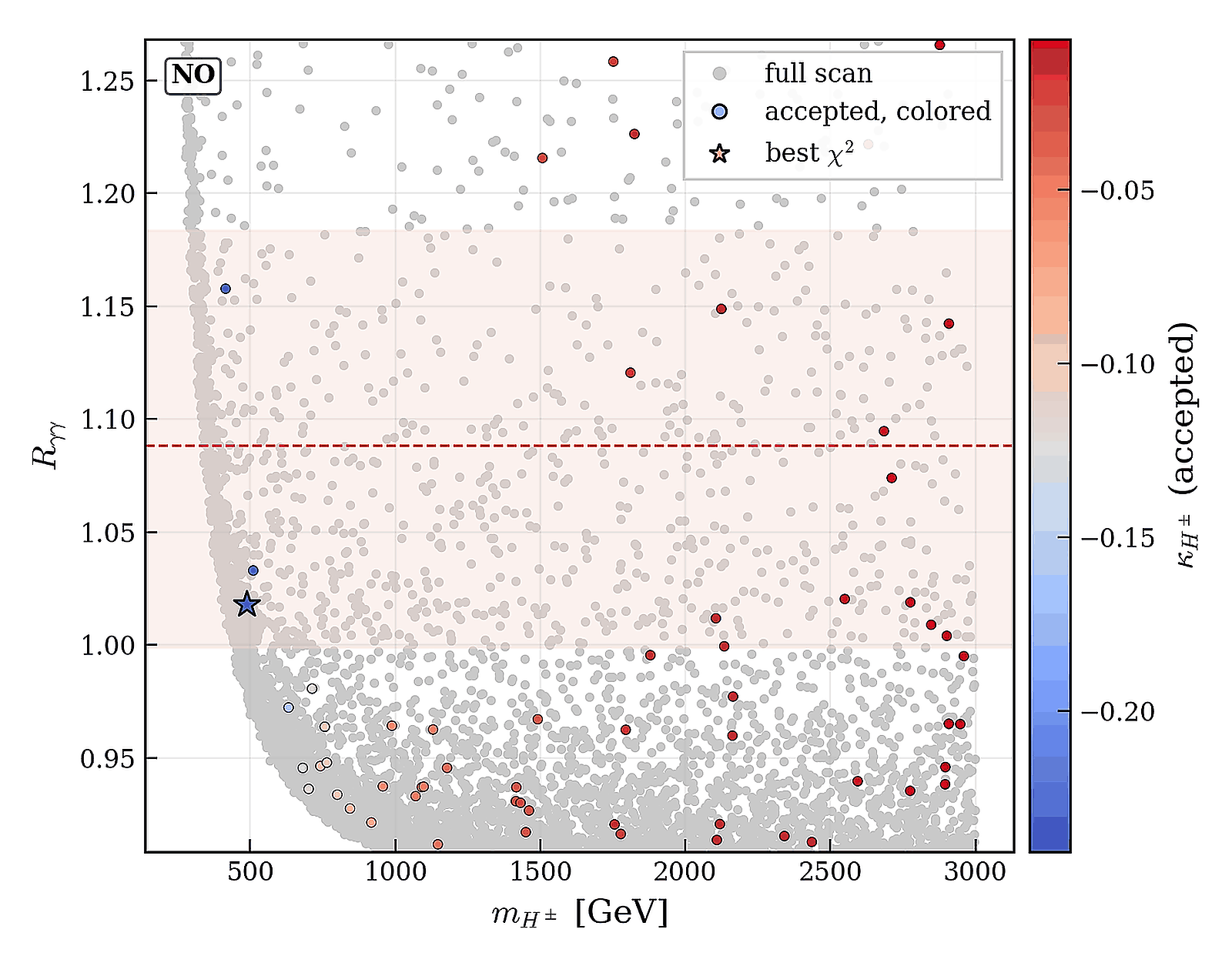}
\includegraphics[width=0.245\linewidth]{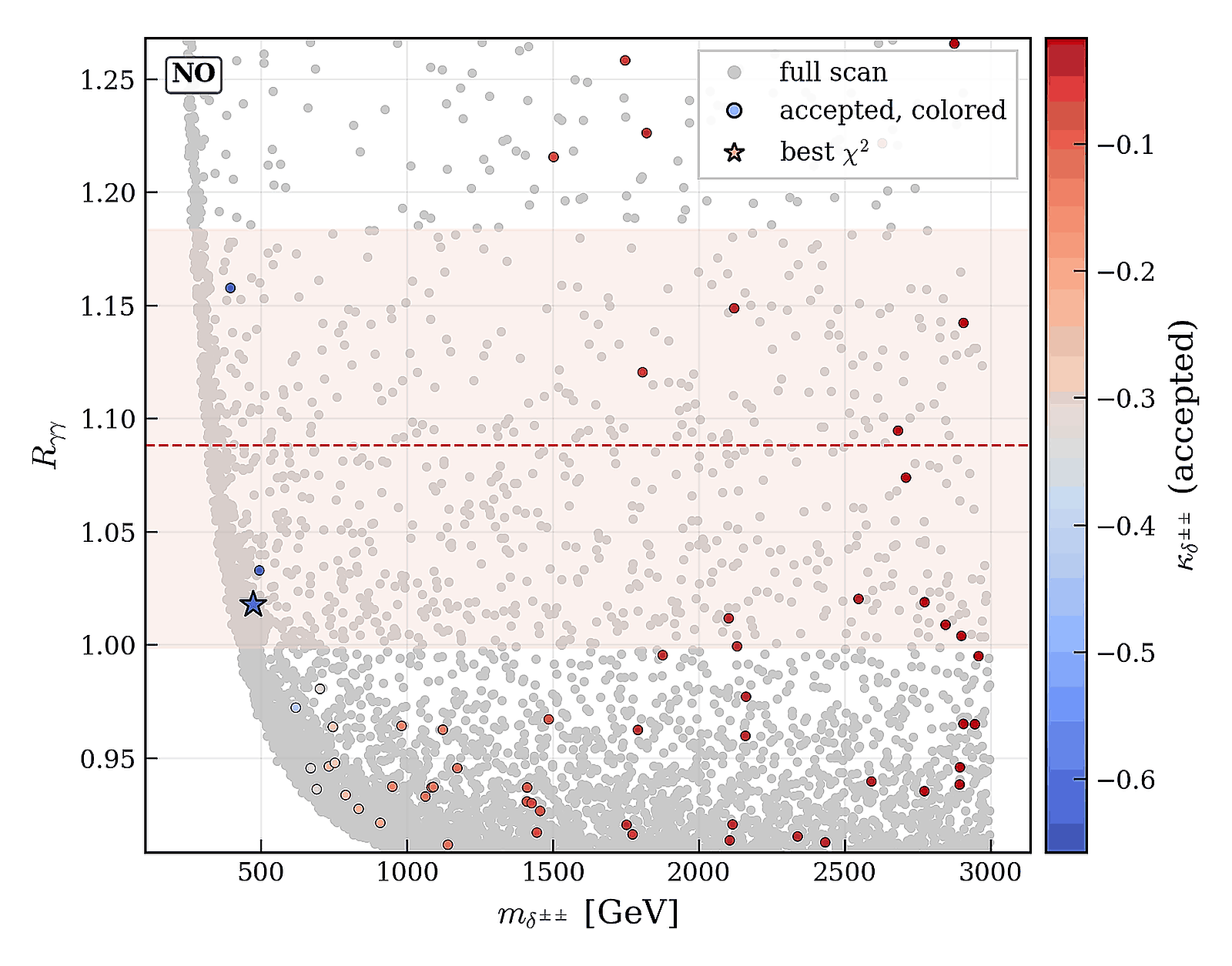}
\includegraphics[width=0.245\linewidth]{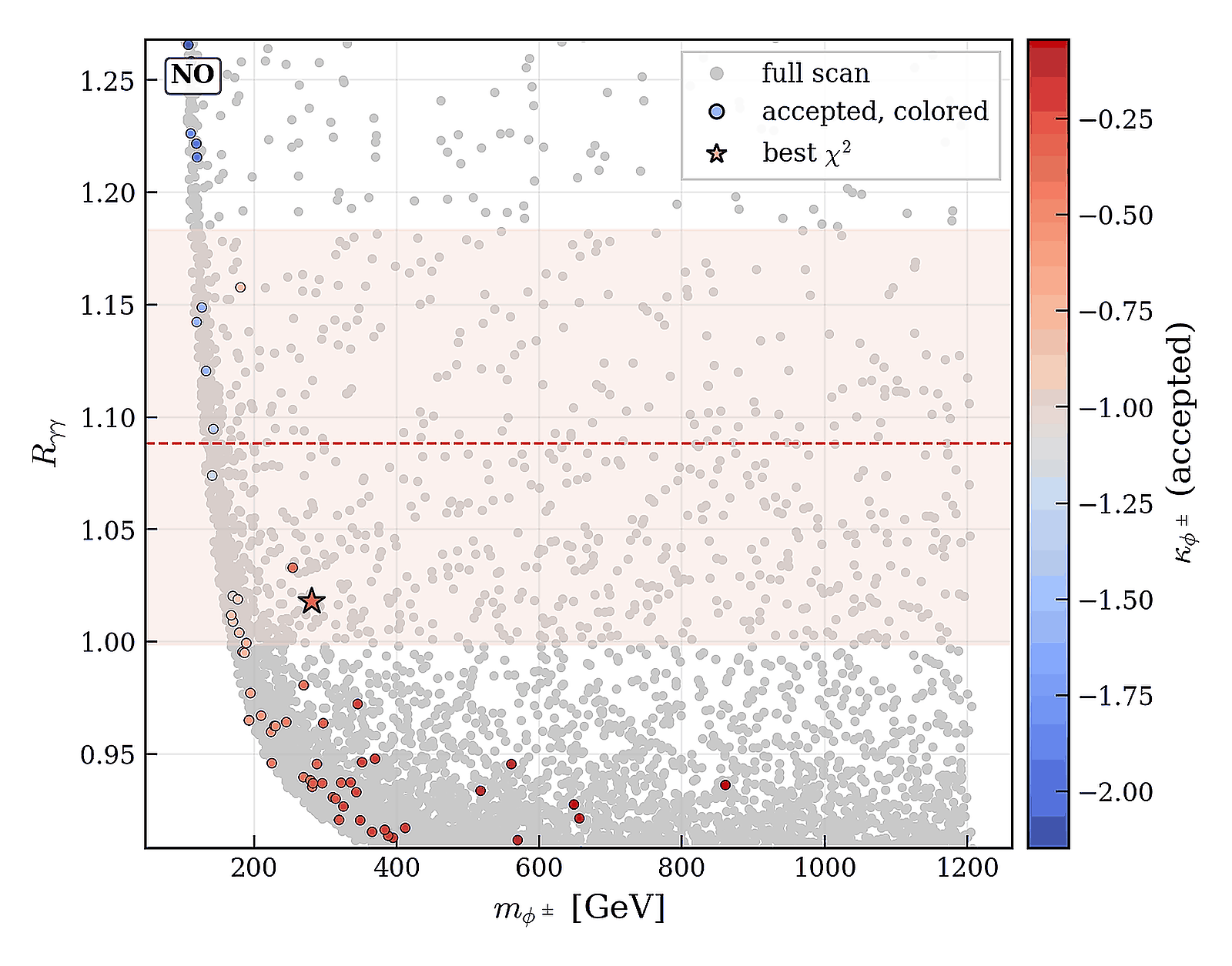}
\includegraphics[width=0.245\linewidth]{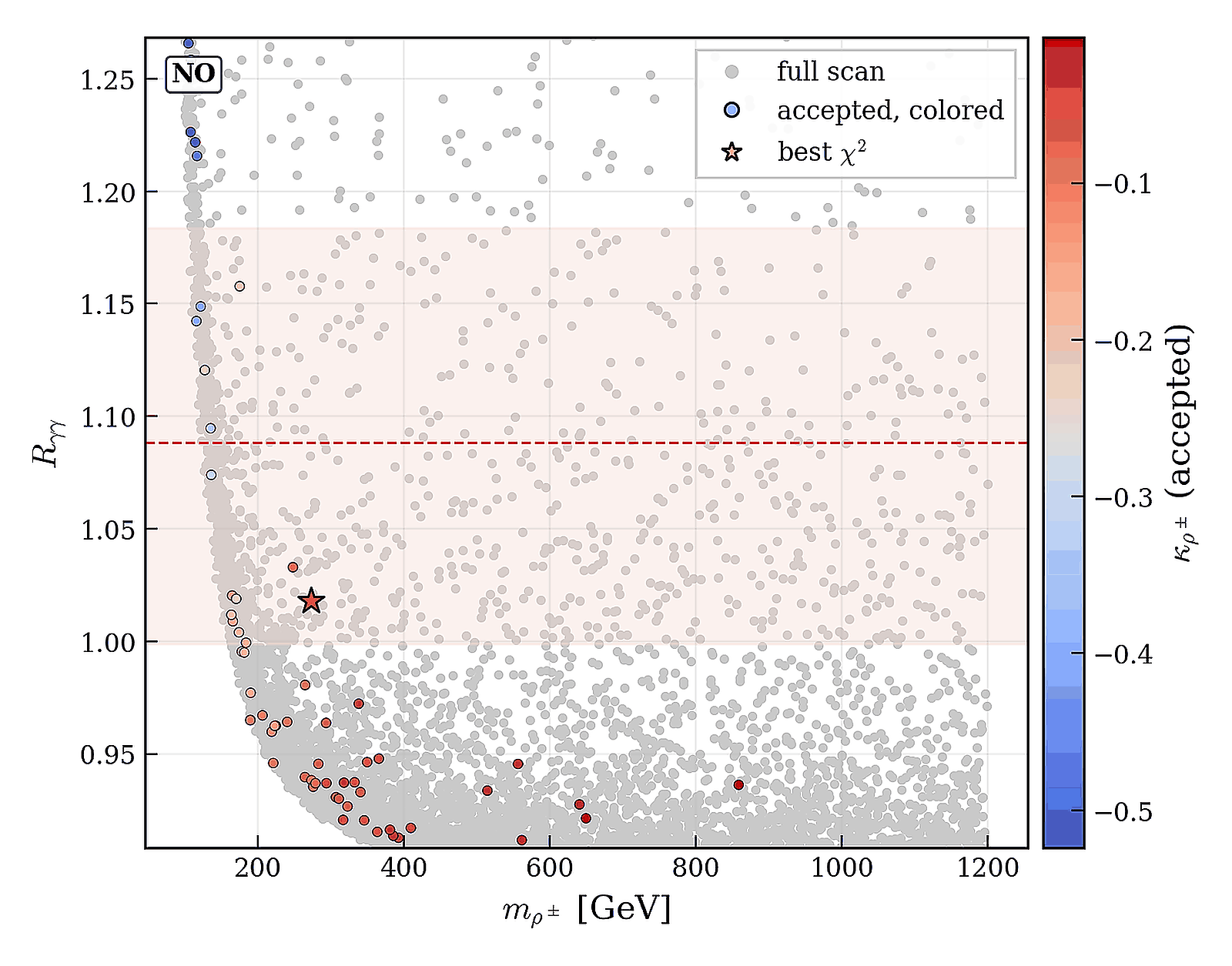}
\caption{From left to right the panels show $R_{\gamma\gamma}$ versus $m_{\delta^{\pm\pm}}$, $m_{H^\pm}$, $m_{\phi^\pm}$, and $m_{\rho^\pm}$. The color palette encodes the associated effective Higgs couplings $\kappa_{\delta^{\pm\pm}}$, $\kappa_{H^\pm}$, $\kappa_{\phi^\pm}$, and $\kappa_{\rho^\pm}$ of the accepted points, while the horizontal band marks the ATLAS preferred range in the diphoton channel~\cite{ATLAS:2022vkf}.}
\label{fig5}
\end{figure}
Figures~\ref{fig4} and~\ref{fig8} summarize the basic DM and EW precision picture. In both orderings the viable relic-density solutions lie along a coannihilation strip compatible with the Planck bound~\cite{Planck:2018vyg}.
\begin{figure}[!]
\centering
\includegraphics[width=0.245\textwidth]{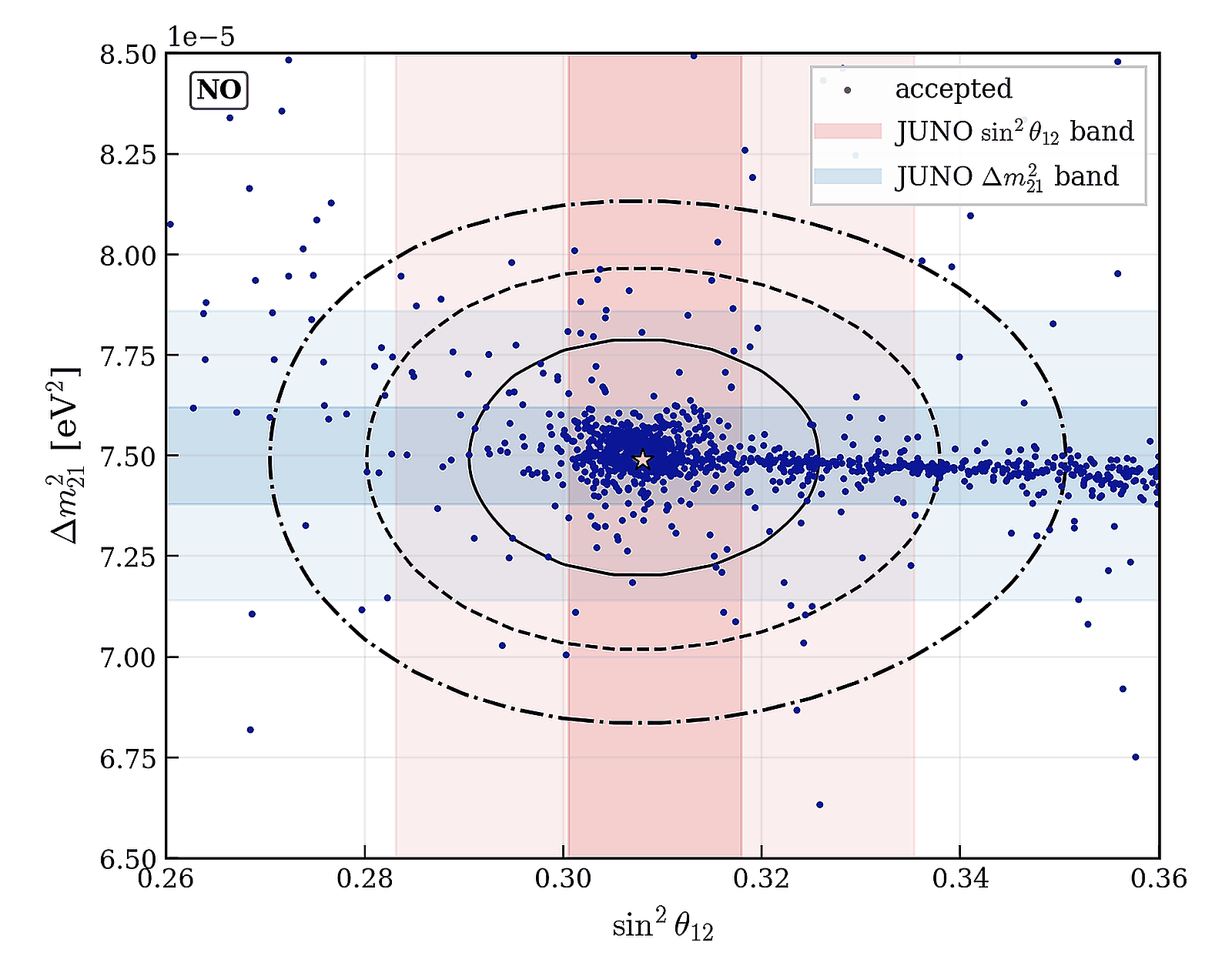}
\includegraphics[width=0.245\linewidth]{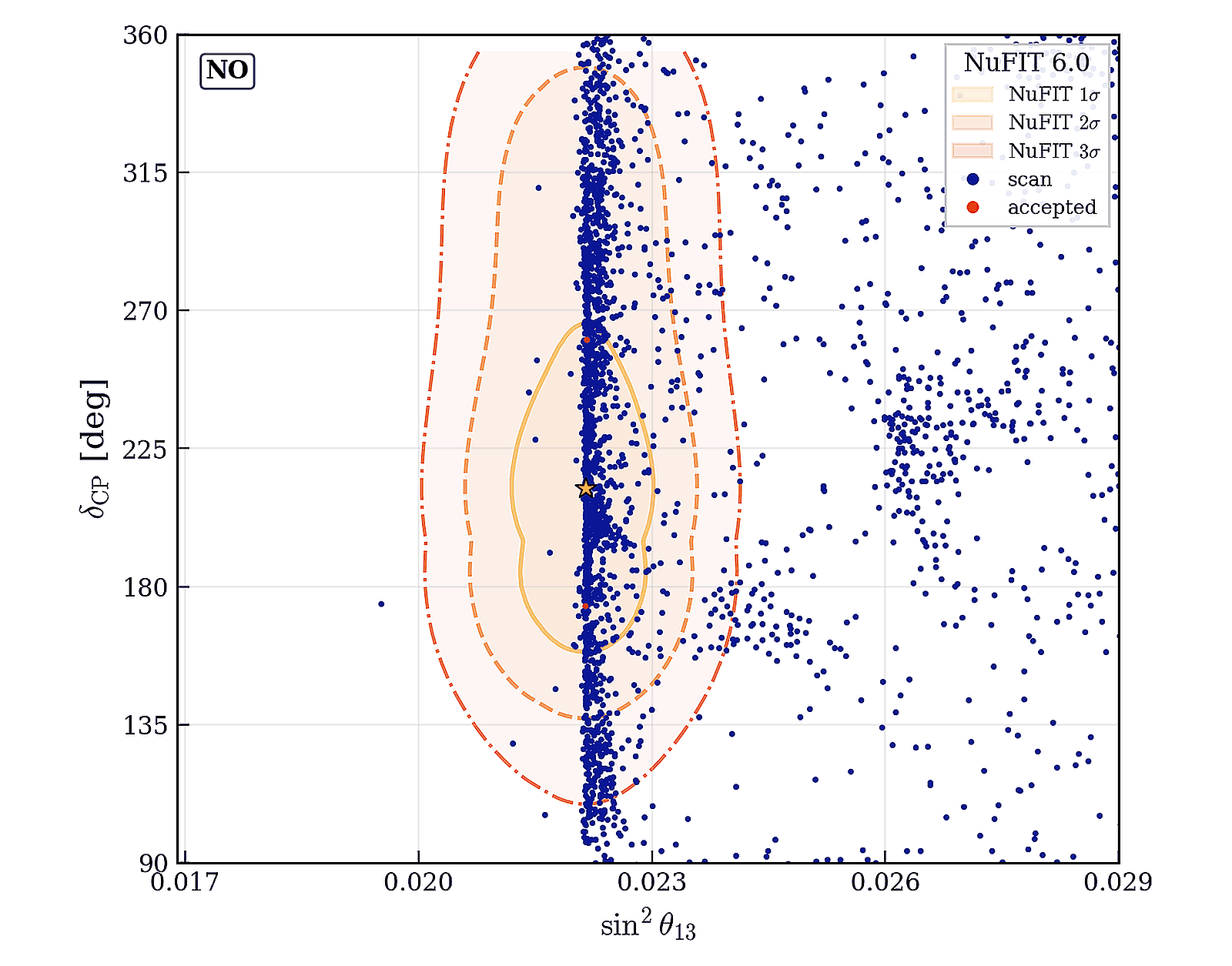}
\includegraphics[width=0.245\linewidth]{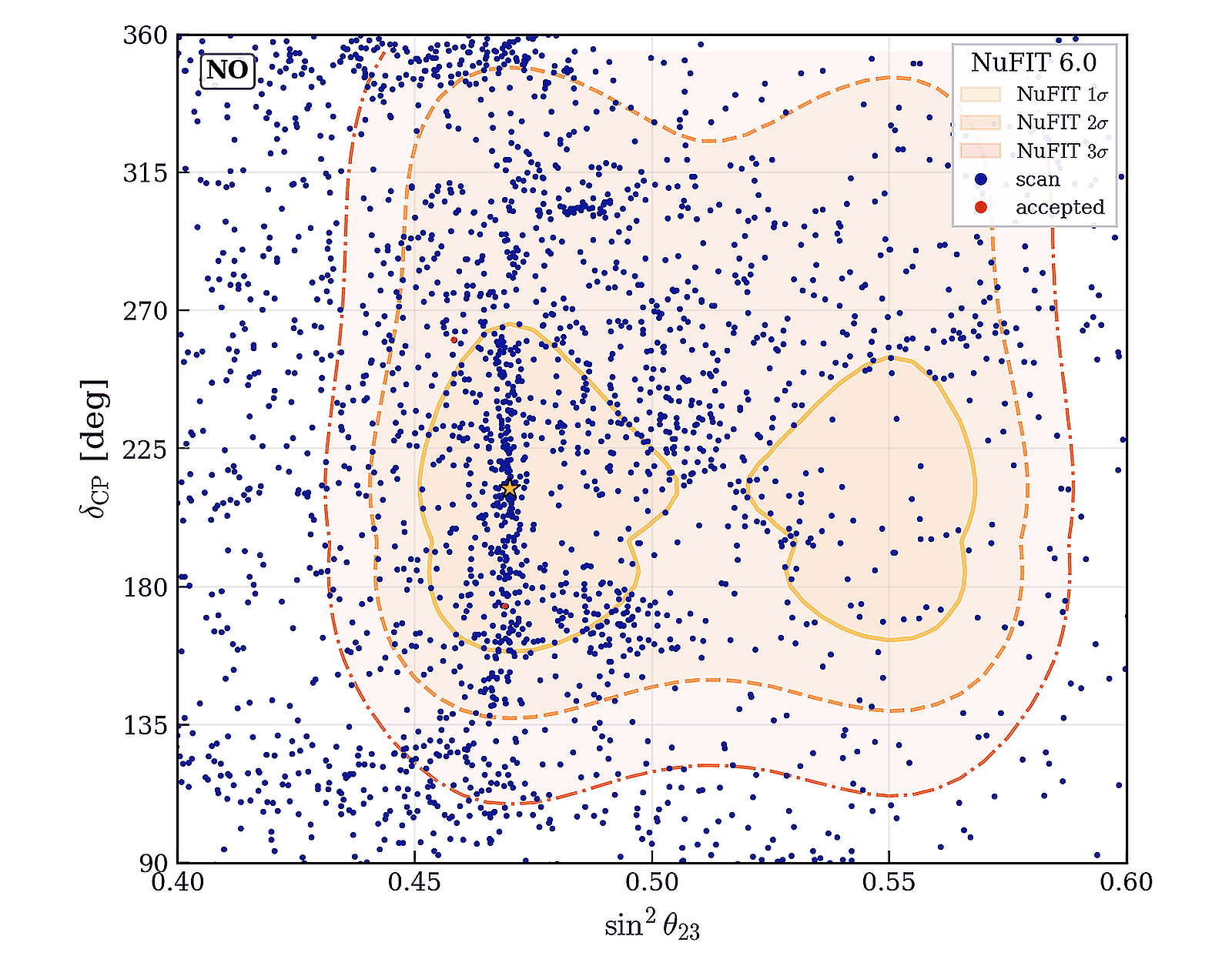}
\includegraphics[width=0.245\linewidth]{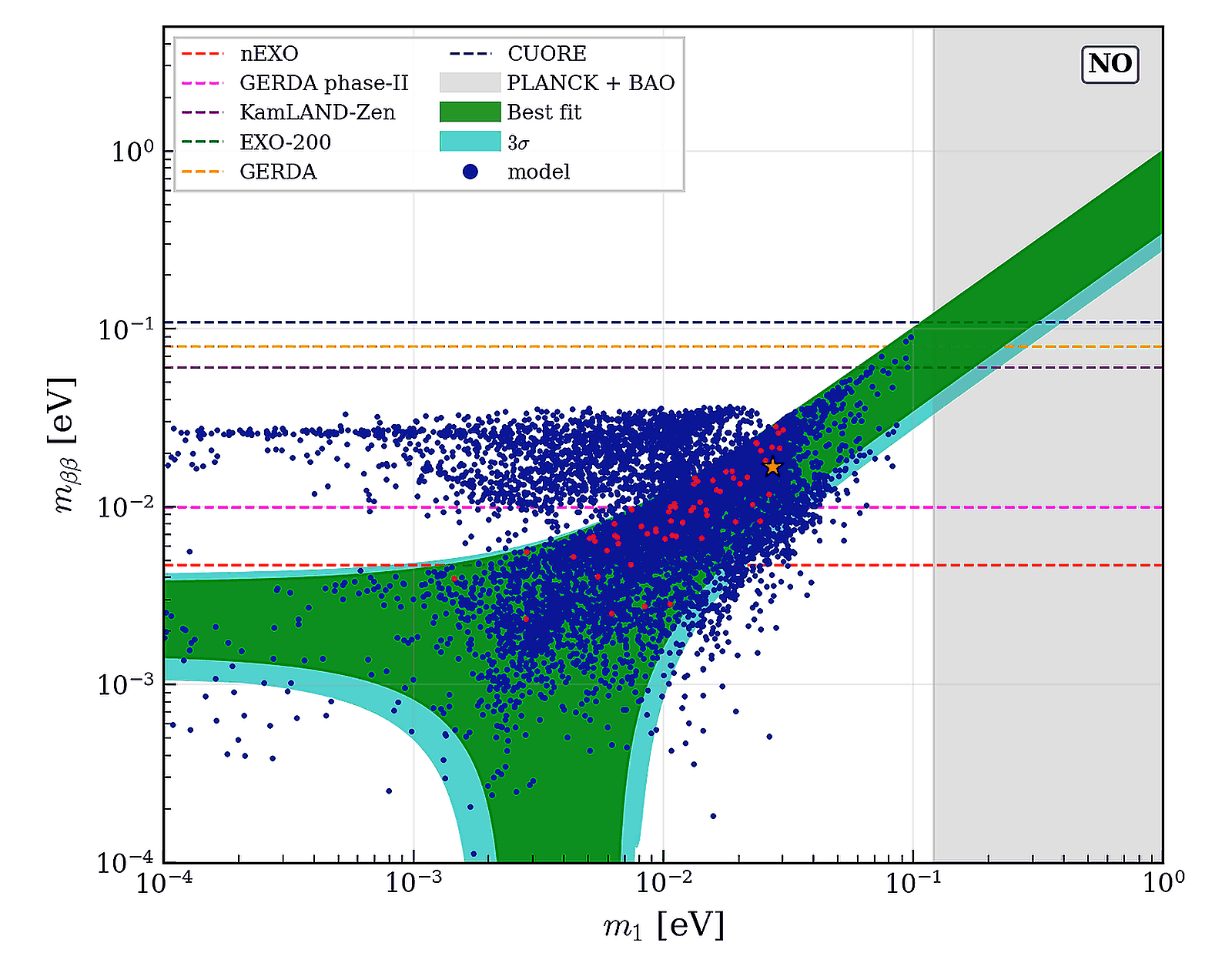}
\includegraphics[width=0.245\linewidth]{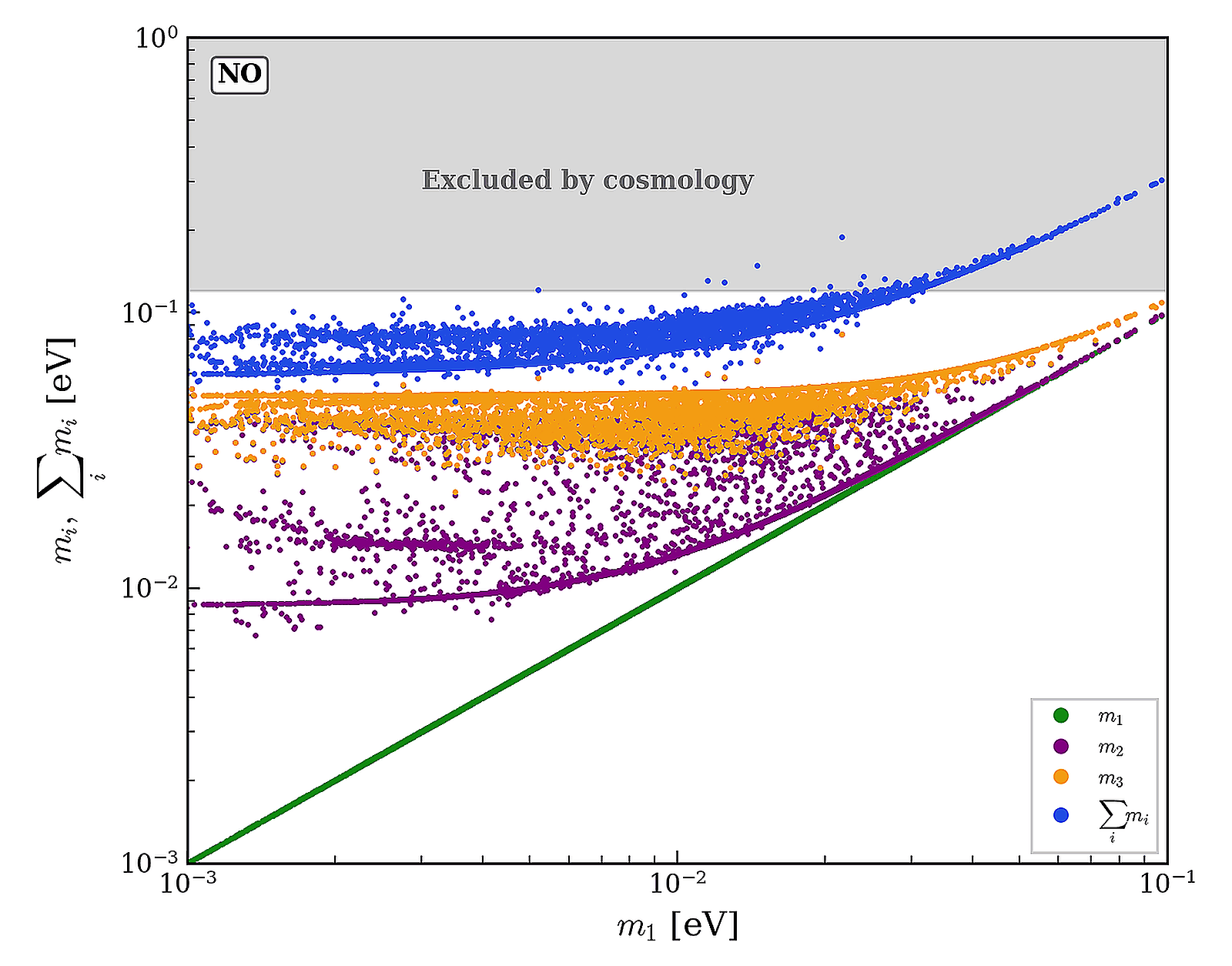}
\includegraphics[width=0.245\linewidth]{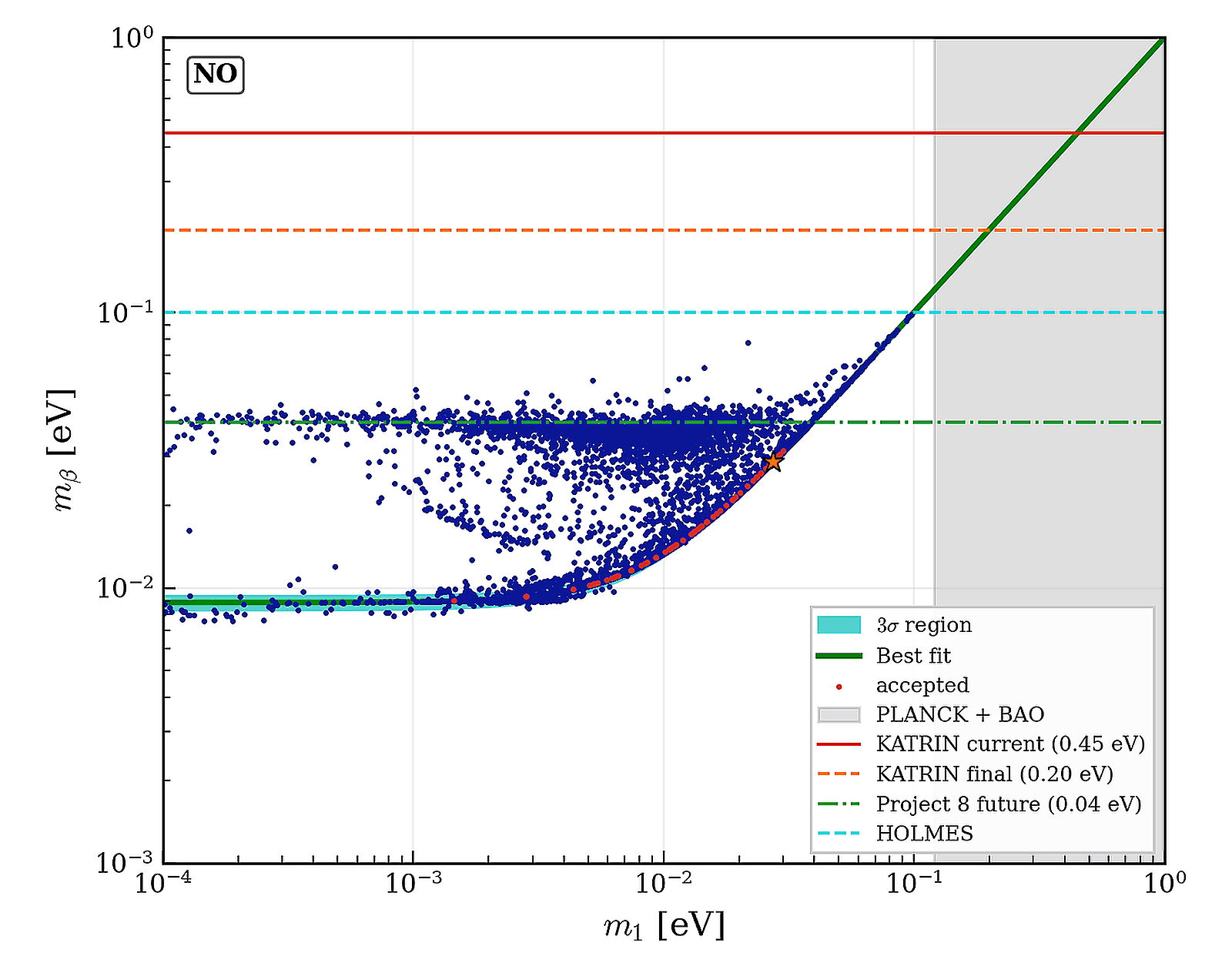}
\includegraphics[width=0.245\linewidth]{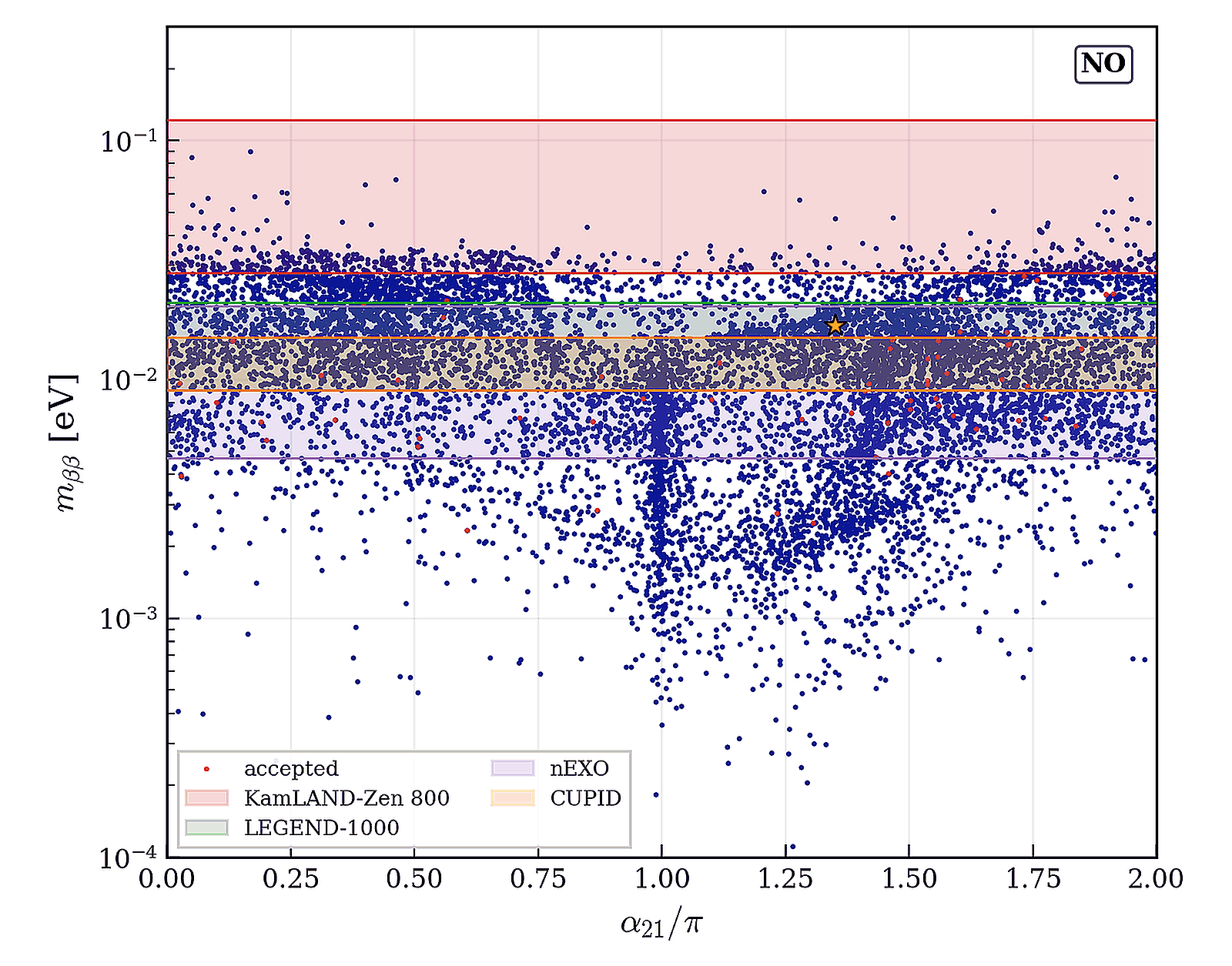}
\includegraphics[width=0.245\linewidth]{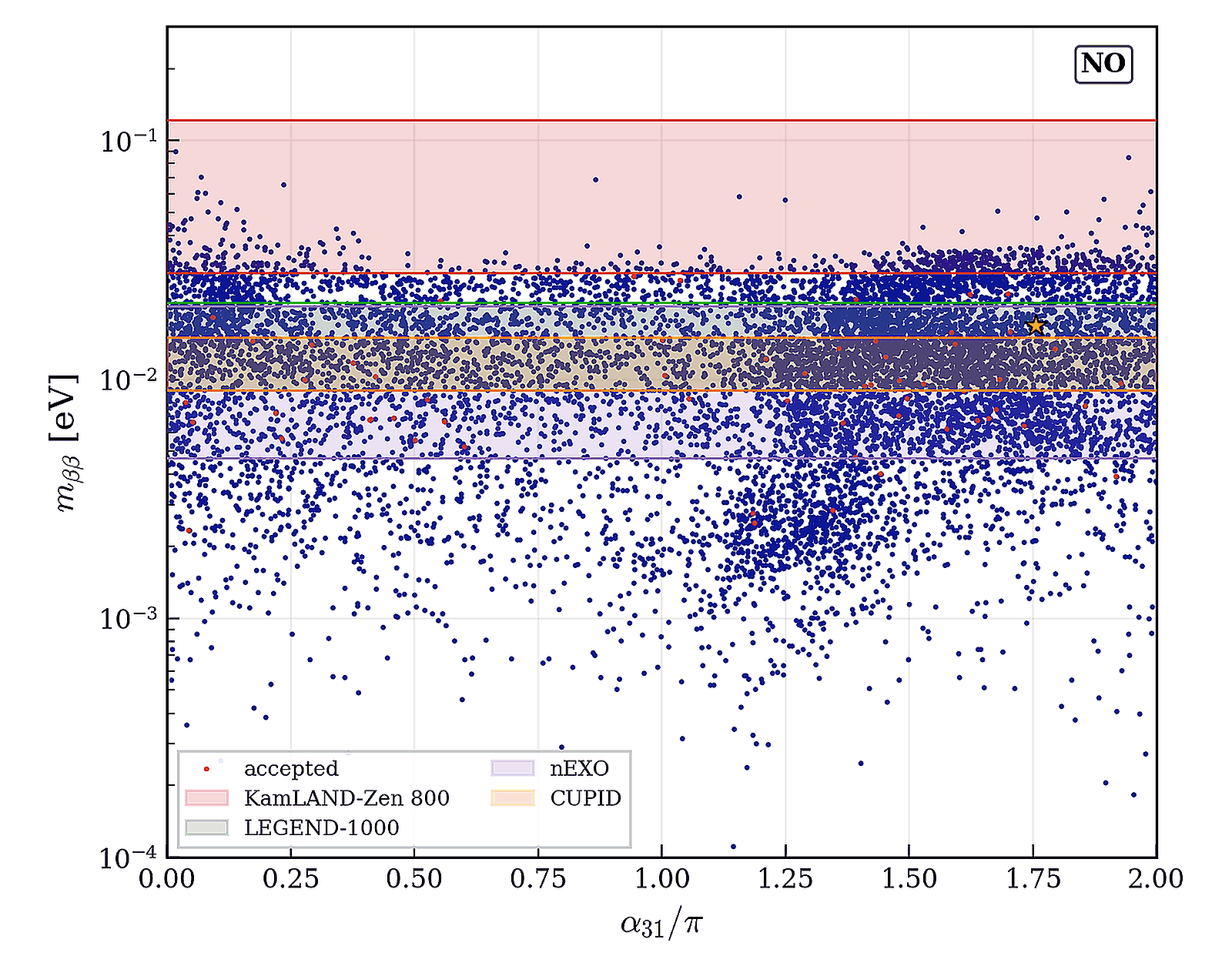}
\caption{Top plots: from left to right, the panels display $(\Delta m_{21}^2,\sin^2\theta_{12})$, $(\delta_{\rm CP},\sin^2\theta_{13})$, $(\delta_{\rm CP},\sin^2\theta_{23})$, $m_{\beta\beta}$ versus $m_1$. Bottom plots: from left to right, the panels show $(m_i,\sum_i m_i)$ versus $m_1$, $m_\beta$ versus $m_1$, and $m_{\beta\beta}$ versus the Majorana phases $\alpha_{21}$ and $\alpha_{31}$. The oscillation overlays follow JUNO~\cite{JUNO:2025gmd} and NuFIT~6.0~\cite{Esteban:2024eli}; the absolute-mass panels compare with cosmology and with current and projected $0\nu\beta\beta$ and beta-decay sensitivities~\cite{Planck:2018vyg,KamLAND-Zen:2016pfg,EXO-200:2019rkq,GERDA:2020xhi,CUORE:2022fgx,LEGEND:2021bnm,nEXO:2021ujk,CUPID:2022olt,KATRIN:2024mass,Project8:2022bla,HOLMES:2016qqz}.}
\label{fig6}
\end{figure}
The accepted NO points extend over the broader interval $97\lesssim M_{N_1}\lesssim 852$~GeV, whereas the accepted IO points are concentrated in the narrower range $127\lesssim M_{N_1}\lesssim 378$~GeV. The SI cross sections remain far below the present XENONnT and LZ limits~\cite{XENON:2023cxc,LZ:2024zvo}, and in practice stay below the xenon neutrino floor across the accepted sample. The oblique shifts lie inside the correlated region used in the scan, with small positive $\Delta T$ and modest excursions in $\Delta S$.

Figures~\ref{fig5} and~\ref{fig9} display the ratio $R_{\gamma \gamma}$ as a function of the charged-scalar masses. The color code gives the corresponding effective Higgs couplings $\kappa_{\delta^{\pm\pm}}$, $\kappa_{H^\pm}$, $\kappa_{\phi^\pm}$, and $\kappa_{\rho^\pm}$, so one can directly see how the sign and size of the charged-scalar contribution vary across the scan. In both orderings the viable points remain close to the SM value and inside the ATLAS preferred band~\cite{ATLAS:2022vkf}. Lighter charged scalars can produce a visible spread in $R_{\gamma\gamma}$, but the correction is never large enough to move the model away from the measured diphoton rate.

Figures~\ref{fig6} and~\ref{fig10} collect the oscillation and absolute-mass observables. The $(\Delta m_{21}^2,\sin^2\theta_{12})$ panels and the $(\delta_{\rm CP},\sin^2\theta_{13,23})$ planes show that the allowed points fall in the regions preferred by JUNO and NuFIT~6.0~\cite{JUNO:2025gmd,Esteban:2024eli}. The clearest difference between NO and IO appears in the absolute neutrino masses. For NO, the allowed points cover a wider range of the lightest mass and give $0.060\lesssim \sum_i m_i\lesssim 0.120$~eV. For IO, the spectrum is more compressed and one finds $0.099\lesssim \sum_i m_i\lesssim 0.119$~eV. This difference is reflected in $m_{\beta\beta}$ and $m_\beta$. In the $m_{\beta\beta}$ panels, the plots are compared with the present limits $m_{\beta\beta}\lesssim 0.061$~eV from KamLAND-Zen, $m_{\beta\beta}\lesssim 0.08$~eV from EXO-200 and GERDA, and $m_{\beta\beta}\lesssim 0.11$~eV from CUORE~\cite{KamLAND-Zen:2016pfg,EXO-200:2019rkq,GERDA:2020xhi,CUORE:2022fgx}. The projected reaches shown in the figure extend to the ${\cal O}(10^{-2})$~eV range for LEGEND-1000 and CUPID, and to about $5\times10^{-3}$~eV for nEXO~\cite{LEGEND:2021bnm,nEXO:2021ujk,CUPID:2022olt}.
\begin{figure}[!]
\centering
\includegraphics[width=0.245\textwidth]{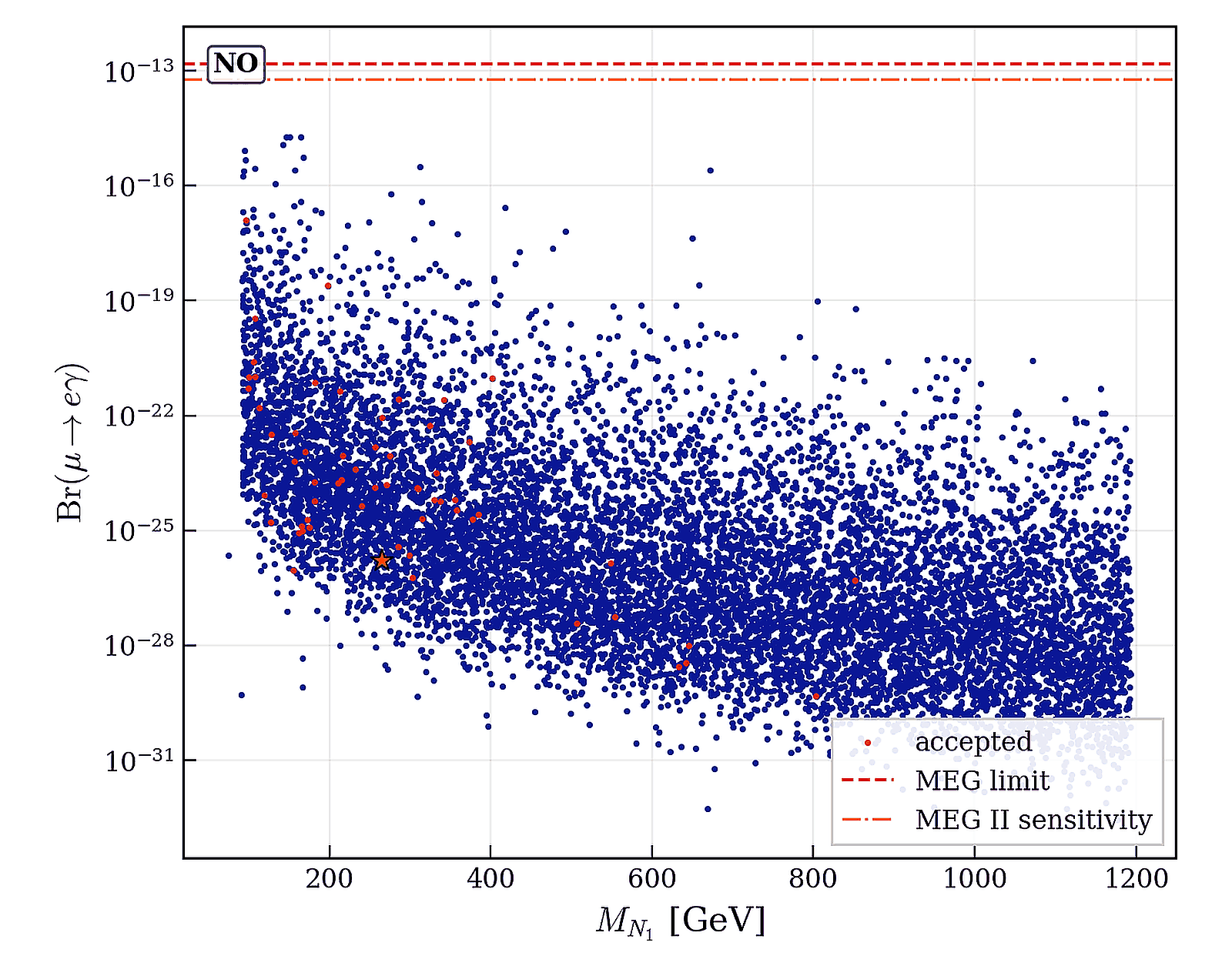}
\includegraphics[width=0.245\linewidth]{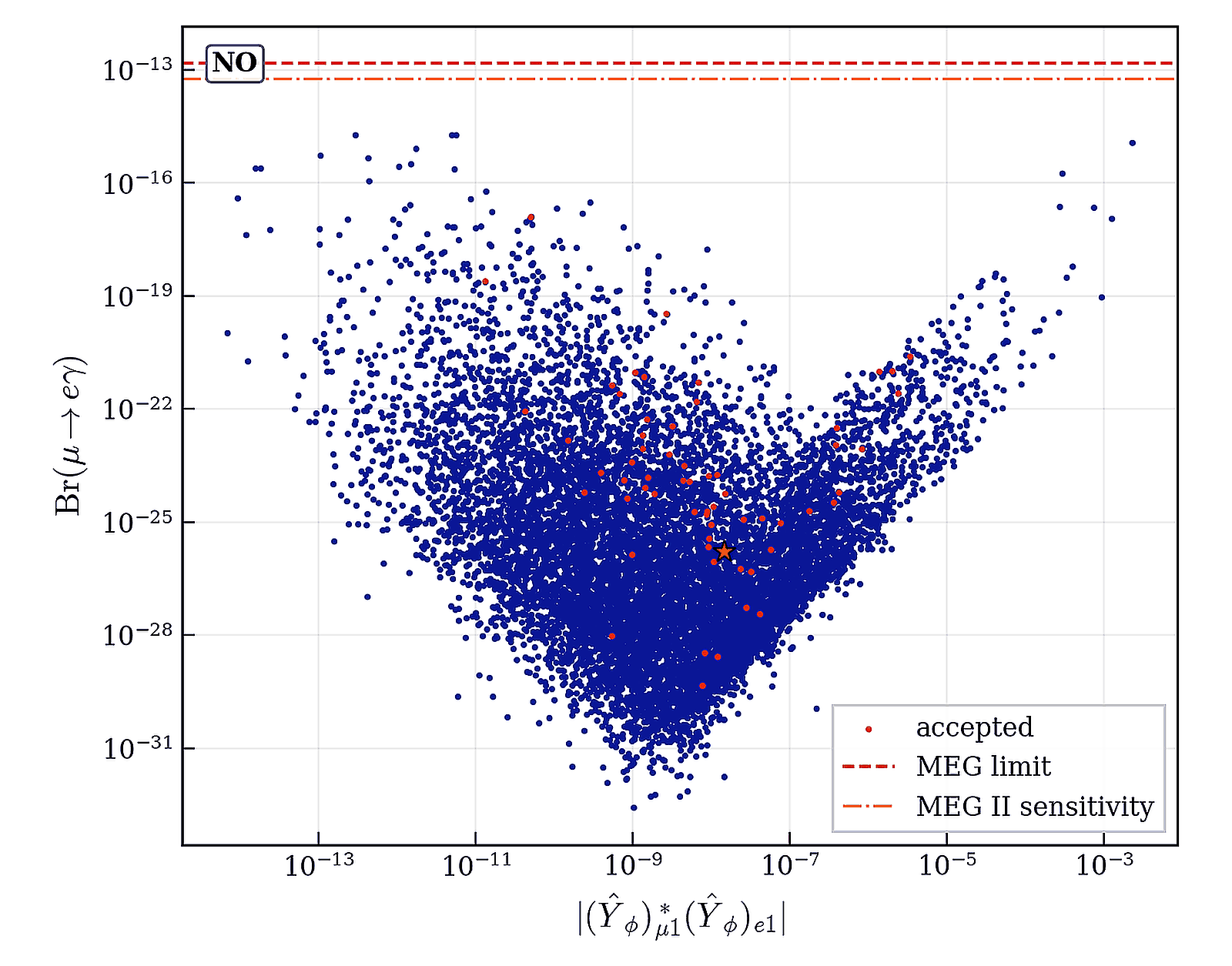}
\includegraphics[width=0.245\linewidth]{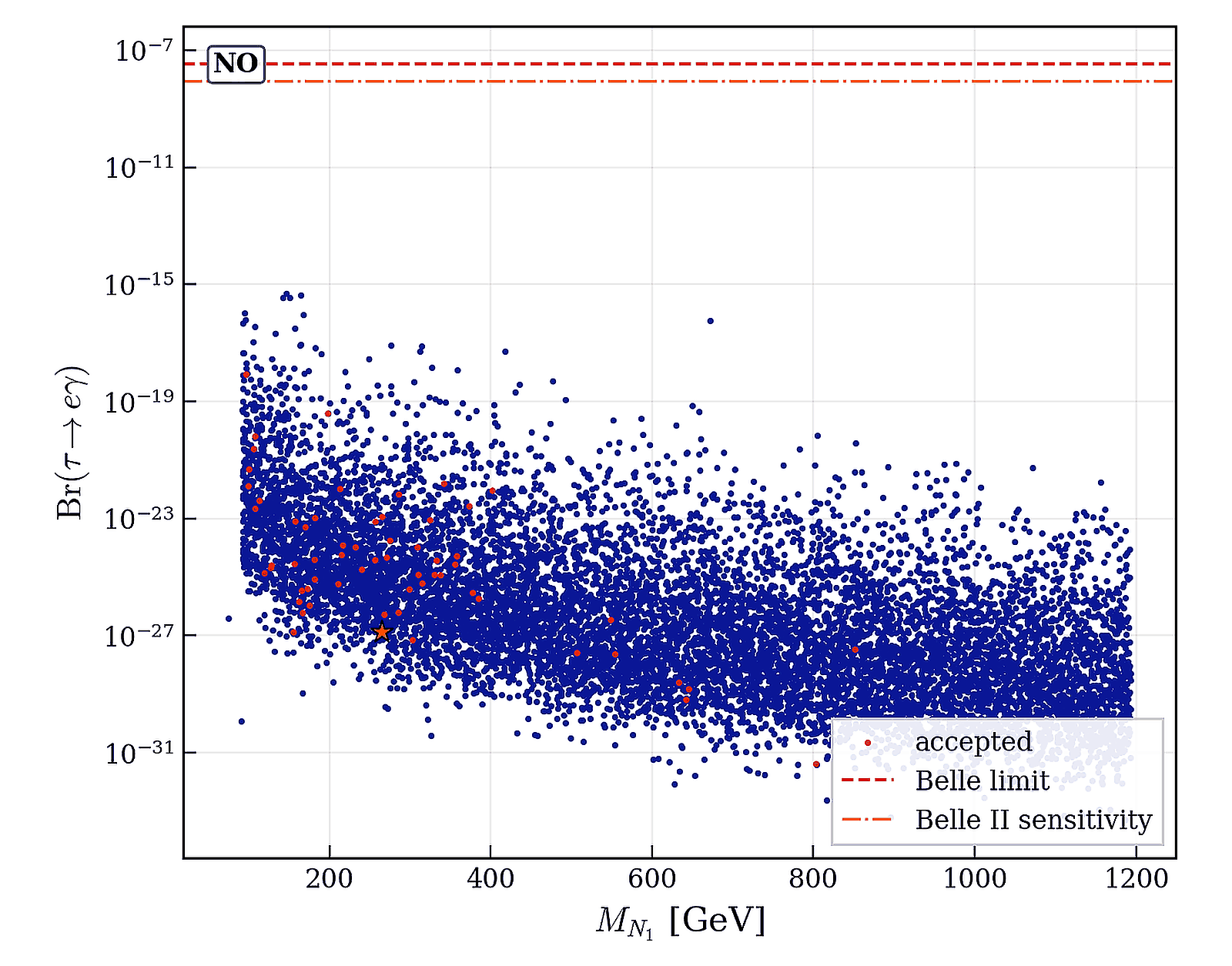}
\includegraphics[width=0.245\linewidth]{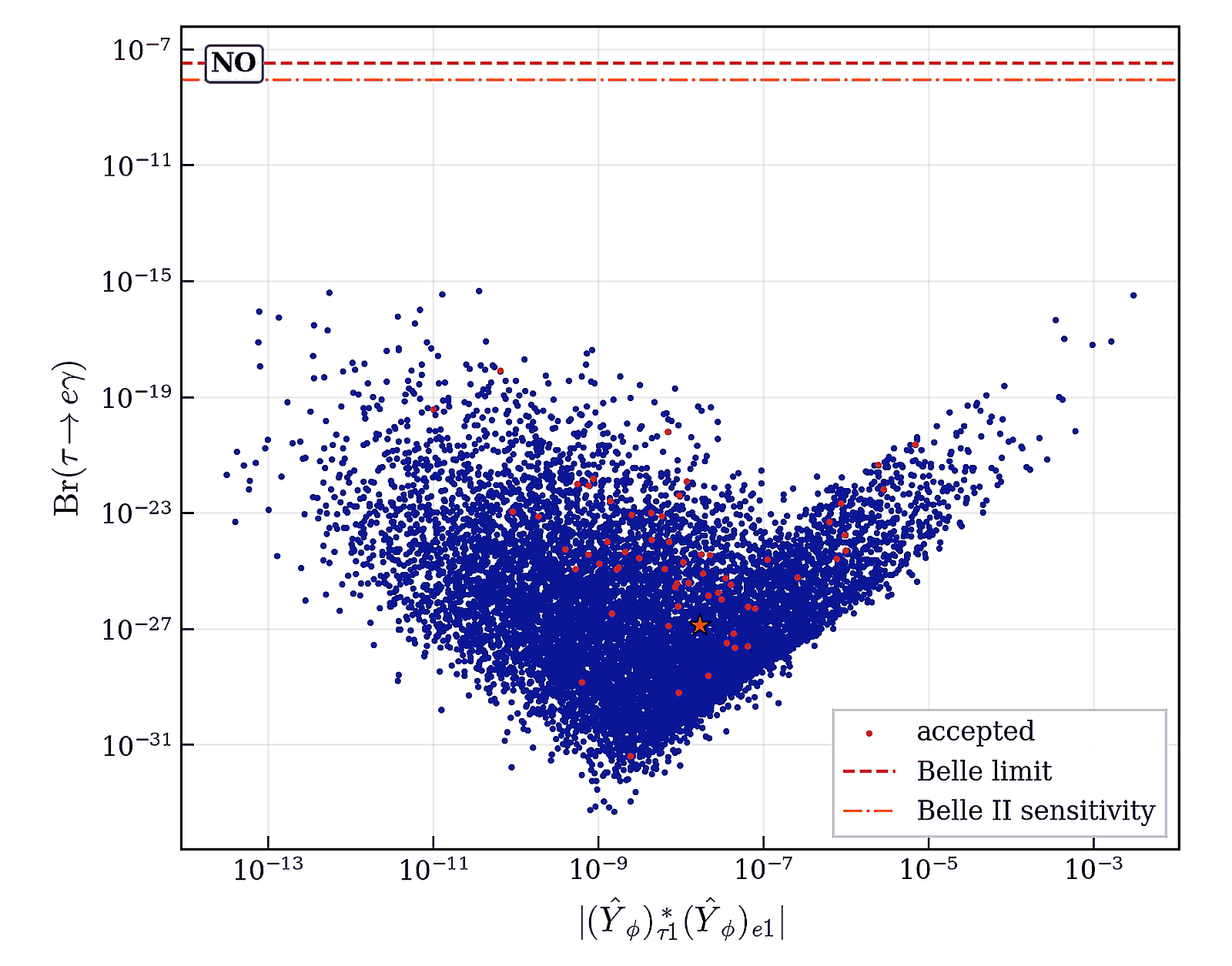}
\includegraphics[width=0.245\linewidth]{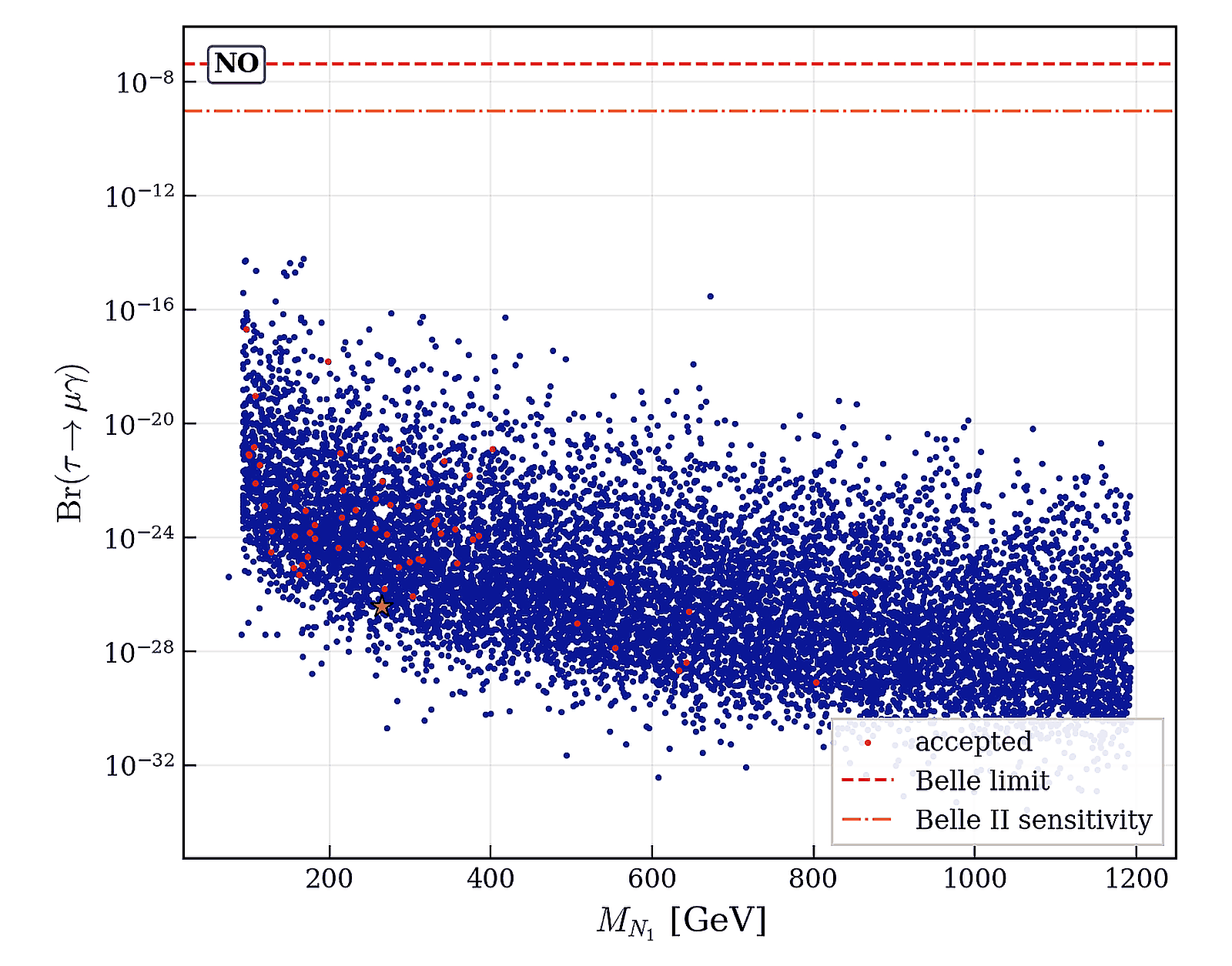}
\includegraphics[width=0.245\linewidth]{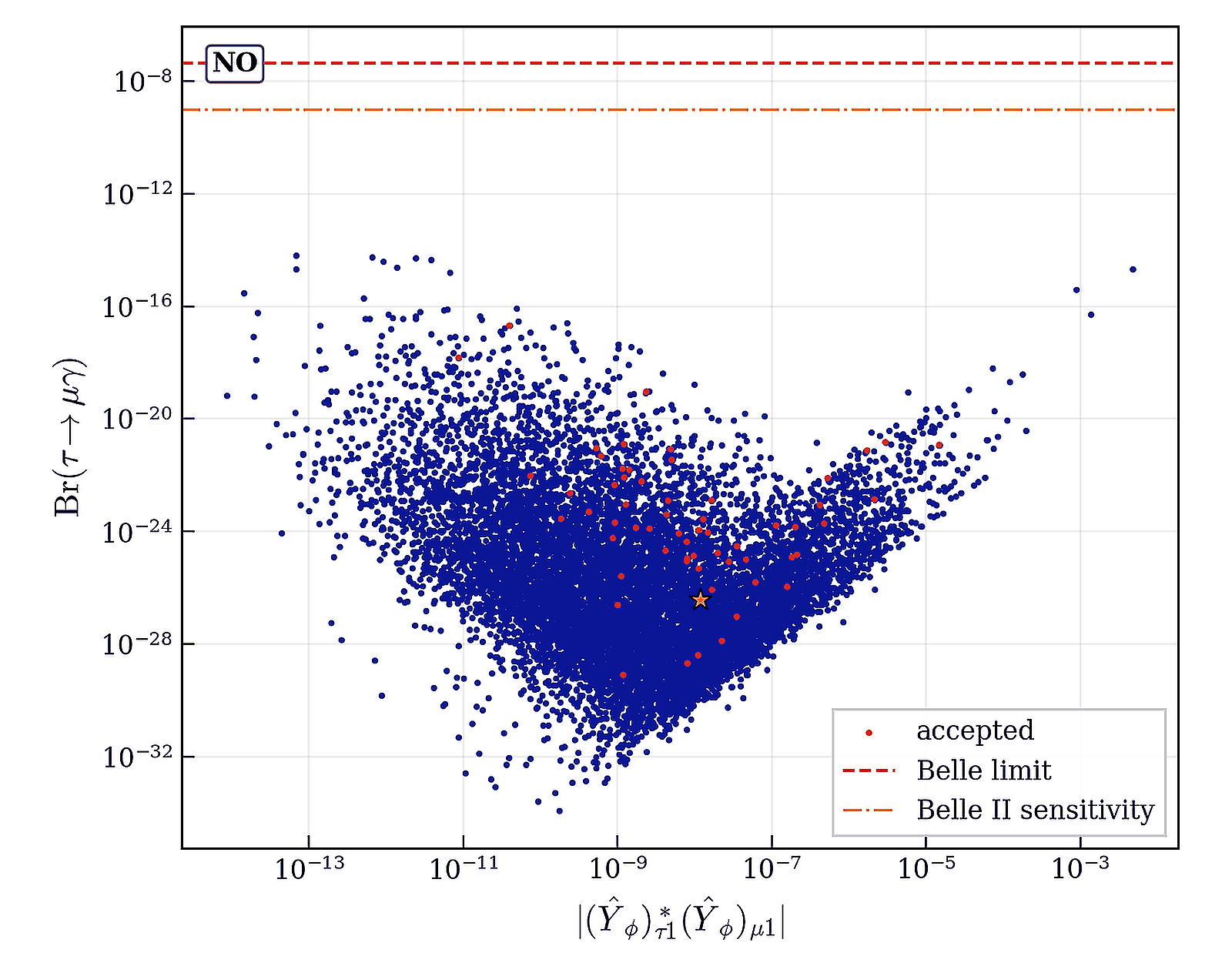}
\includegraphics[width=0.245\linewidth]{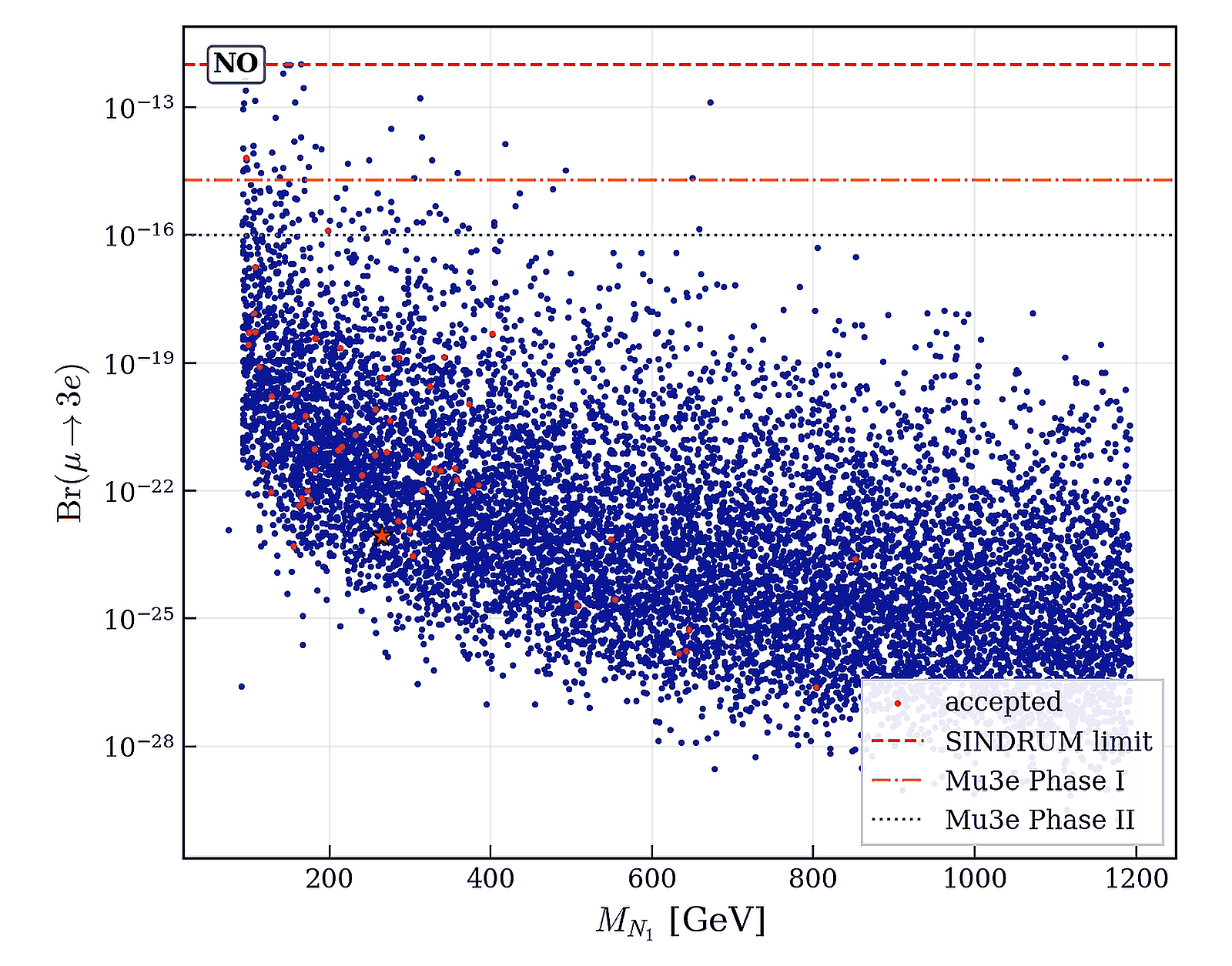}
\includegraphics[width=0.245\linewidth]{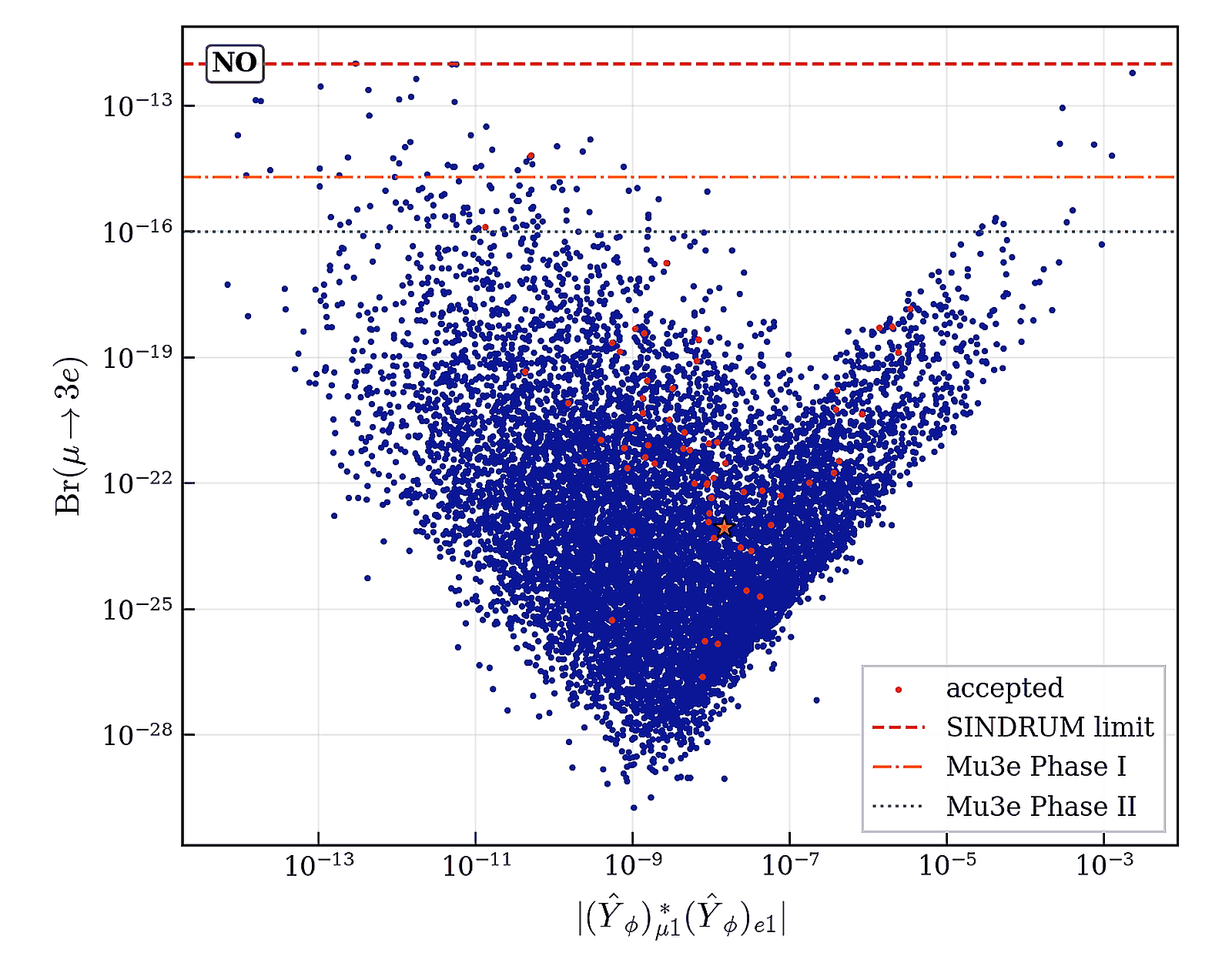}
\includegraphics[width=0.245\linewidth]{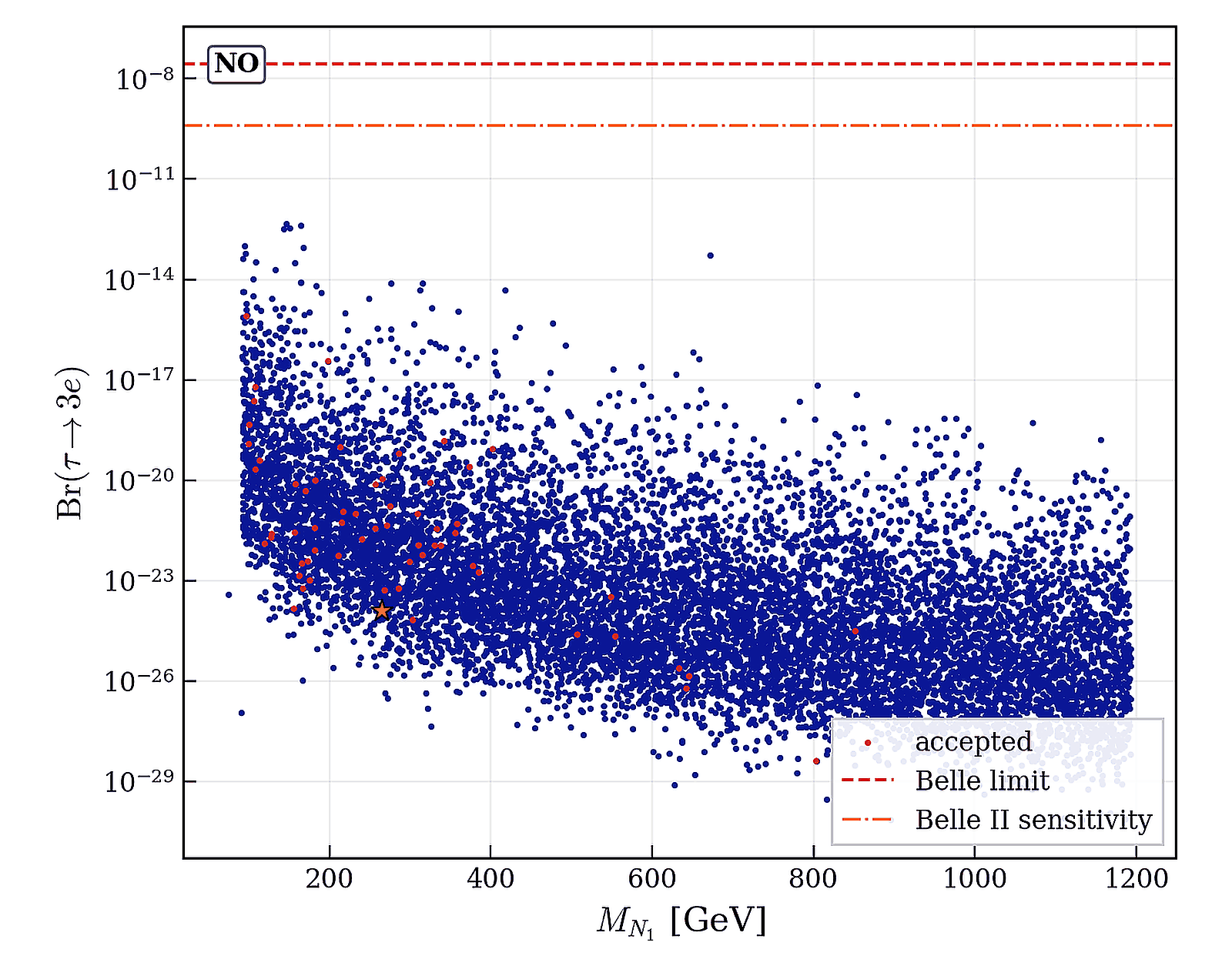}
\includegraphics[width=0.245\linewidth]{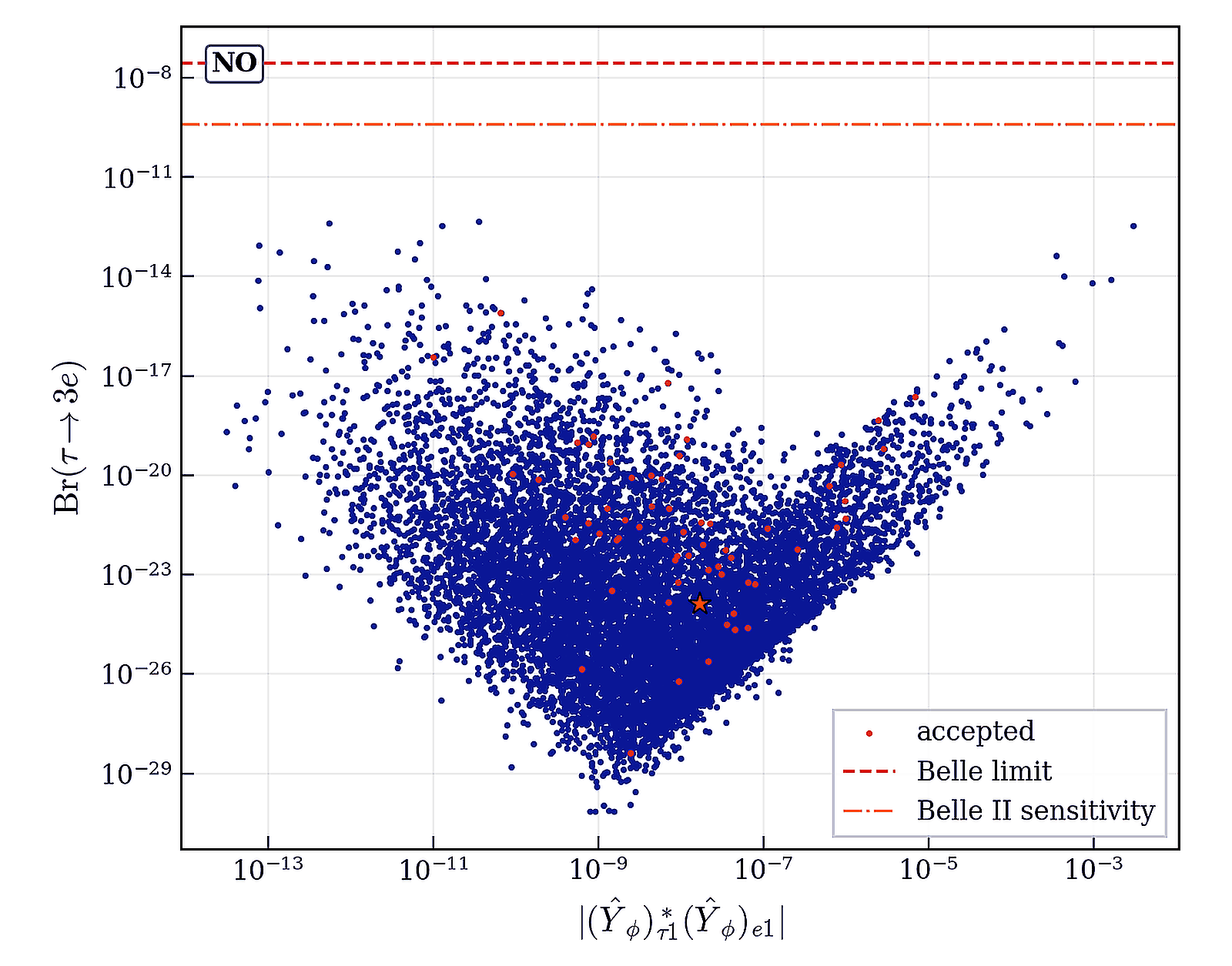}
\includegraphics[width=0.245\linewidth]{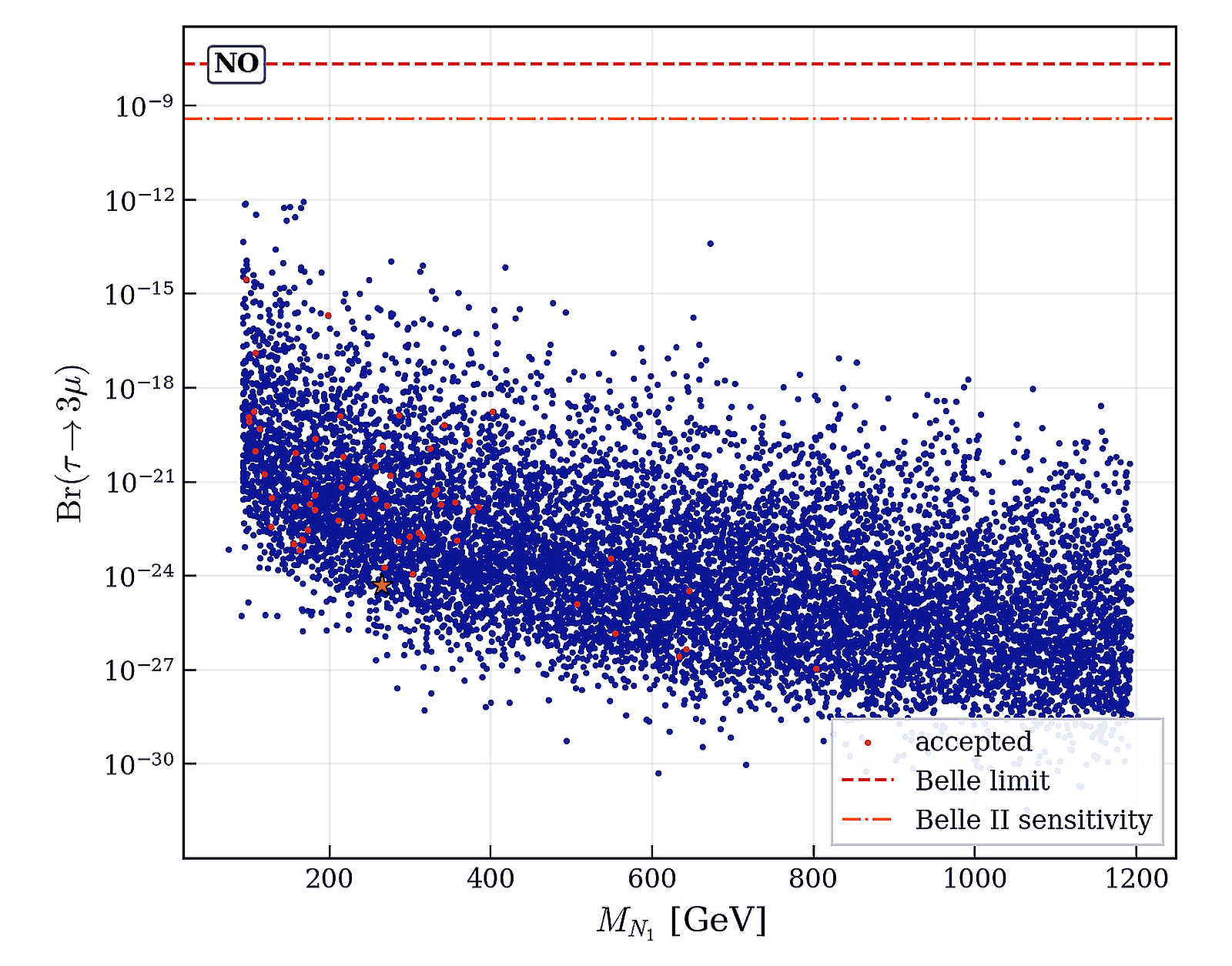}
\includegraphics[width=0.245\linewidth]{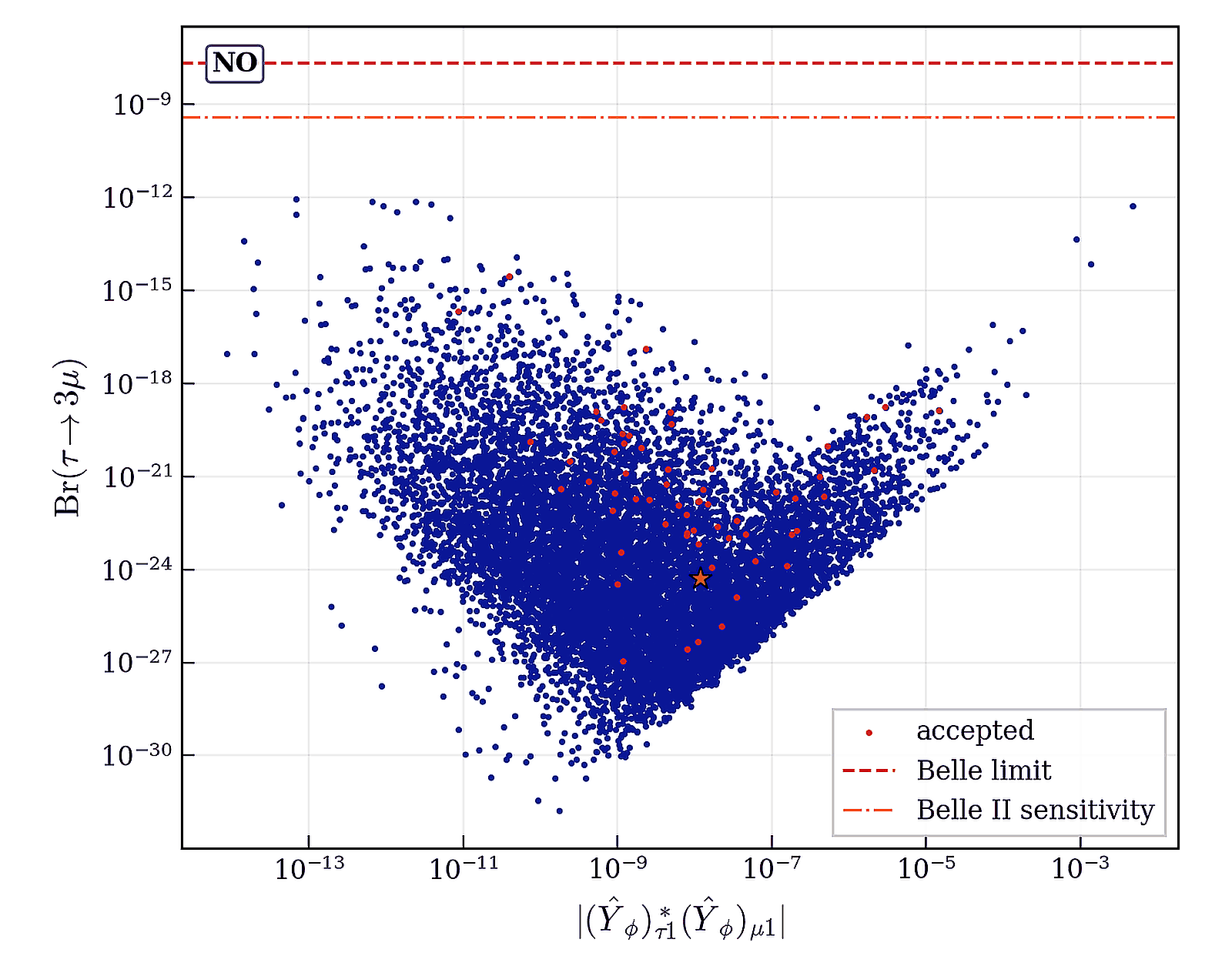}
\includegraphics[width=0.245\linewidth]{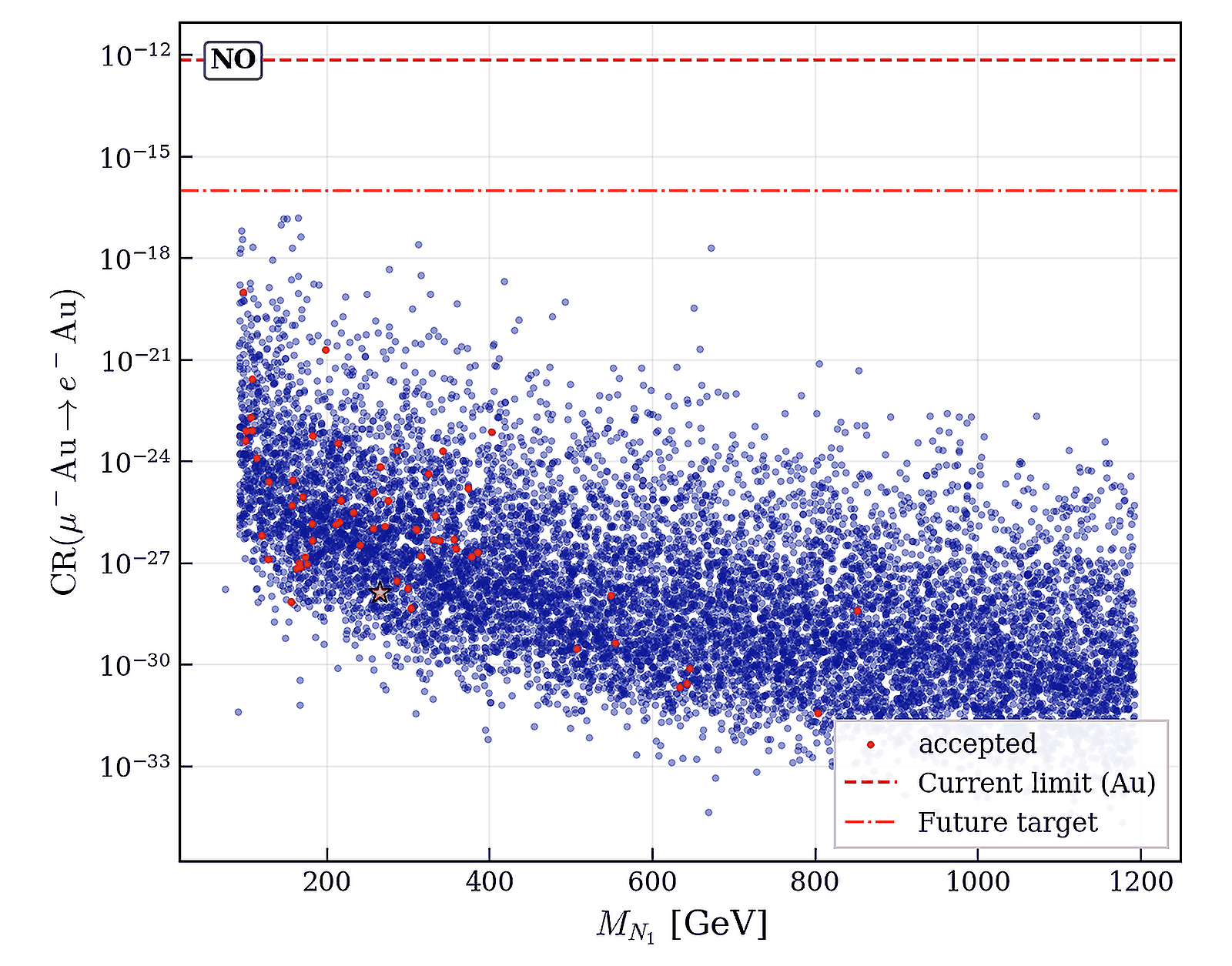}
\includegraphics[width=0.245\linewidth]{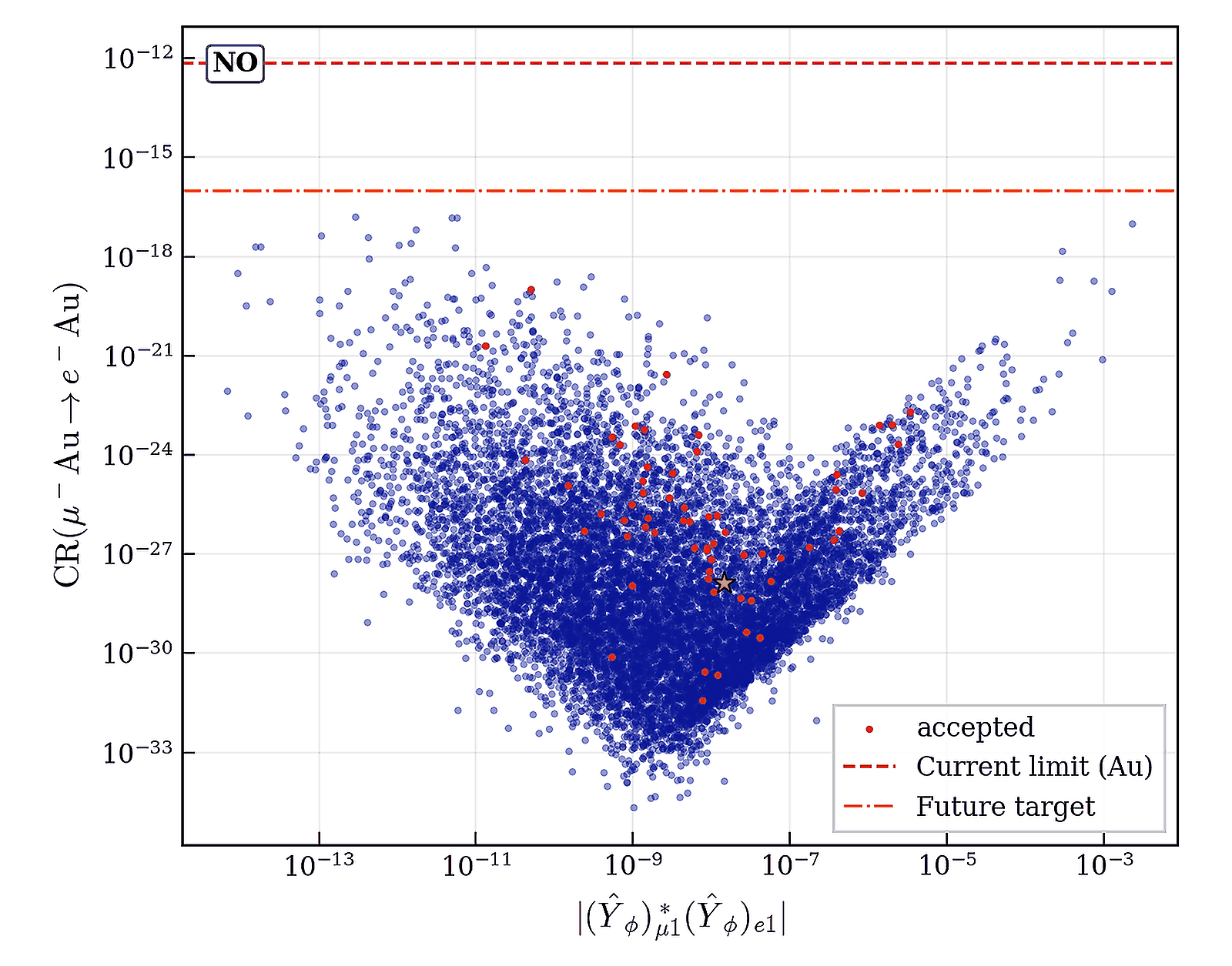}
\caption{From left to right and top to bottom, the panels show radiative decays, three-body decays, and $\mu$--$e$ conversion in gold as functions of $M_{N_1}$ and of the relevant mass-basis Yukawa products. The horizontal lines correspond to current limits from MEG~II, Belle, SINDRUM, and SINDRUM~II, together with projected reaches from MEG~II, Belle~II, and Mu3e~\cite{Baldini:2018uhj,MEGII:2025muegamma,Belle:2021taugamma,Belle:2010tau3l,SINDRUM:1988mu3e,SINDRUMII:2006au,Belle-II:2018jsg,Blondel:2021fji}. The $10^{-16}$ line in the conversion panels is shown only as a reference target sensitivity scale.}
\label{fig7}
\end{figure}
The IO points therefore sit much closer to the current and future neutrinoless double-beta decay sensitivity.
\begin{figure}[!]
\centering
\includegraphics[width=0.32\linewidth]{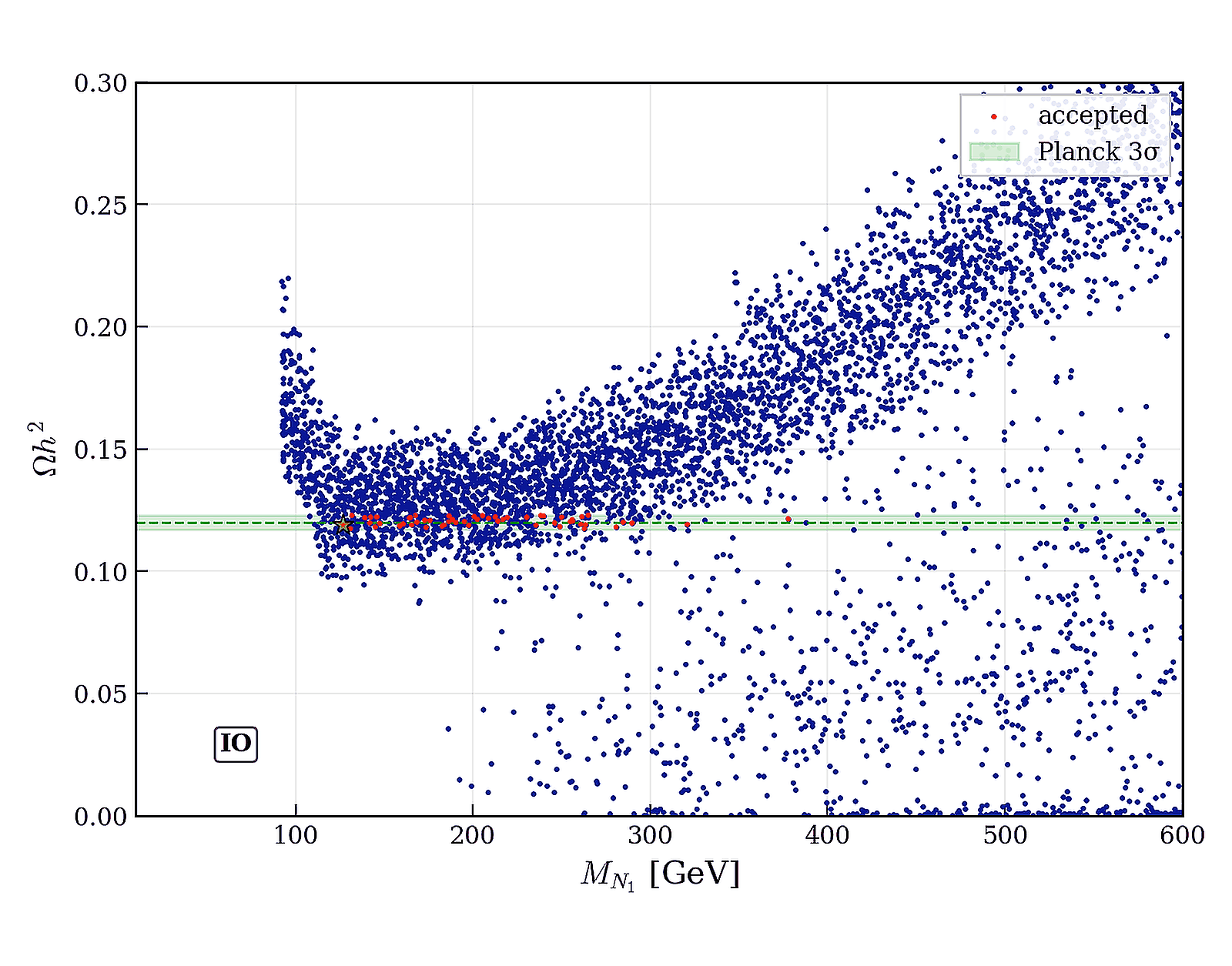}
\includegraphics[width=0.31\linewidth]{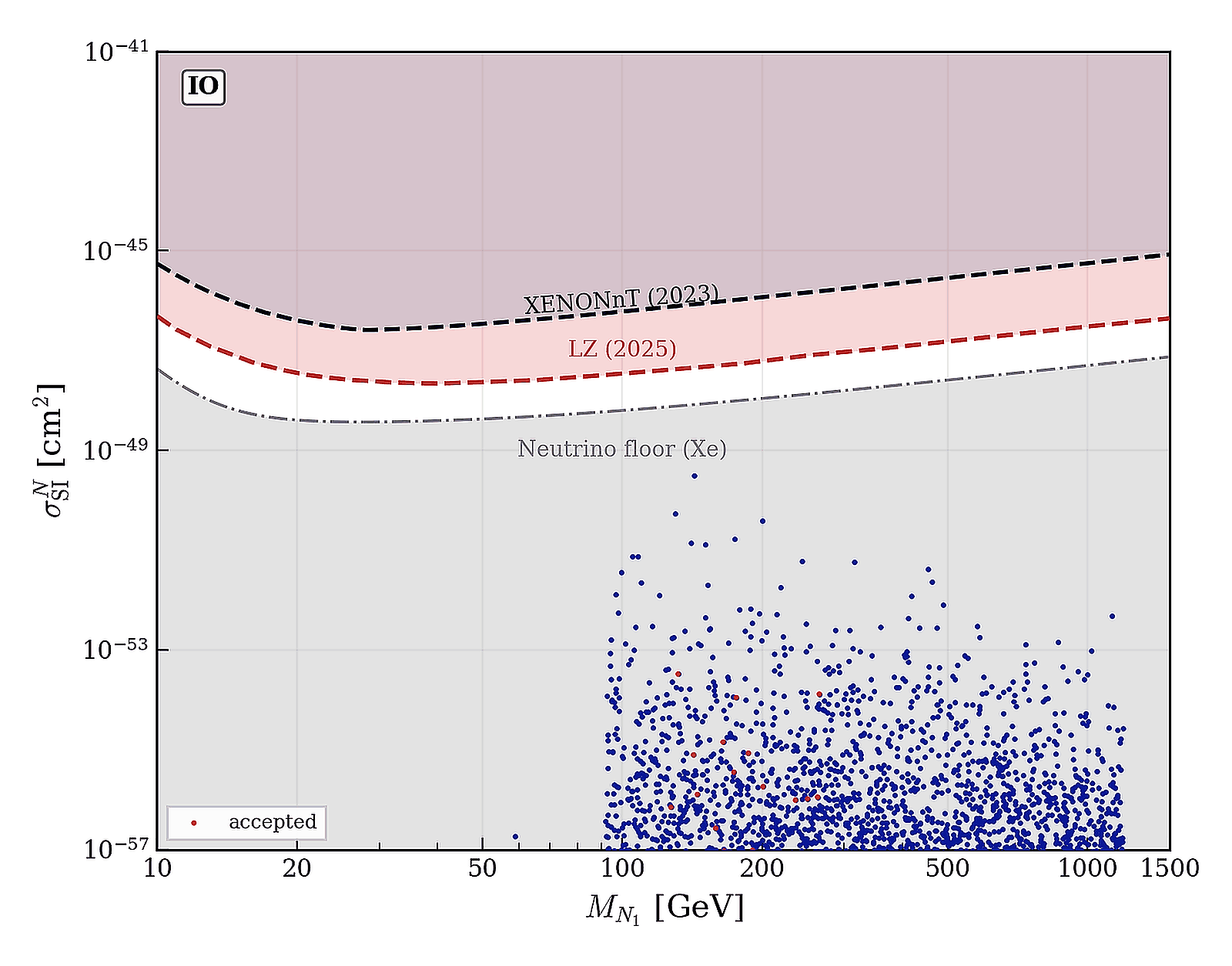}
\includegraphics[width=0.3\linewidth]{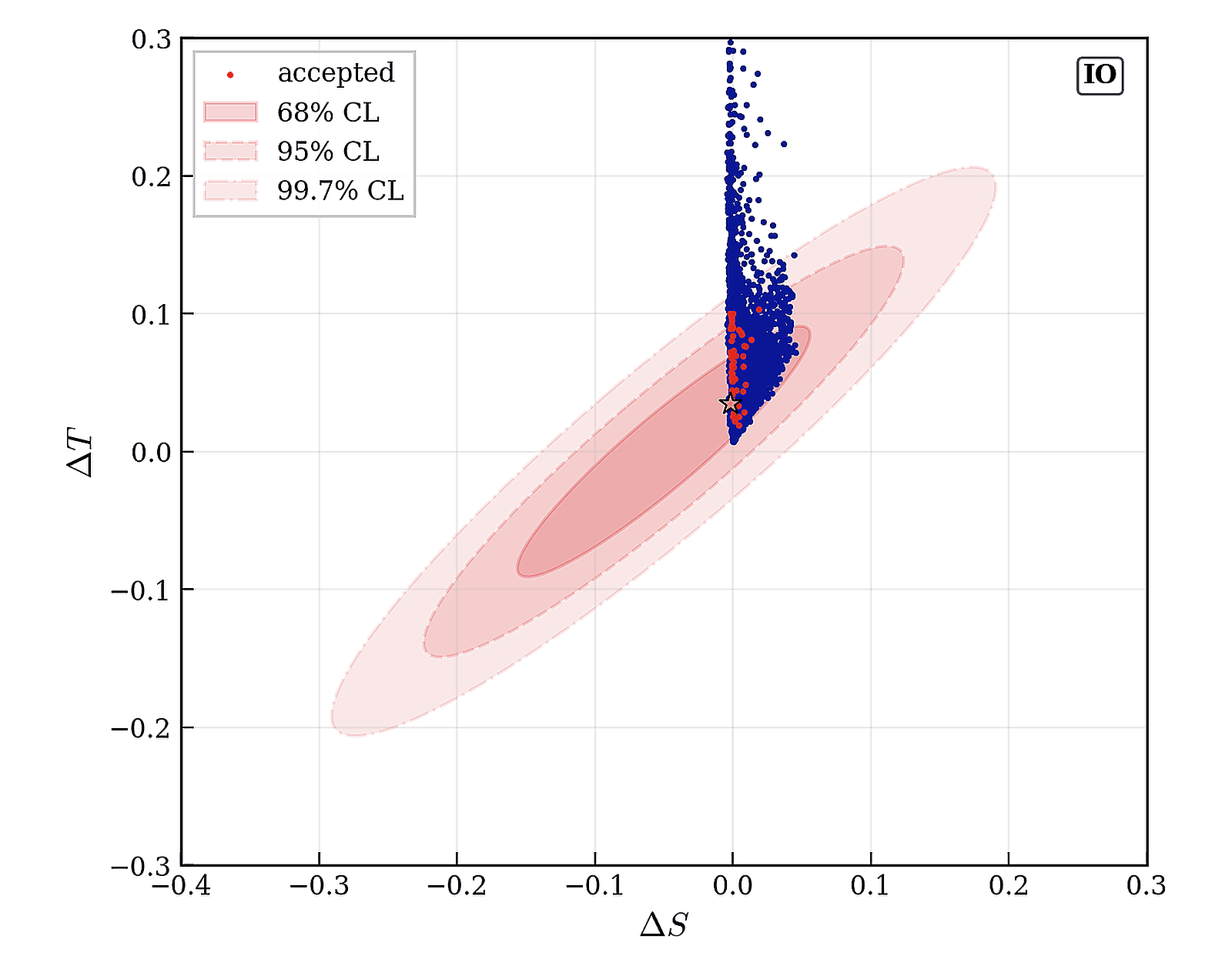}
\caption{Same as figure \ref{fig4}, but for IO.}
\label{fig8}
\end{figure}
The same behaviour is seen for the kinematic mass $m_\beta$. Both orderings remain well below the present KATRIN bound, $m_\beta<0.45$~eV, and also below its design sensitivity of about $0.20$~eV~\cite{KATRIN:2024mass}. The HOLMES target around $0.10$~eV and the Project~8 goal at the few$\times 10^{-2}$~eV level are shown for comparison~\cite{Project8:2022bla,HOLMES:2016qqz}.
\begin{figure}[!]
\centering
\includegraphics[width=0.245\textwidth]{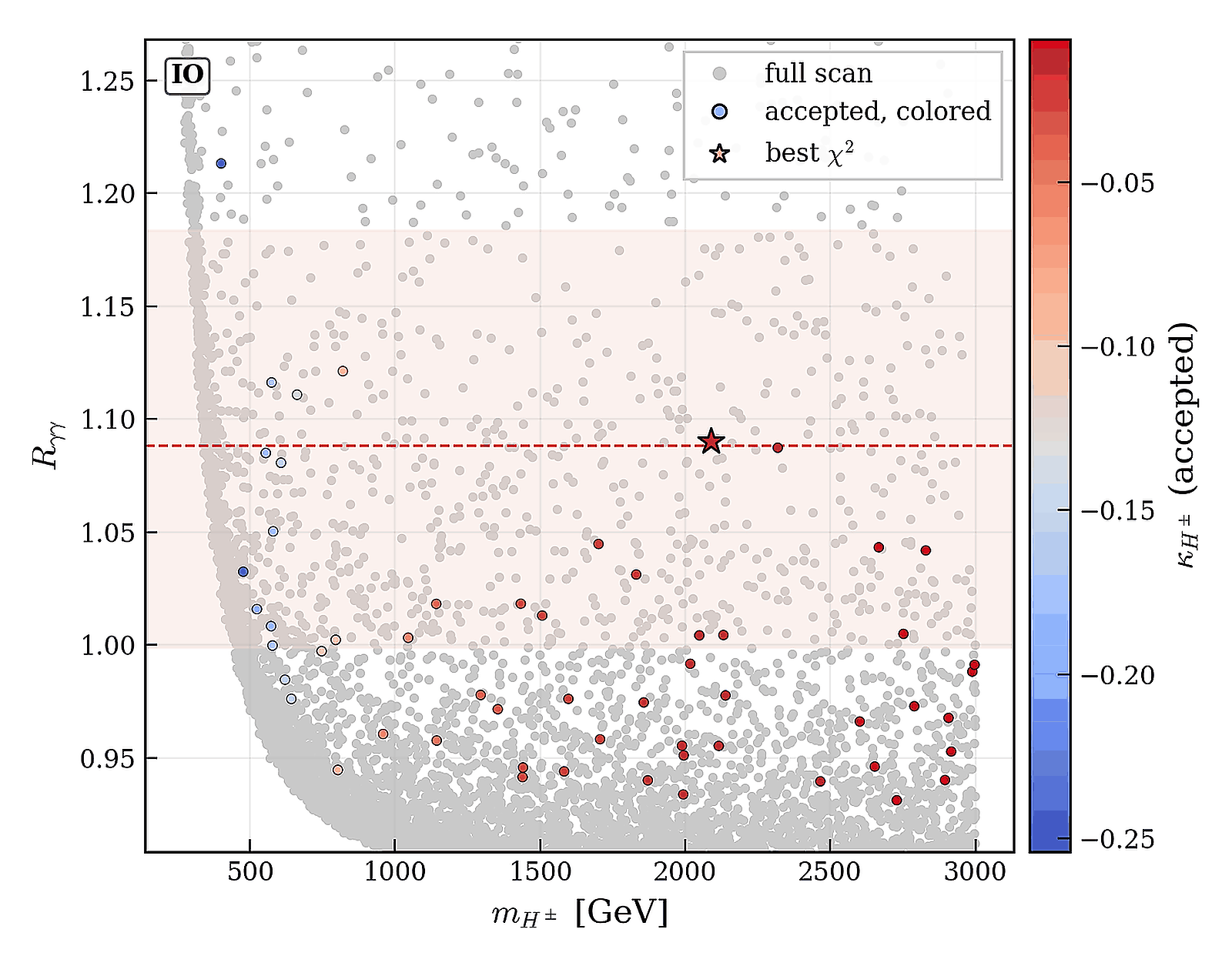}
\includegraphics[width=0.245\linewidth]{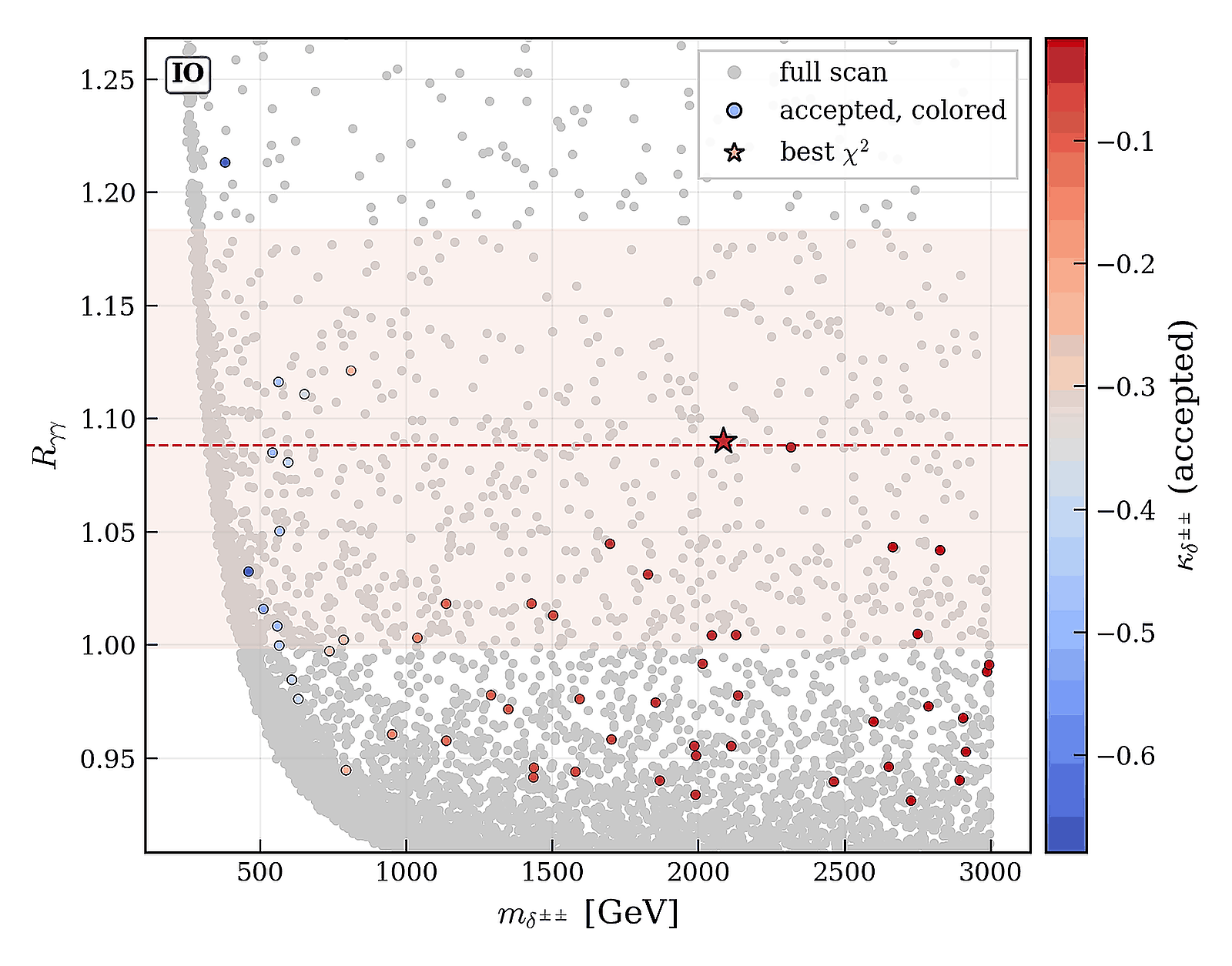}
\includegraphics[width=0.245\linewidth]{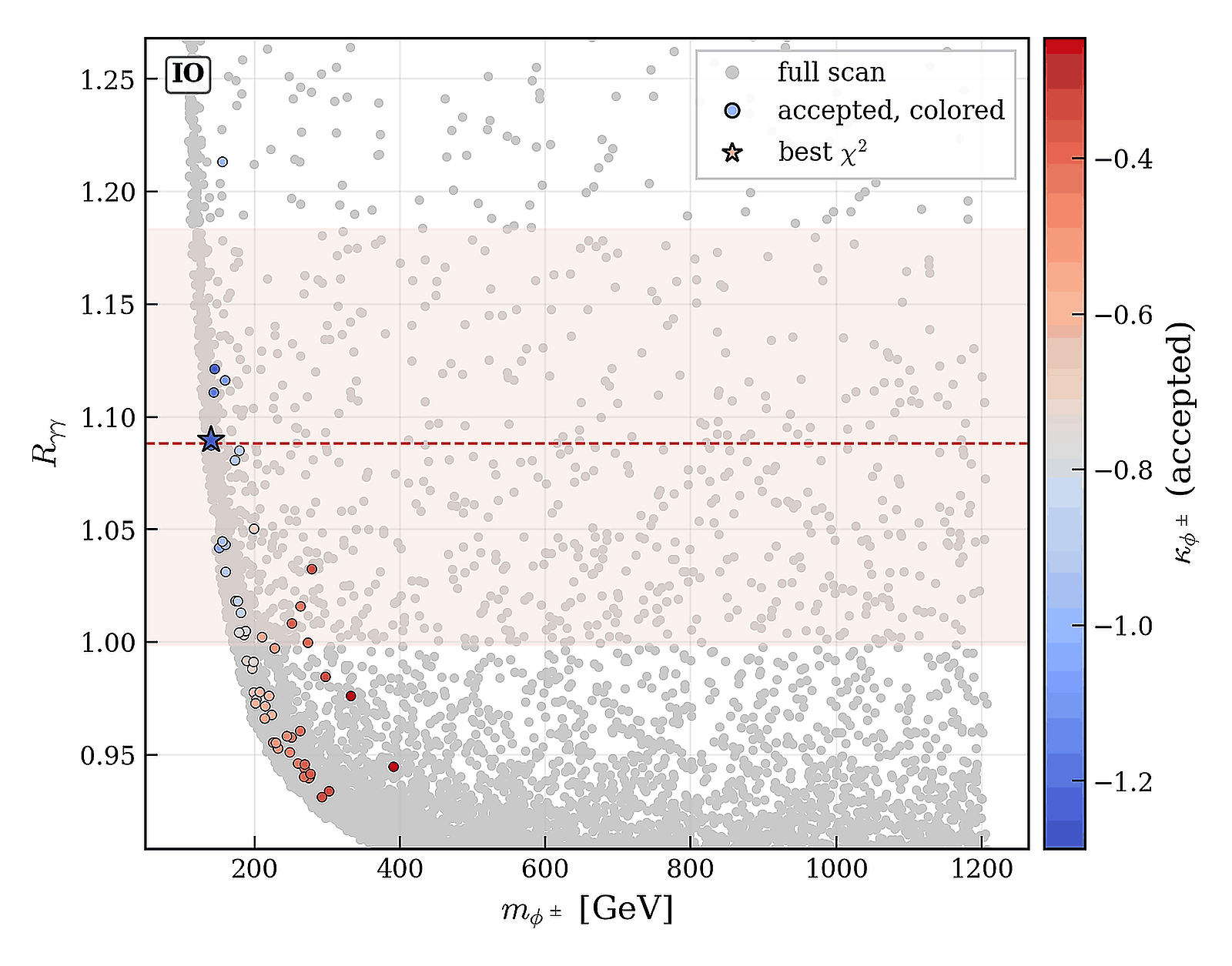}
\includegraphics[width=0.245\linewidth]{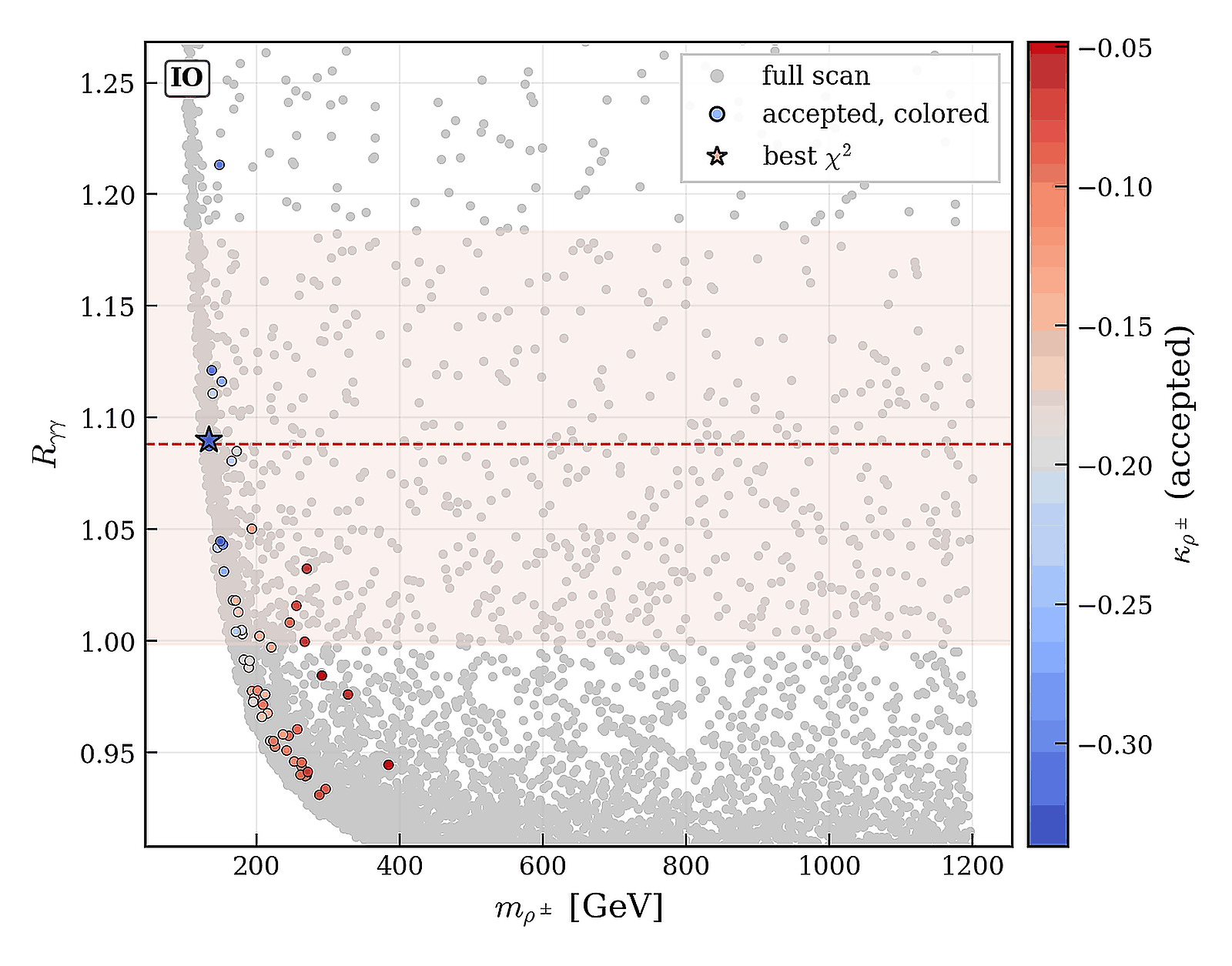}
\caption{Same as figure \ref{fig5}, but for IO.}
\label{fig9}
\end{figure}
Figures~\ref{fig7} and~\ref{fig11} show the cLFV observables. The present bounds used in the scan are summarized in Table~\ref{tab:clfv}. The plots show that all allowed points lie below the current limits for the radiative decays, the three-body decays, and $\mu$--$e$ conversion in gold. The radiative and three-body branching ratios follow the expected dependence on the relevant mass-basis Yukawa combinations, and the conversion rate in gold shows the same pattern. 
\begin{figure}[!]
\centering
\includegraphics[width=0.245\textwidth]{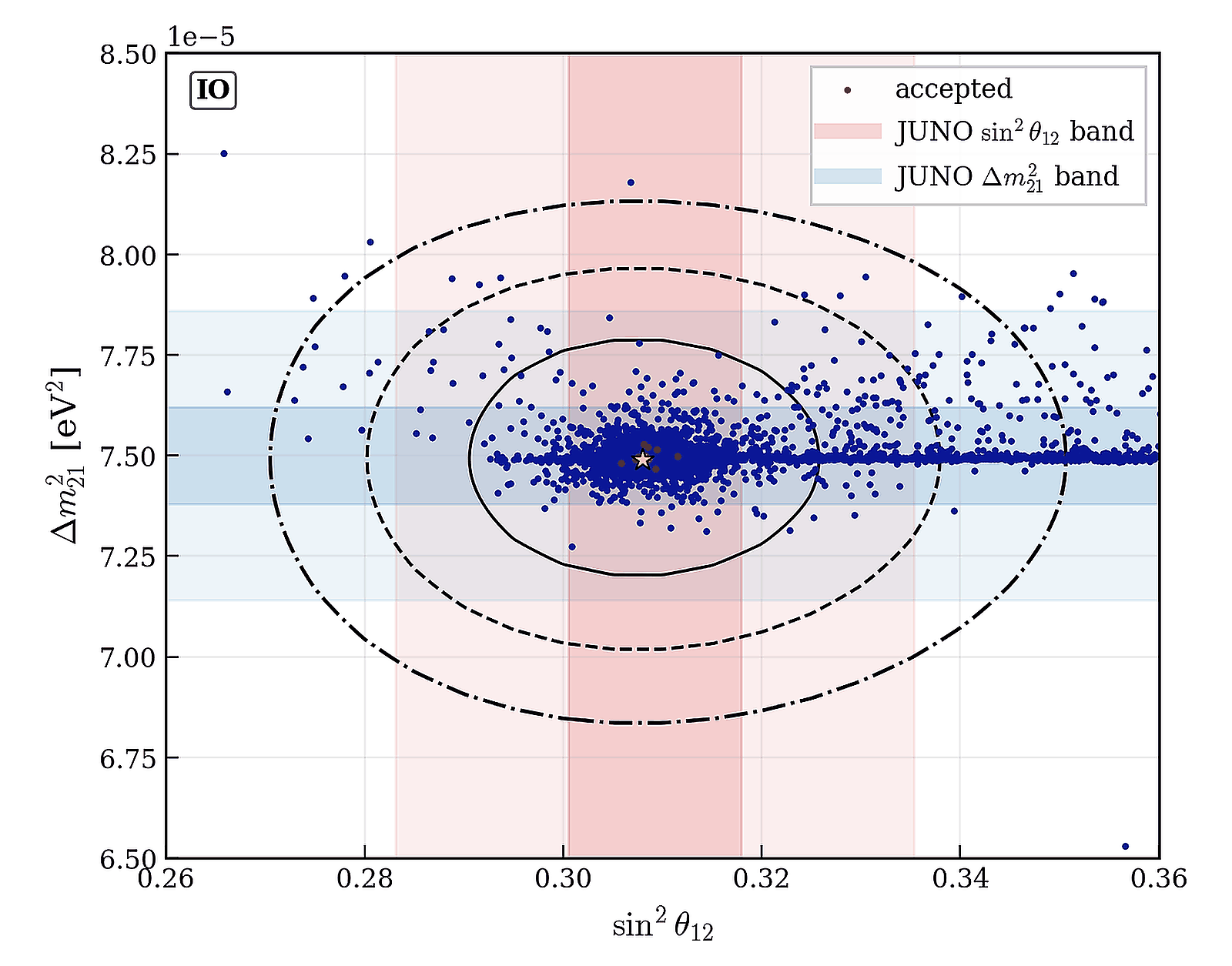}
\includegraphics[width=0.245\linewidth]{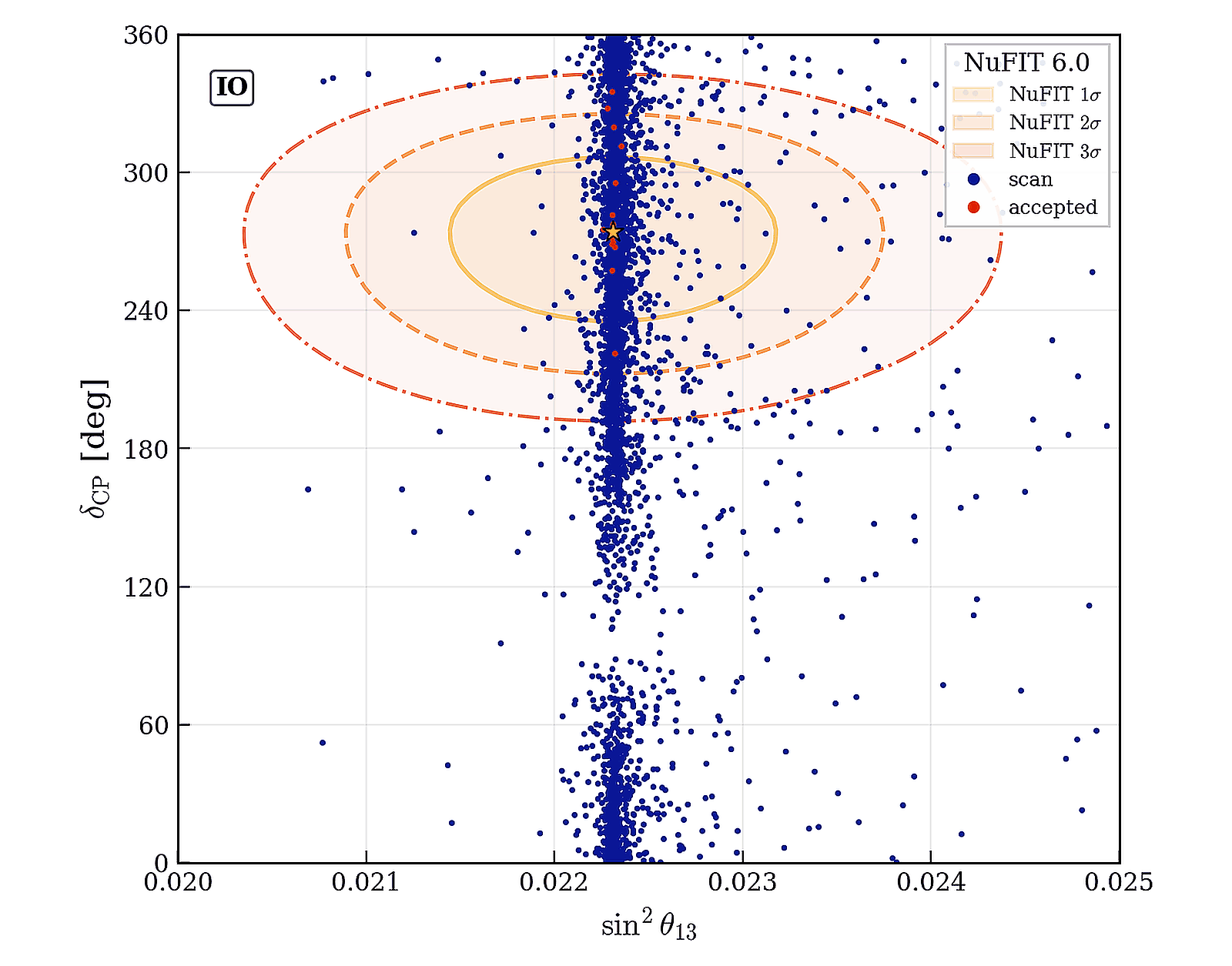}
\includegraphics[width=0.245\linewidth]{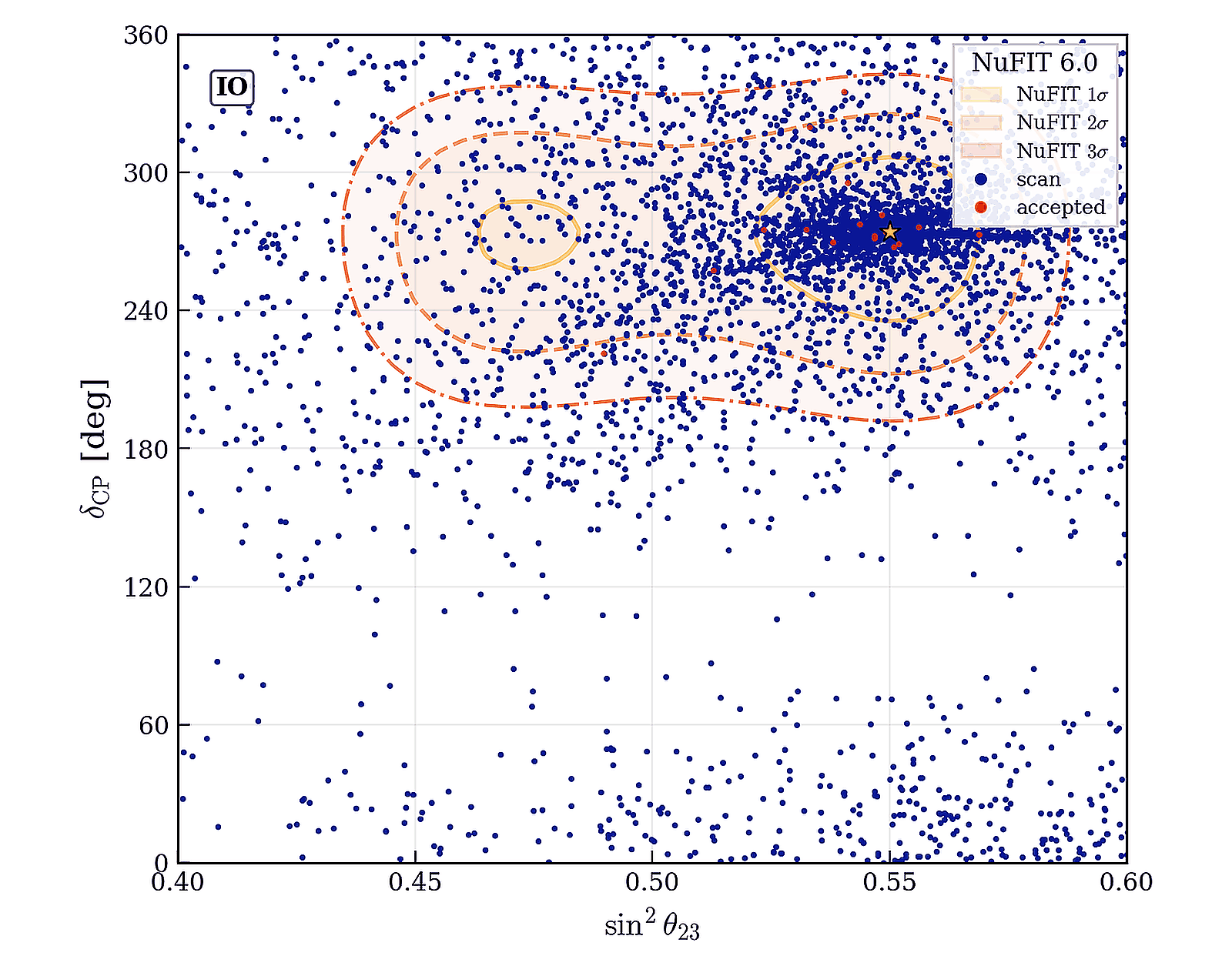}
\includegraphics[width=0.245\linewidth]{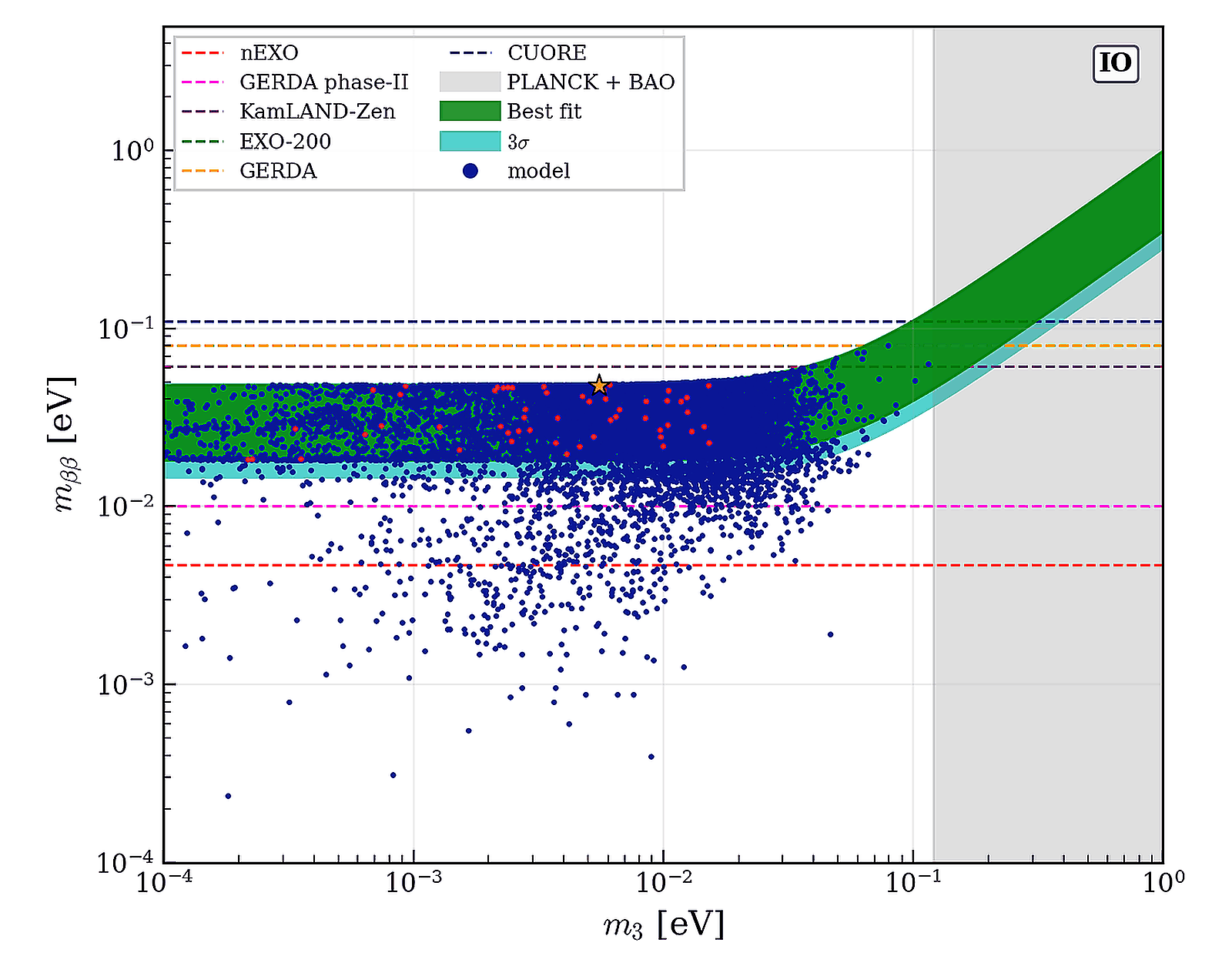}
\includegraphics[width=0.245\linewidth]{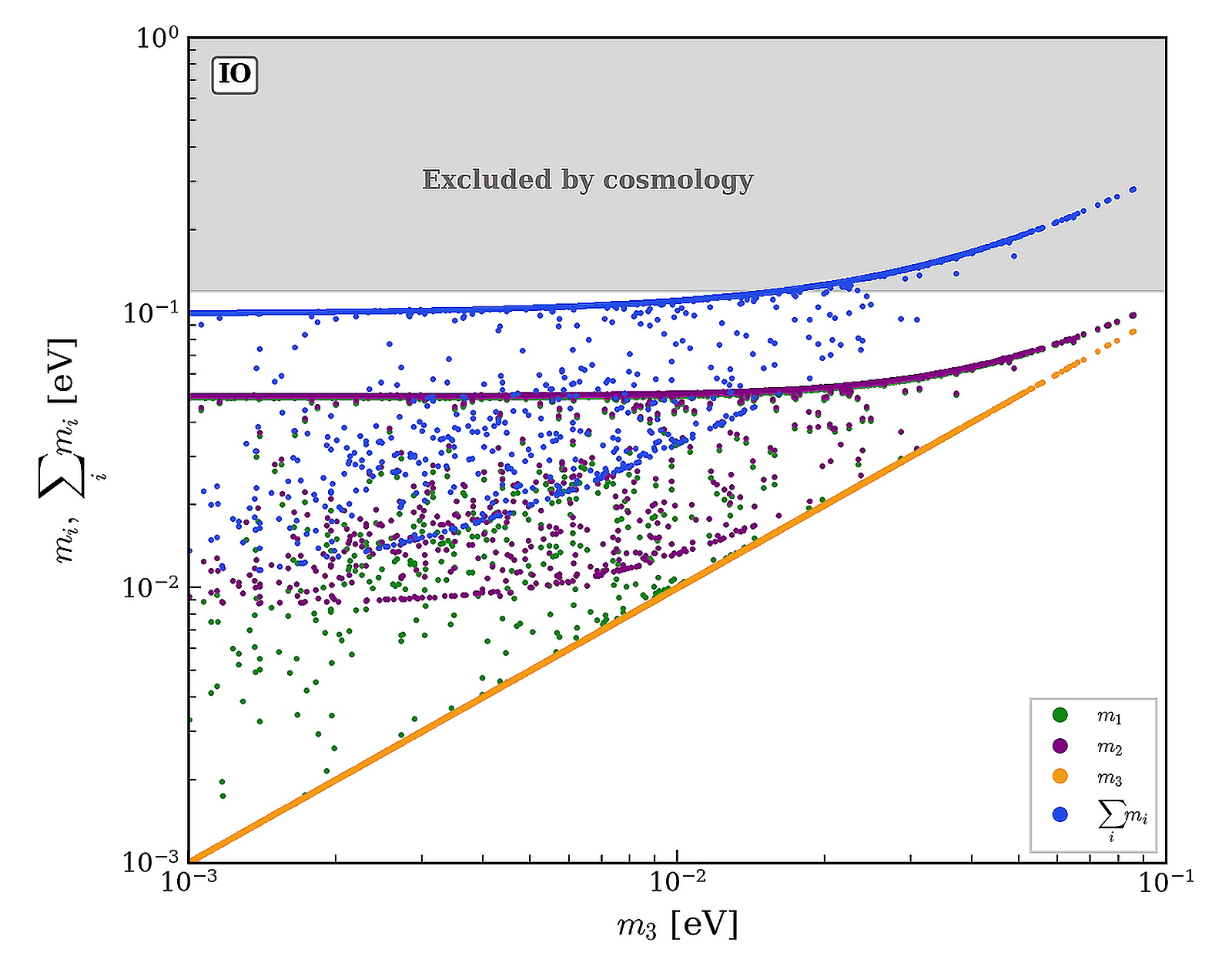}
\includegraphics[width=0.245\linewidth]{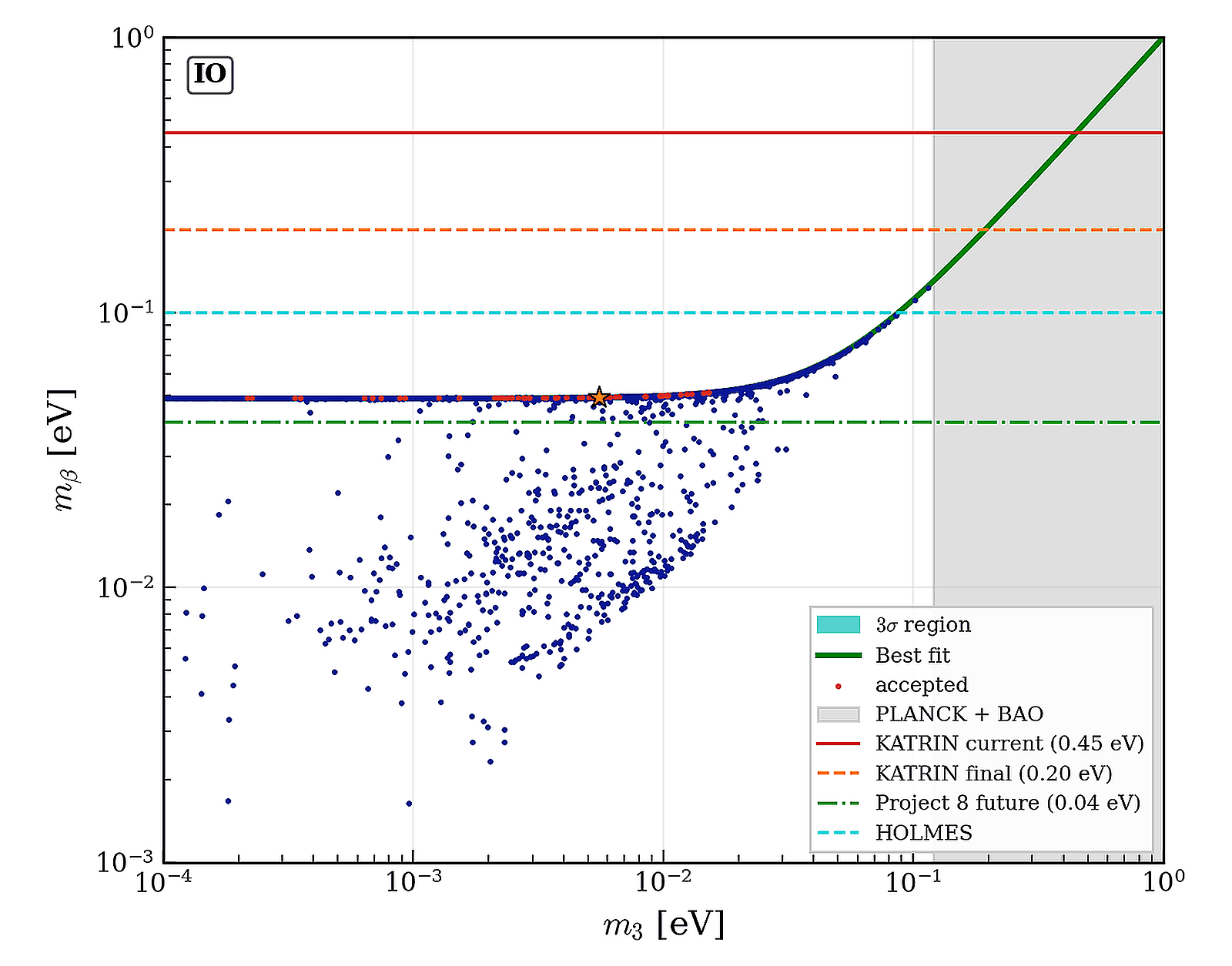}
\includegraphics[width=0.245\linewidth]{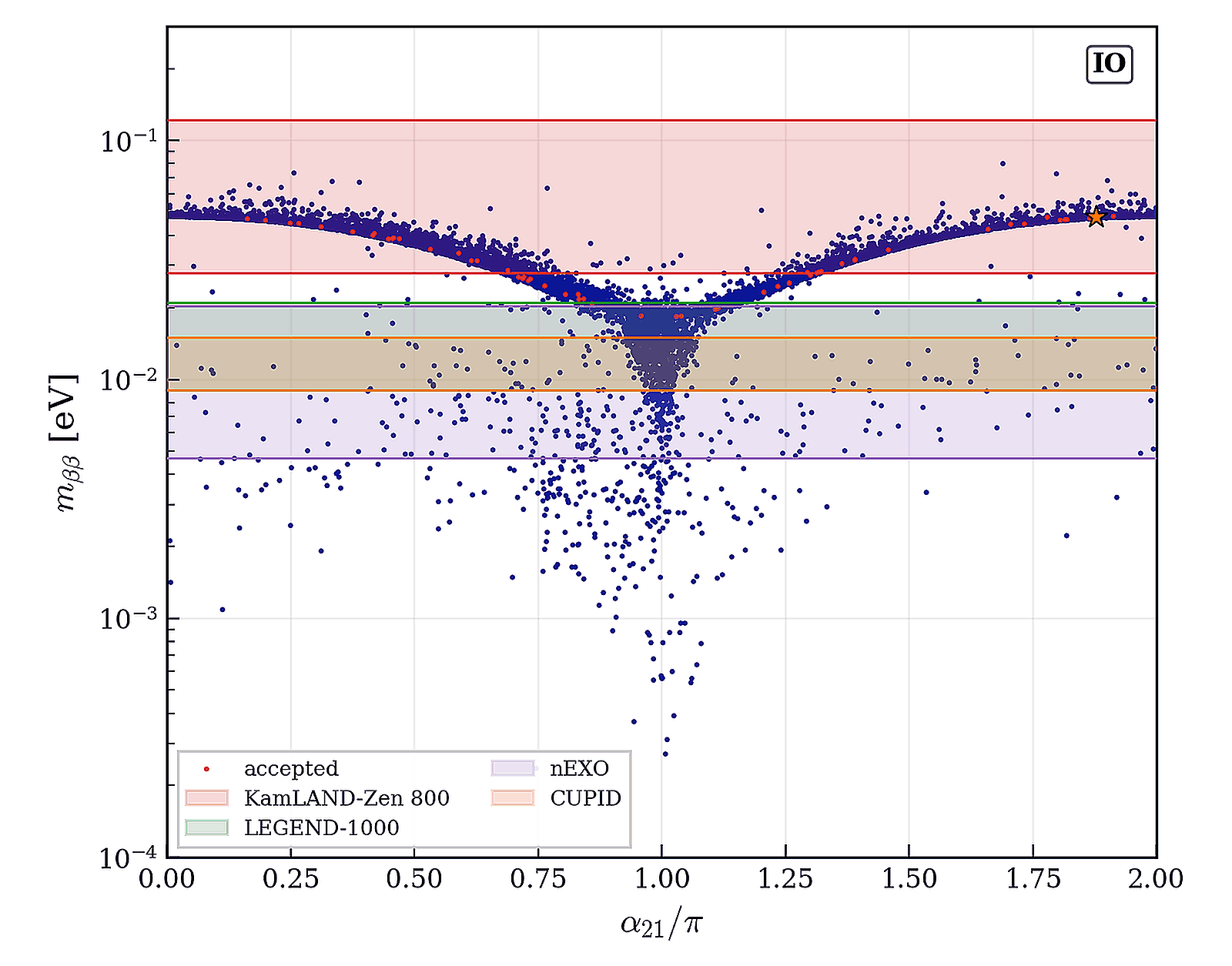}
\includegraphics[width=0.245\linewidth]{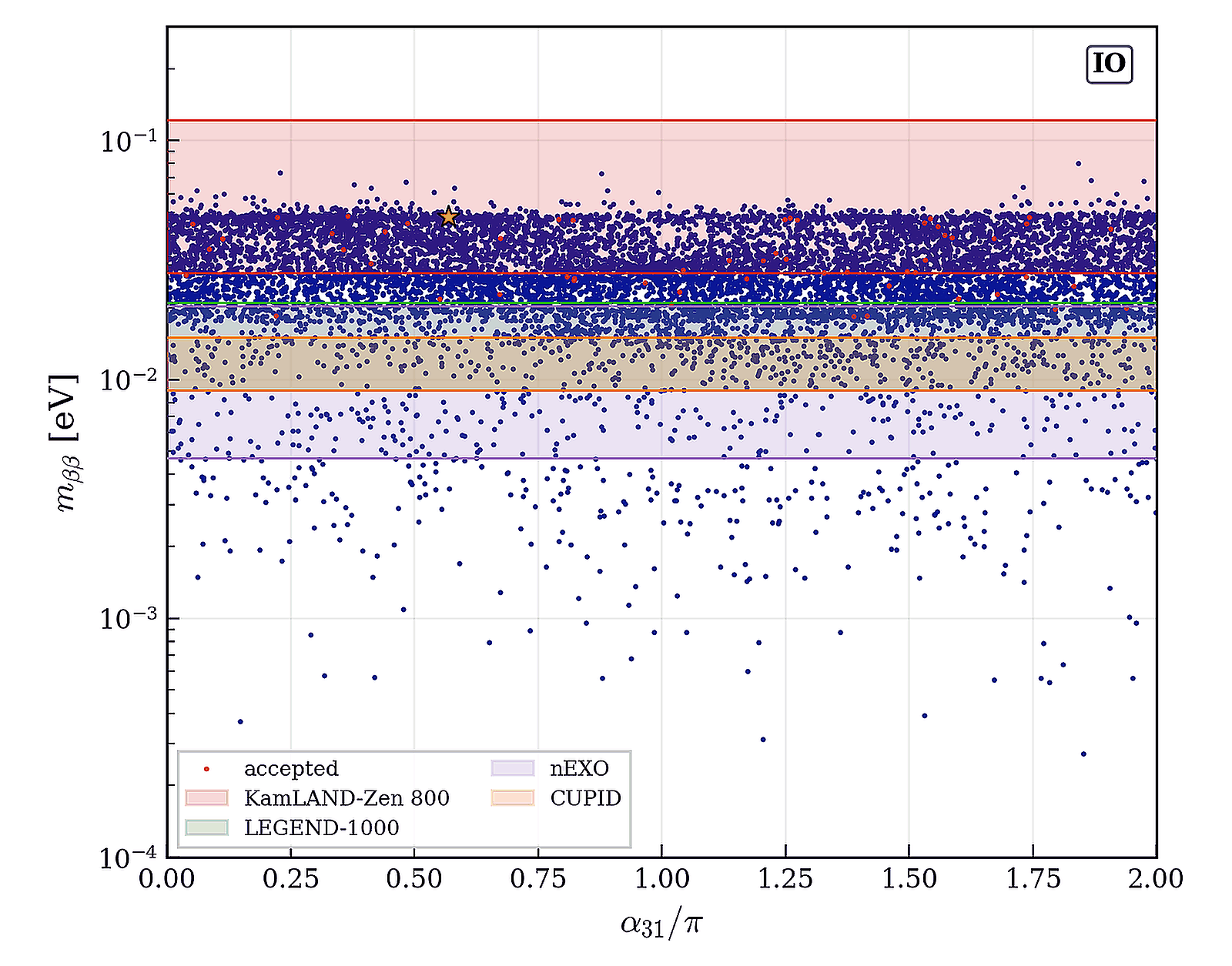}
\caption{Same as figure \ref{fig6}, but for IO.}
\label{fig10}
\end{figure}
At the same time, the figures make clear that cLFV is not what selects the surviving points in the present scan. The more restrictive conditions come from relic density, neutrino data, electroweak precision observables, and the diphoton rate. Still, the projected sensitivities shown in the plots indicate that future searches can test part of the remaining parameter space, especially through $\mu\to e\gamma$ at the level of $6\times10^{-14}$, $\mu\to 3e$ down to $2\times10^{-15}$ in Mu3e phase~I and $10^{-16}$ in phase~II, and the Belle~II reach for the $\tau$ modes near $10^{-9}$--$10^{-10}$~\cite{Baldini:2018uhj,Belle-II:2018jsg,Blondel:2021fji}.
\begin{figure}[!]
\centering
\includegraphics[width=0.245\textwidth]{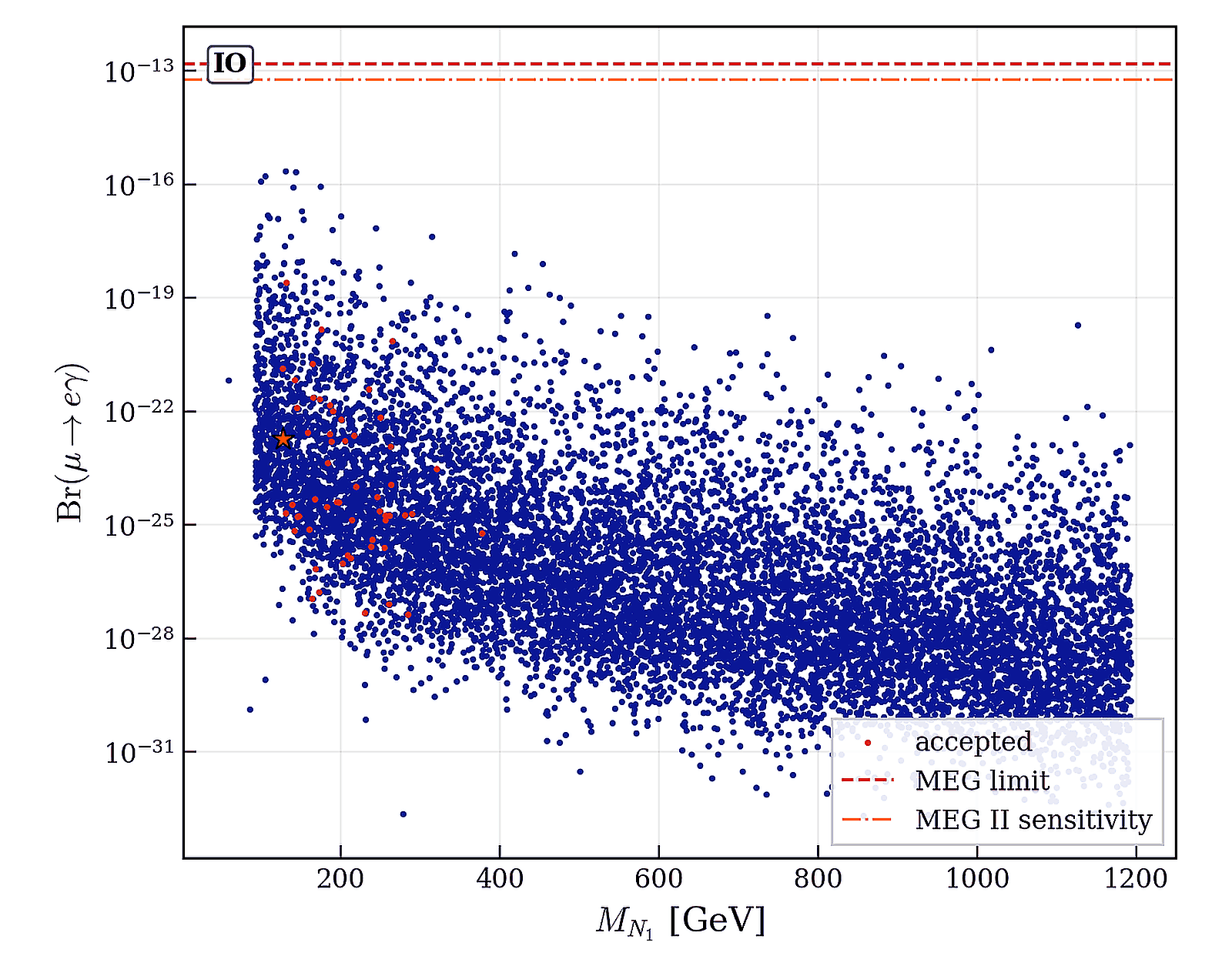}
\includegraphics[width=0.245\linewidth]{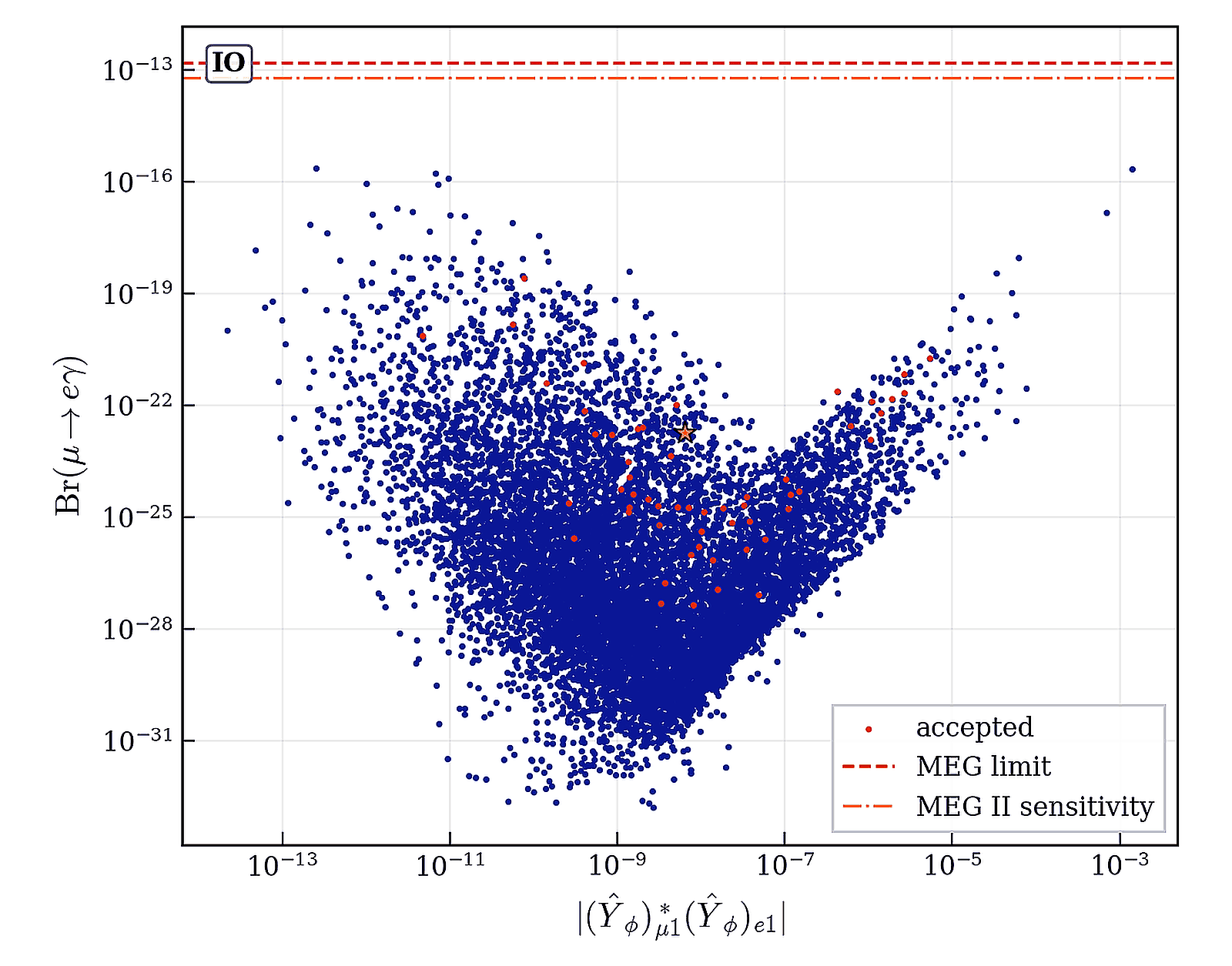}
\includegraphics[width=0.245\linewidth]{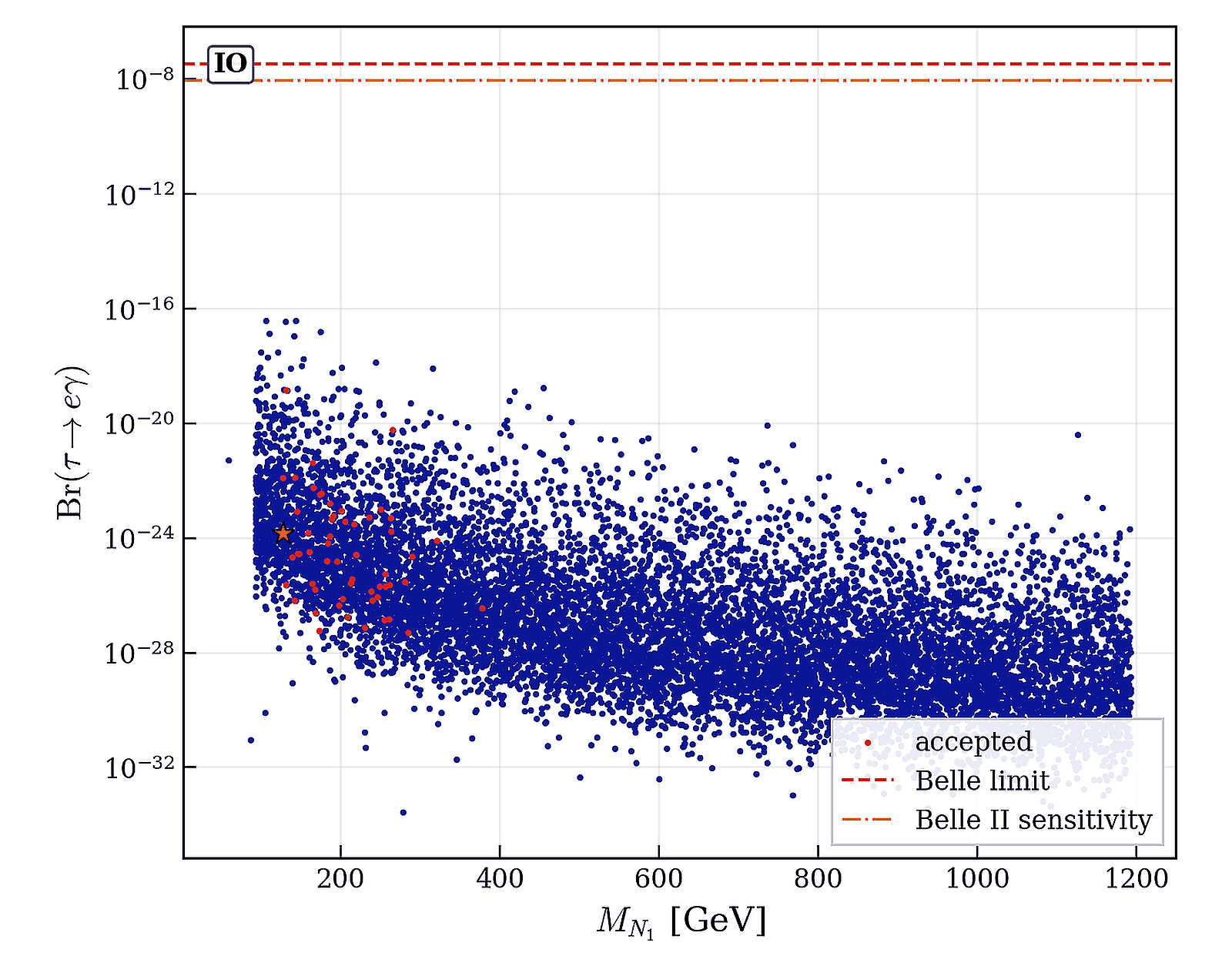}
\includegraphics[width=0.245\linewidth]{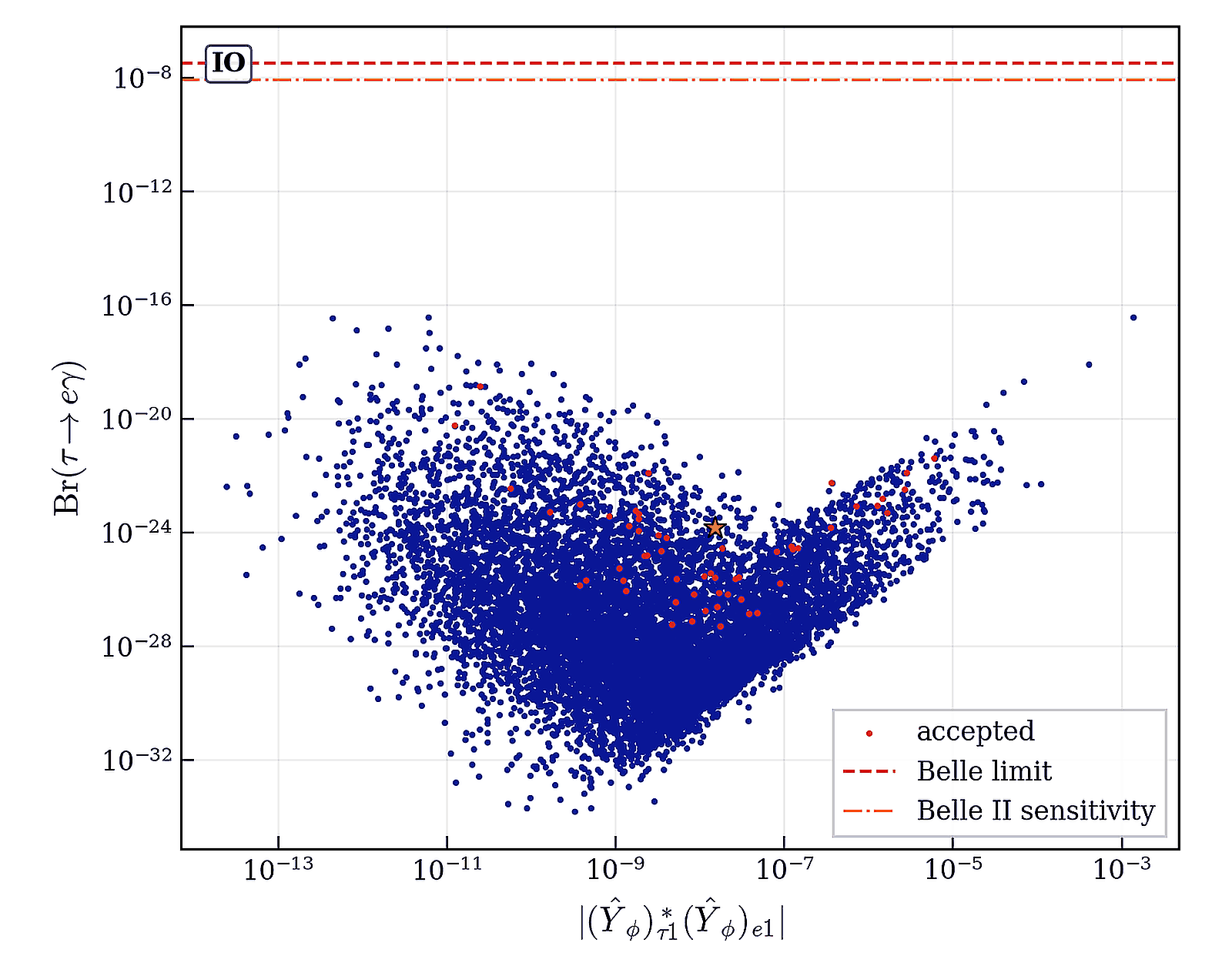}
\includegraphics[width=0.245\linewidth]{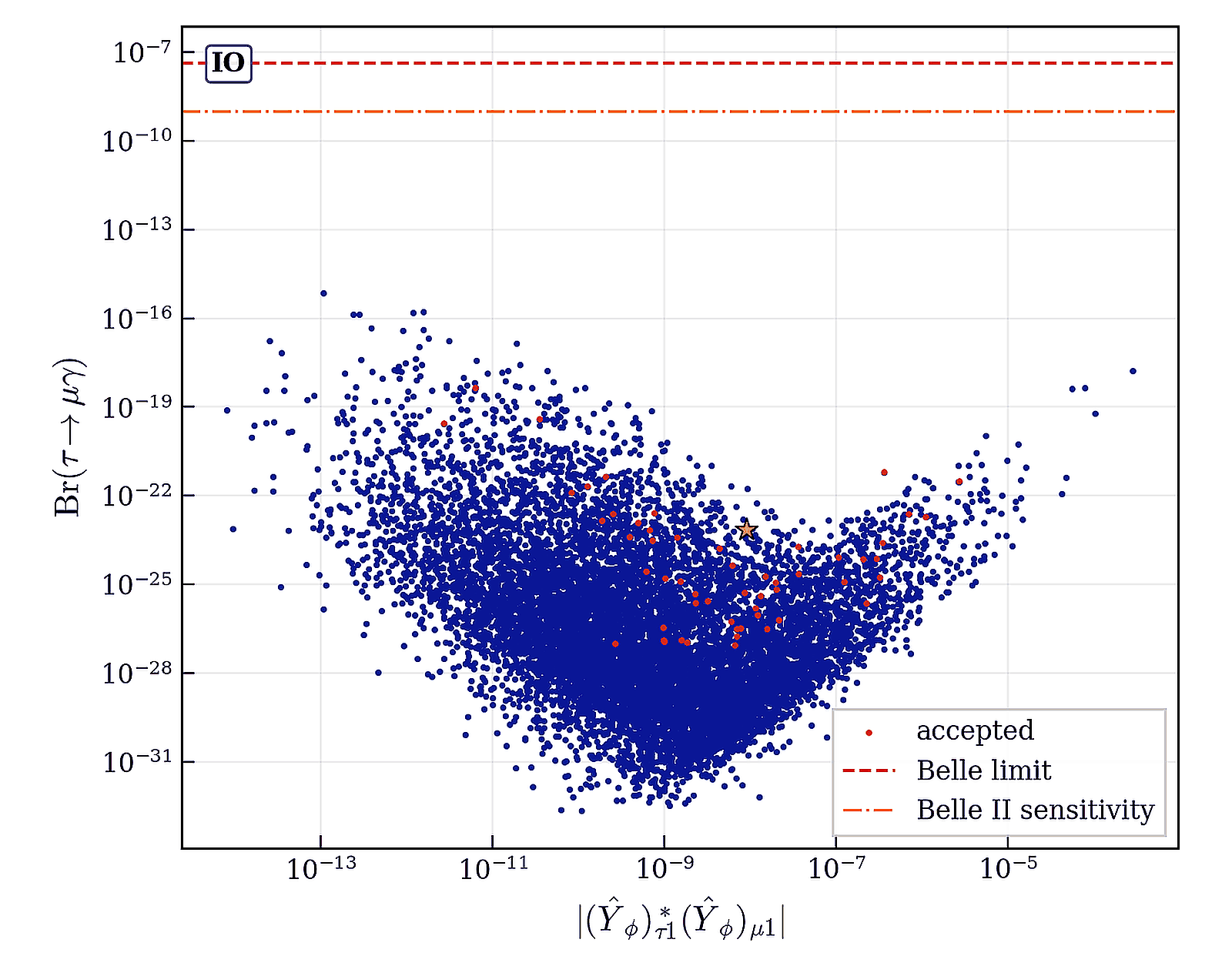}
\includegraphics[width=0.245\linewidth]{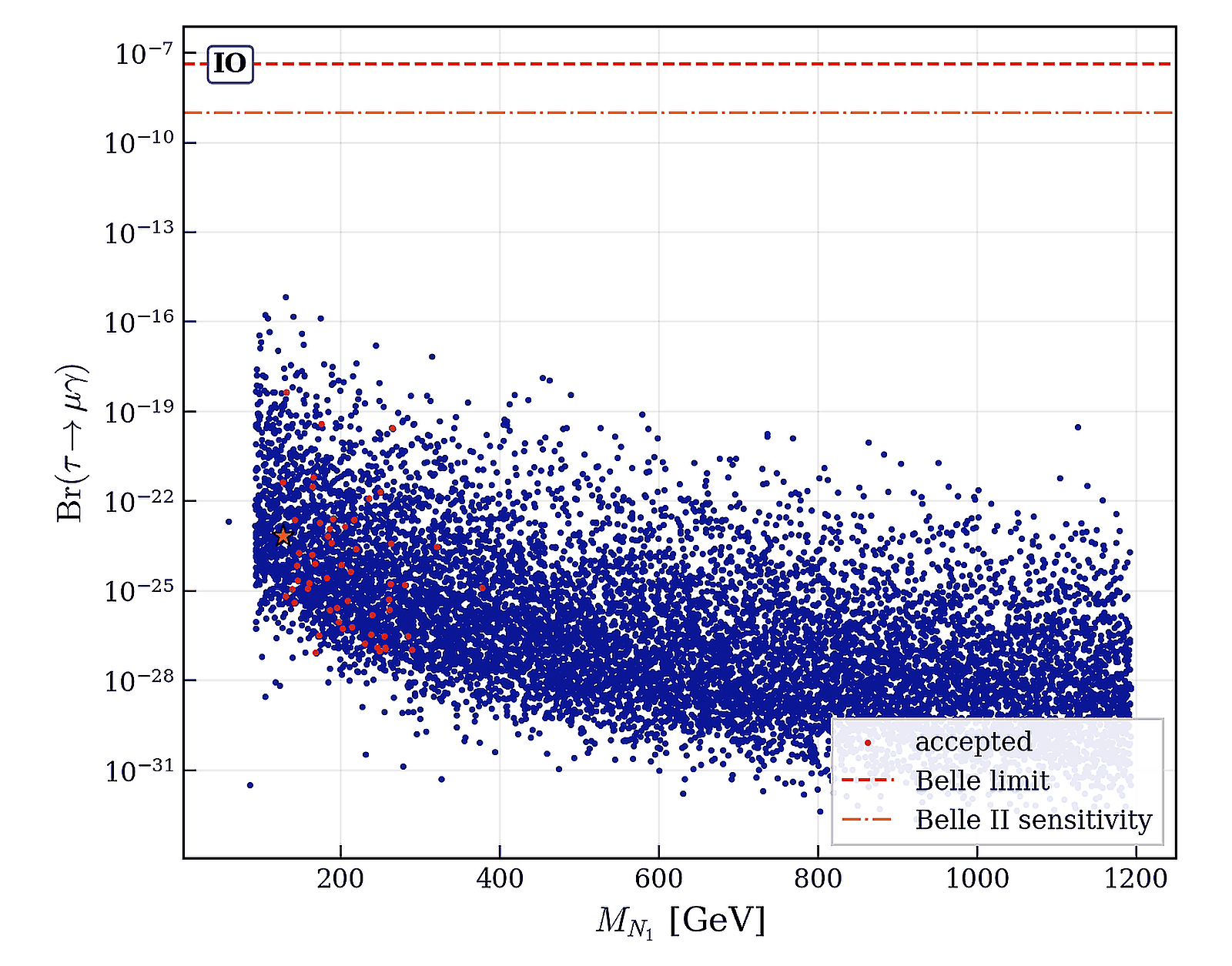}
\includegraphics[width=0.245\linewidth]{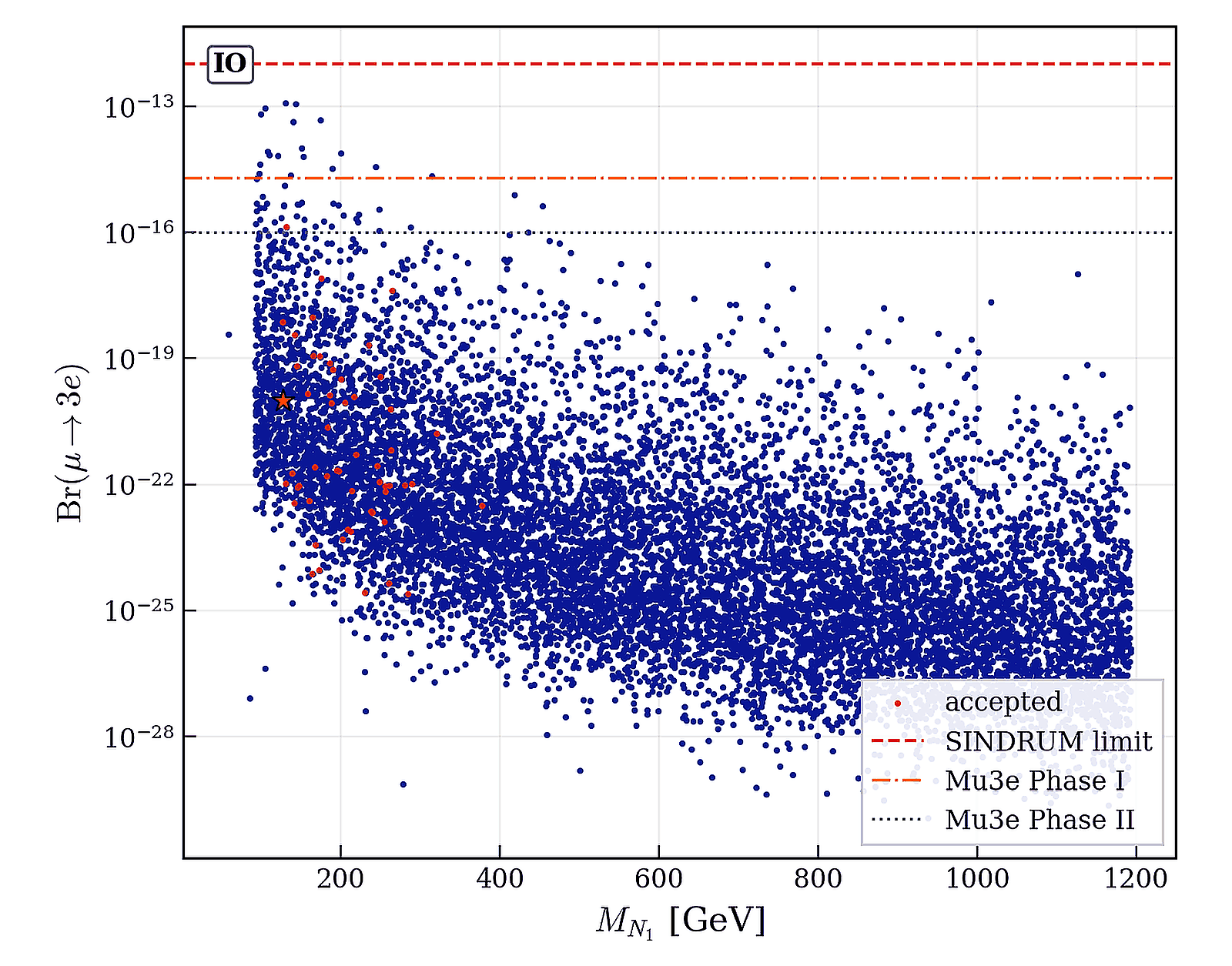}
\includegraphics[width=0.245\linewidth]{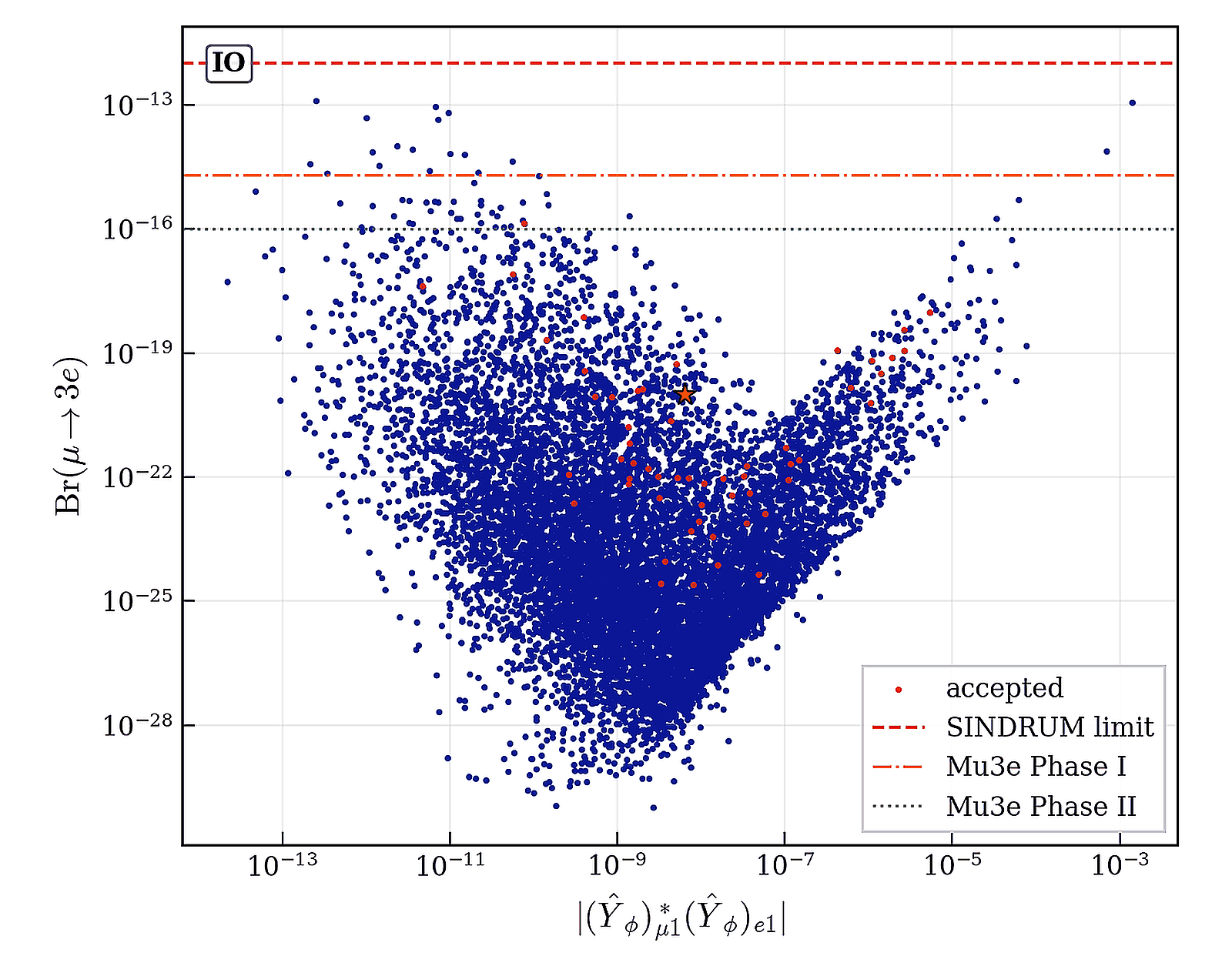}
\includegraphics[width=0.245\linewidth]{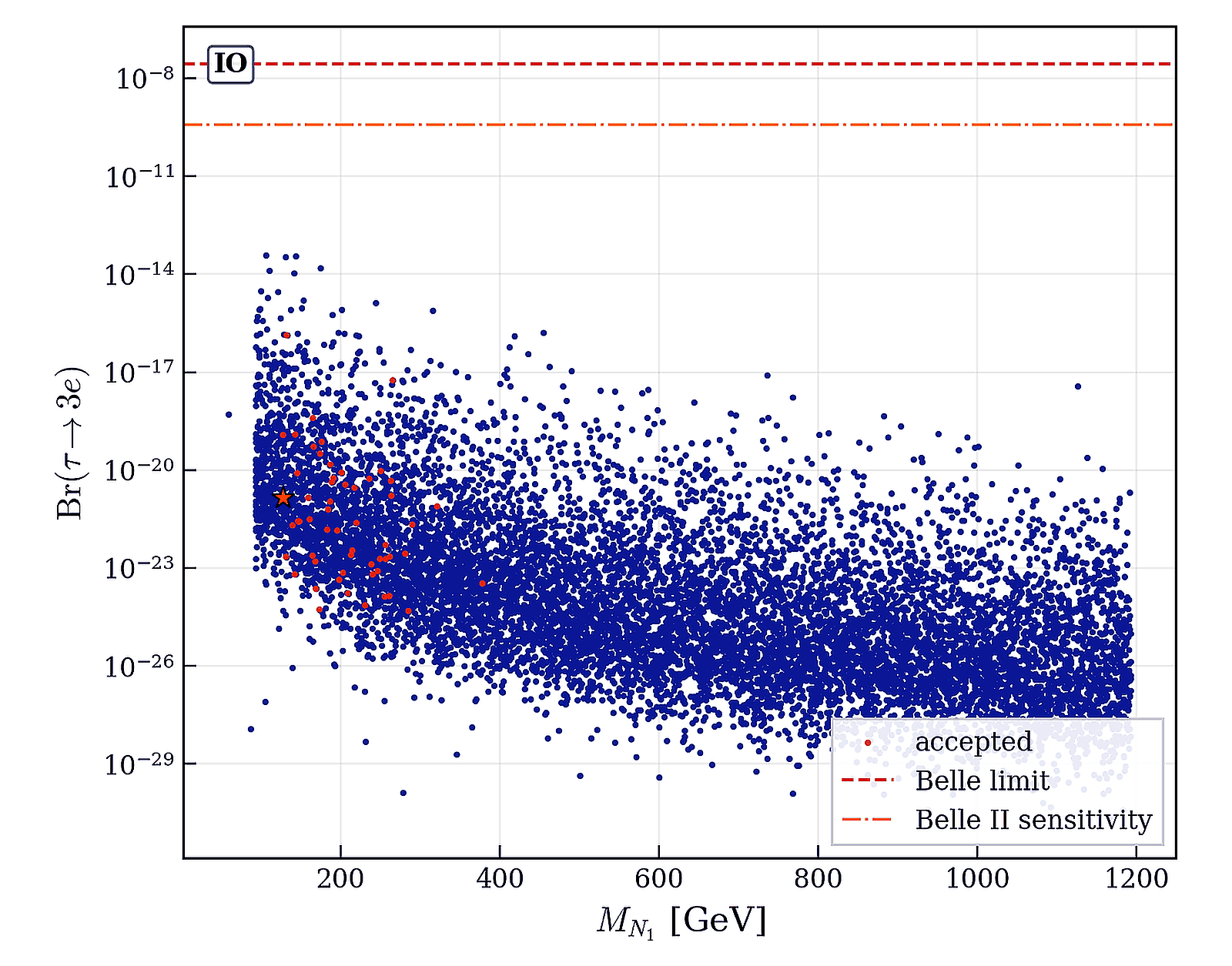}
\includegraphics[width=0.245\linewidth]{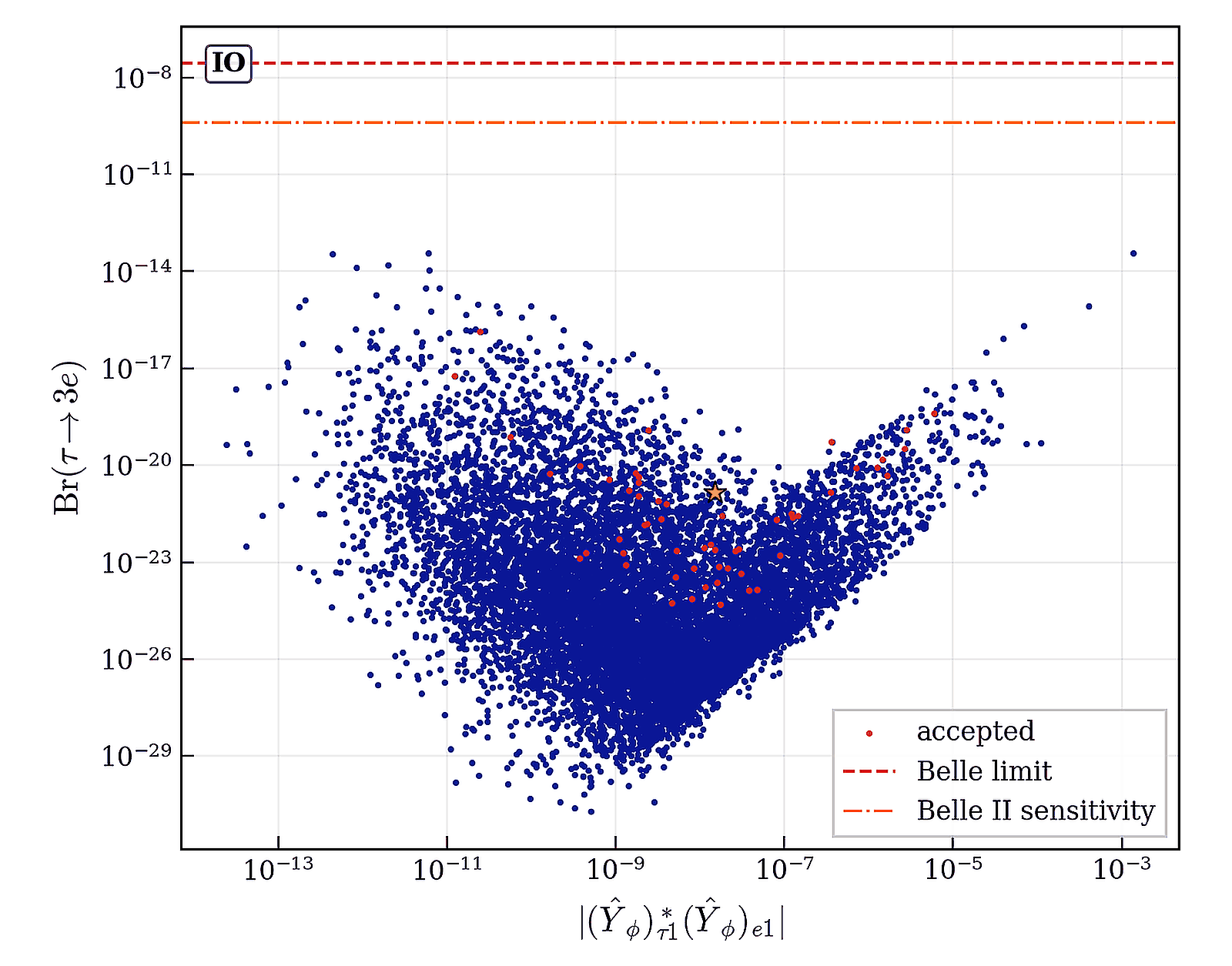}
\includegraphics[width=0.245\linewidth]{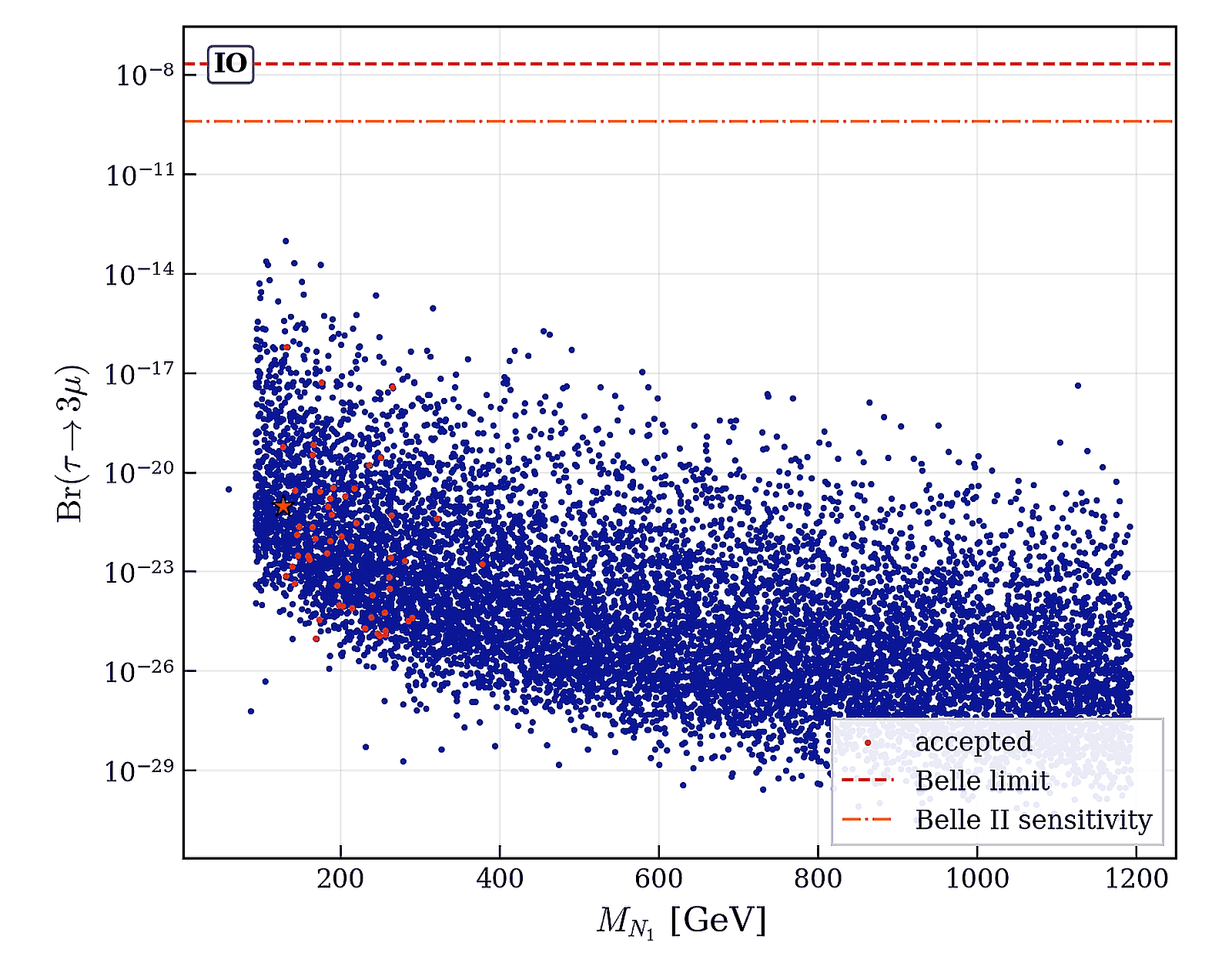}
\includegraphics[width=0.245\linewidth]{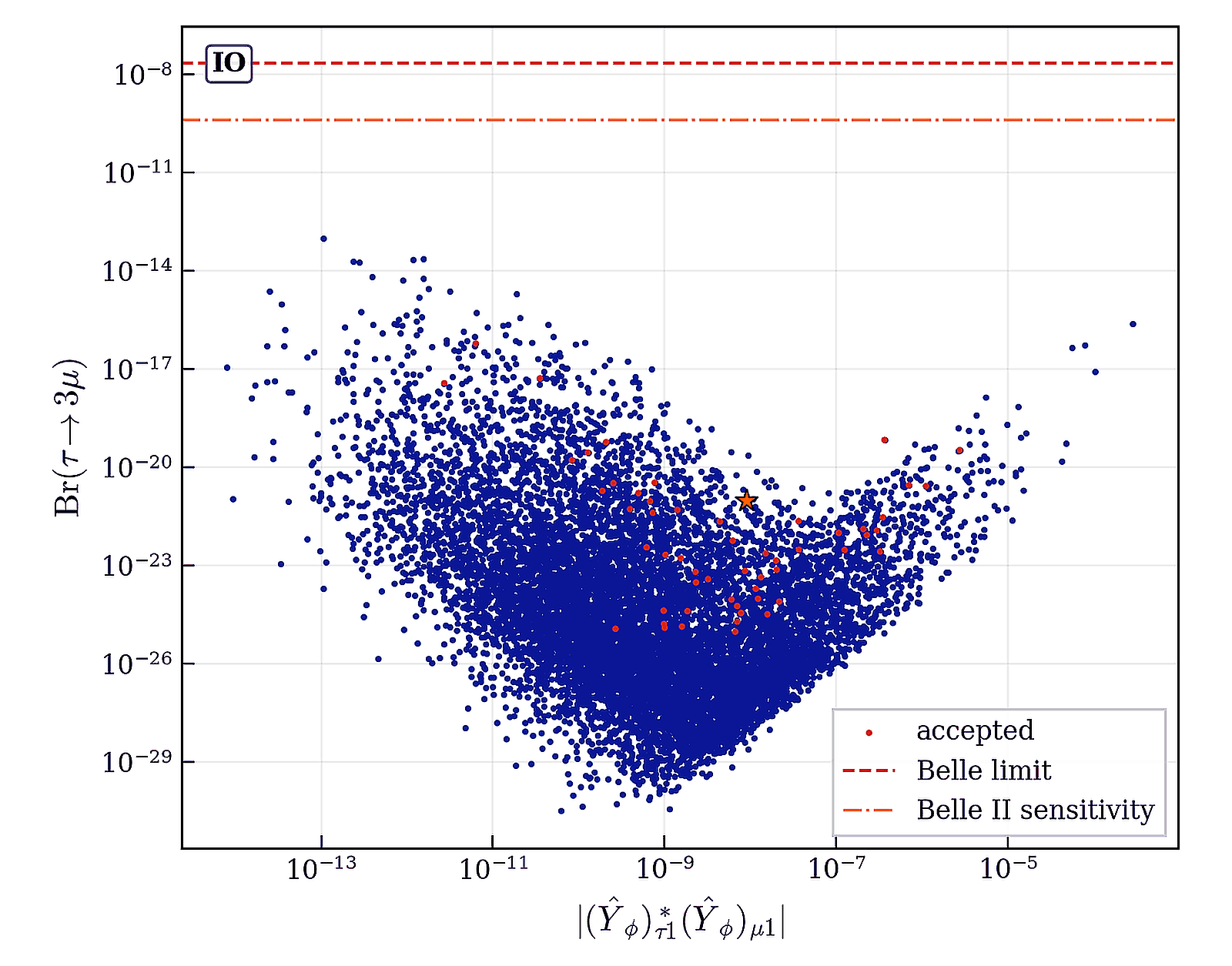}
\includegraphics[width=0.245\linewidth]{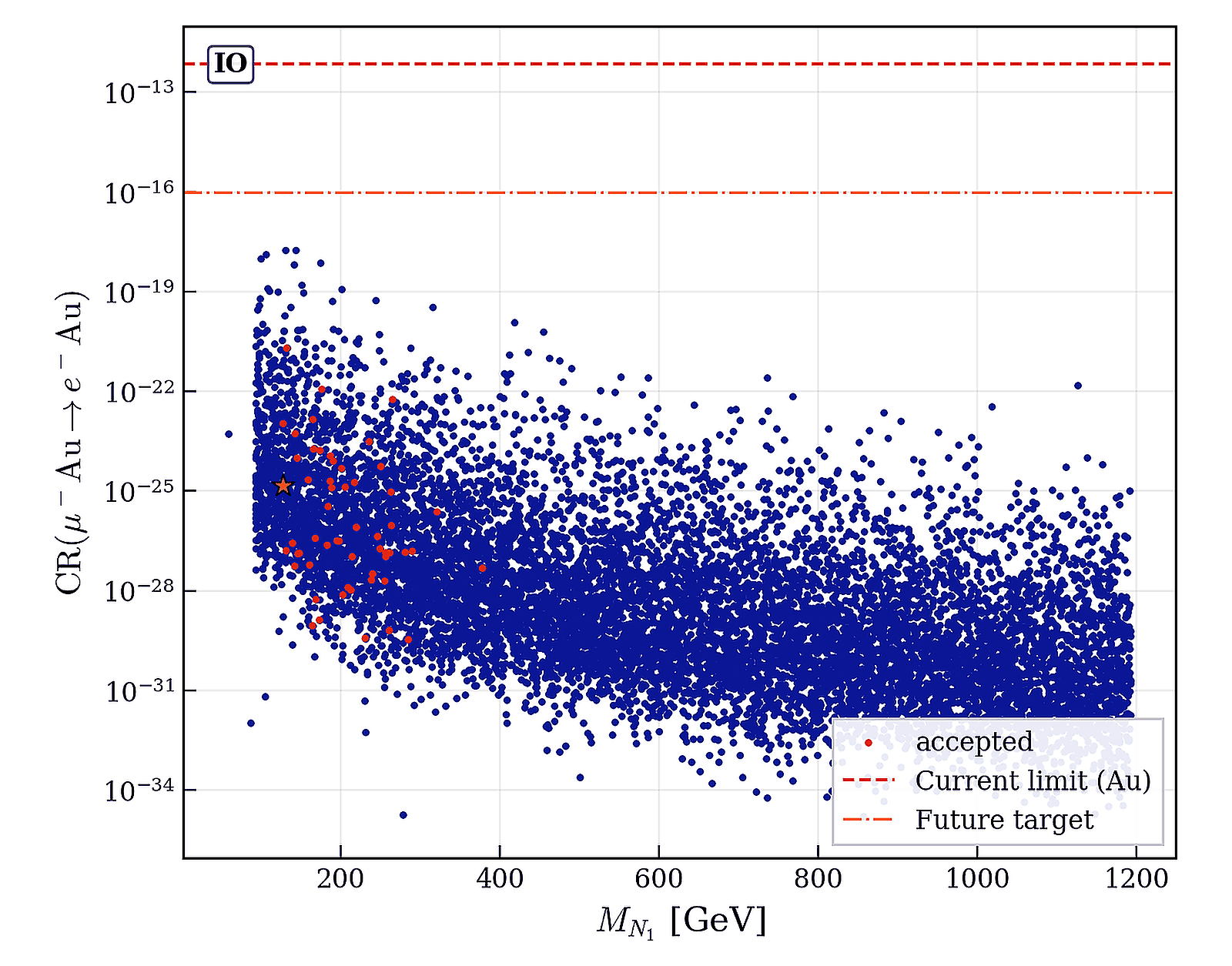}
\includegraphics[width=0.245\linewidth]{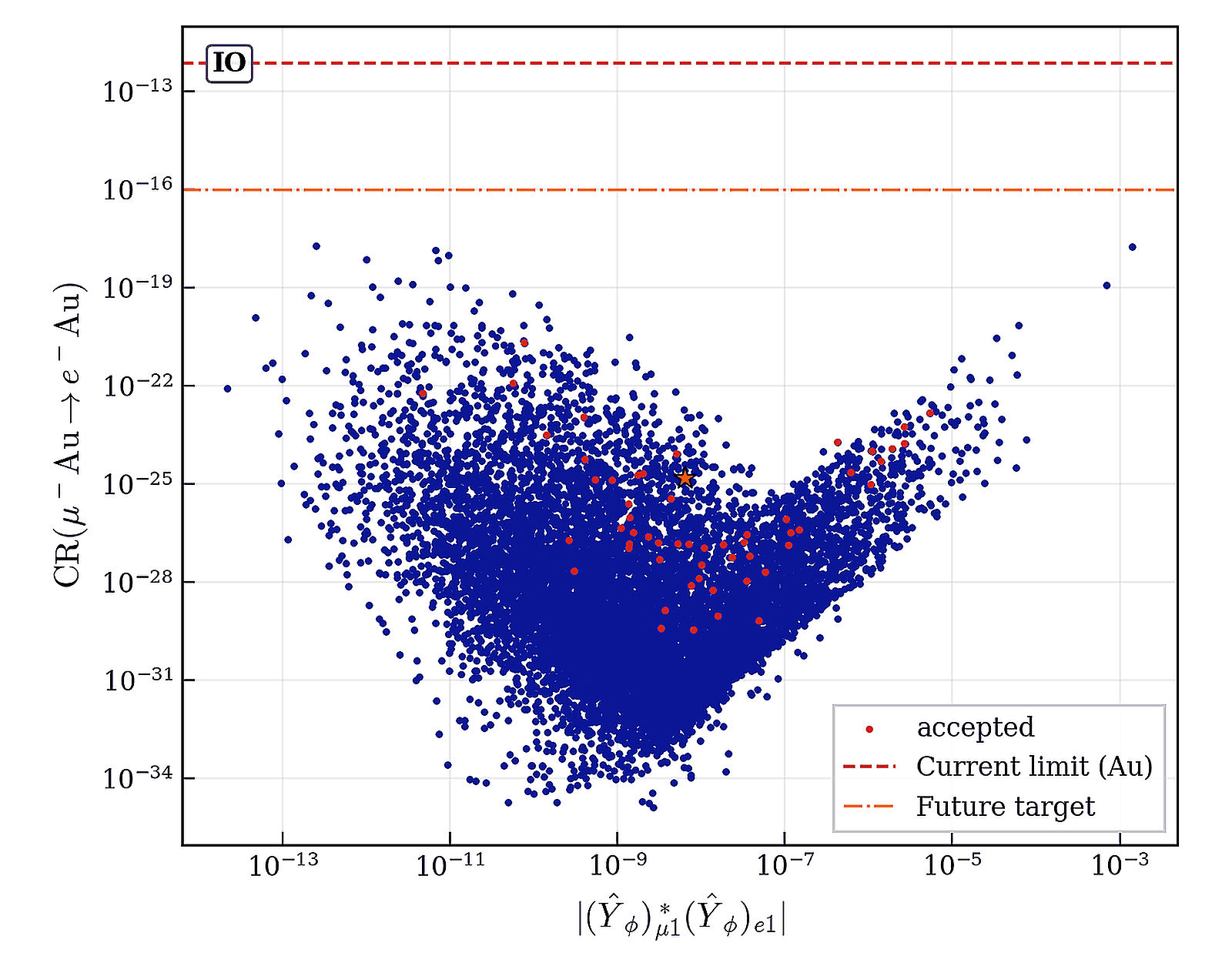}
\caption{Same as figure \ref{fig7}, but for IO.}
\label{fig11}
\end{figure}
The overall similarity between NO and IO follows from their common modular structure, one-loop neutrino-mass mechanism, and compressed odd-sector spectrum, which is consistent with efficient coannihilation. Their differences arise primarily from the light-neutrino spectrum. In IO, the neutrino masses are necessarily more quasi-degenerate, pushing $\sum_i m_i$, $m_\beta$, and $m_{\beta\beta}$ closer to their current experimental and cosmological sensitivities and leaving a narrower viable interval for $M_{N_1}$. By contrast, NO permits a broader branch with heavier $N_1$ and a wider scalar-mass range while remaining compatible with the same collider, electroweak-precision, and flavor constraints.
Table~\ref{tab:summaryNOIO} collects the corresponding ranges of the accepted points for the two mass orderings. Several features are common to both: the points consistent with the observed relic abundance lie along a compressed coannihilation strip, the loop-induced Higgs-mediated scalar contribution to direct detection remains below the xenon neutrino floor, and the absolute neutrino-mass observables are confined to narrow intervals. The principal qualitative difference is that the light-neutrino spectrum in IO is more quasi-degenerate, placing $\sum_i m_i$ and $m_{\beta\beta}$ closer to the present cosmological bound and the sensitivity of upcoming $0\nu\beta\beta$ experiments, respectively. These correlations constitute characteristic phenomenological predictions of the model within the parameter space explored in the present scan.
\begin{table}[!]
\centering
\begin{tabular}{lcc}
\hline\hline
Observable / feature & NO (accepted) & IO (accepted) \\
\hline
DM mass window $M_{N_1}$ [GeV] & $97 \lesssim M_{N_1} \lesssim 852$ & $127 \lesssim M_{N_1} \lesssim 378$ \\
$\sum_i m_i$ [eV] & $0.060 \lesssim \sum_i m_i \lesssim 0.120$ & $0.099 \lesssim \sum_i m_i \lesssim 0.119$ \\
$0\nu\beta\beta$ reach (qualitative) & typically farther from reach & typically closer to reach \\
Relic-density mechanism & coannihilation strip & coannihilation strip \\
Direct detection (SI) & loop-induced, strongly suppressed & loop-induced, strongly suppressed \\
cLFV probes & can test a subset of points & can test a subset of points \\
\hline\hline
\end{tabular}
\caption{Summary of characteristic viable ranges and qualitative predictions for the NO and IO schemes obtained in our scan (accepted points). The IO is typically more constrained in the absolute neutrino mass observables and lies closer to present and future $0\nu\beta\beta$ sensitivity.}
\label{tab:summaryNOIO}
\end{table}
\section{Summary and conclusions}
\label{sec6}
We have presented a nonholomorphic modular-invariant realization of the one-loop \texttt{T4-2-i} topology based on the double-cover modular group $T’$. The modular assignments organize the Yukawa sector, forbid the tree-level type-I operator $\overline L\widetilde HN$, and generate predictive charged-lepton and neutrino-mass textures. The tree-level $LL\Delta$ interaction is excluded from the lepton-number-conserving Yukawa sector, while the modular-invariant Majorana mass matrix provides the soft $\Delta L=2$ breaking required for neutrino-mass generation at one loop. The modular field assignments also yield an exact parity that stabilizes the lightest odd state.

To assess the phenomenological viability of this construction, we fitted the charged-lepton masses and neutrino-oscillation data while imposing constraints from cosmology, electroweak precision observables, the Higgs diphoton rate, relic abundance, cLFV processes, and direct detection. Although the numerical analysis included both scalar and fermionic DM candidates, the phenomenological discussion focused on the Majorana fermion $N_1$. The viable fermionic-DM points exhibit a compressed odd-sector spectrum characteristic of efficient coannihilation with the inert scalar partners. The loop-induced Higgs-mediated scalar contribution to direct detection is strongly suppressed, while a complete recoil analysis including the anapole and $Z$-mediated interactions is deferred to future work.

The numerical analysis yields viable solutions for both neutrino-mass orderings, with broadly similar dark-matter, electroweak, collider, and flavor patterns owing to their common modular structure and radiative mass-generation mechanism. Their principal difference originates from the light-neutrino spectrum: IO is more quasi-degenerate and predicts larger, more narrowly distributed values of $\sum_i m_i$, $m_\beta$, and $m_{\beta\beta}$. Consequently, $\sum_i m_i$ and $m_{\beta\beta}$ lie closer to current cosmological constraints and the sensitivity of upcoming $0\nu\beta\beta$ experiments, respectively, while the viable fermionic-DM mass interval is also more restricted. By contrast, NO permits wider ranges for the lightest neutrino mass and $M_{N_1}$, together with a somewhat broader viable scalar spectrum.

Overall, these results show that the \texttt{T4-2-i} topology can be consistently embedded in a nonsupersymmetric $T’$ modular framework involving even- and odd-weight polyharmonic Maa{\ss} forms, while accommodating radiative neutrino masses and a viable fermionic DM candidate. Future cosmological neutrino mass measurements, $0\nu\beta\beta$ searches, direct-detection experiments, and cLFV searches will further probe the remaining parameter space.
\section*{Acknowledgements}
The work of M.A.L. and S.N. is supported by the United Arab Emirates University (UAEU) under UPAR Grant No. 12S162.
\appendix
\section{Homogeneous finite modular group of level 3}
\label{app1}
The finite homogeneous modular group $\Gamma^{\prime}_N$ is the quotient of the two-dimensional special linear group $SL(2,\mathbb{Z})$ by the principal congruence subgroup of level N: $\Gamma^{\prime}_N = SL(2,\mathbb{Z})/\Gamma(N)$. For level $N=3$, $\Gamma^{\prime}_3$ is isomorphic to the group $T'$ which serves as a double cover of the tetrahedral group $\Gamma_3 \cong A_4$--the alternating group of degree 4 and order 12. The structure of $T'$ is determined through three fundamental generators denoted as $S$, $T$, and $R$ that satisfy the following relations \cite{Liu:2019khw}
\begin{equation}
    S^2 = R, \quad (ST)^3 = T^3 = R^2 = 1, \quad RT = TR
\end{equation}
The group $T'$, often called the binary tetrahedral group, has order 24 and can be viewed as a central extension of $A_4$ by $Z_2$. The generator $R$, which lies in the center of $T'$, commutes with all group elements, and by setting $R=1$, the central extension collapses, and one recovers the alternating group $A_4$. The elements of $T'$ are organized into seven conjugacy classes given by \cite{Ding:2023htn}
\begin{eqnarray}
    1C_1 &=& 1, \quad 1C_2 = R, \quad 4C_3 = T^{-1},~ST^{-1}R,~T^{-1}SR,~TSTR, \quad 4C_3' = T,~TS,~ST,~T^{-1}ST^{-1}, \nonumber \\
    6C_4 &=& S,~T^{-1}ST,~TST^{-1},~SR,~T^{-1}STR,~TST^{-1}R, \quad 4C_6 = TR,~TSR,~STR,~T^{-1}ST^{-1}R, \\
    4C_6' &=& ST^{-1},~T^{-1}S,~TST,~T^{-1}R. \nonumber
\end{eqnarray}
Here, the notation $kC_m$ denotes a conjugacy class of order $m$ containing $k$ elements. Since the number of conjugacy classes equals the number of irreducible representations, $T'$ has seven irreducible representations, with their squared dimensions summing to the group's order: $\sum_{i=1}^7 d_i^2 = |T'| = 24$, where $d_i$ are the representations dimensions. In addition to the triplet $\mathbf{3}$ and the three one-dimensional representations $\mathbf{1}$, $\mathbf{1'}$, and $\mathbf{1''}$ shared with the $A_4$ group, $T'$ includes three two-dimensional spinor representations: $\mathbf{2}$, $\mathbf{2'}$, and $\mathbf{2''}$. In our benchmark models, we adopted the basis in Ref. \cite{Ding:2022aoe}, where the generators $S$, $T$, and $R$ are represented by the following matrices
\begin{eqnarray}
    \mathbf{1}&:& S = 1, \quad T = 1, \quad R = 1, \nonumber \\
    \mathbf{1'}&:& S = 1, \quad T = \omega, \quad R = 1, \nonumber \\
    \mathbf{1''}&:& S = 1, \quad T = \omega^2, \quad R = 1, \nonumber \\
    \mathbf{2}&:& S =\frac{-i}{\sqrt3} \begin{pmatrix}
        1 && \sqrt2 \\
        \sqrt2 && -1
    \end{pmatrix}, \quad T = \begin{pmatrix}
        \omega && 0 \\
        0 && 1
    \end{pmatrix}, \quad R = \begin{pmatrix}
        -1 && 0 \\
        0 && -1
    \end{pmatrix}, \nonumber \\
    \mathbf{2'}&:& S =\frac{-i}{\sqrt3} \begin{pmatrix}
        1 && \sqrt2 \\
        \sqrt2 && -1
    \end{pmatrix}, \quad T = \begin{pmatrix}
        \omega^2 && 0 \\
        0 && \omega
    \end{pmatrix}, \quad R = \begin{pmatrix}
        -1 && 0 \\
        0 && -1
    \end{pmatrix}, \\
    \mathbf{2''}&:& S =\frac{-i}{\sqrt3} \begin{pmatrix}
        1 && \sqrt2 \\
        \sqrt2 && -1
    \end{pmatrix}, \quad T = \begin{pmatrix}
        1 && 0 \\
        0 && \omega^2
    \end{pmatrix}, \quad R = \begin{pmatrix}
        -1 && 0 \\
        0 && -1
    \end{pmatrix}, \nonumber \\
    \mathbf{3}&:& S =\frac{1}{3} \begin{pmatrix}
        -1 && 2 && 2 \\
        2 && -1 && 2 \\
        2 && 2 && -1
    \end{pmatrix}, \quad T = \begin{pmatrix}
        1 && 0  && 0 \\
        0 && \omega && 0 \\
        0 && 0 && \omega^2
    \end{pmatrix},  \quad R = \begin{pmatrix}
        1 && 0 && 0 \\
        0 && 1 && 0 \\
        0 && 0 && 1
    \end{pmatrix}, \nonumber
\end{eqnarray}
with the cube root of unity $\omega = e^{i 2 \pi /3}$. The tensor products between the irreducible representations of $T'$ are given by
\begin{eqnarray}
    \mathbf{1} &\otimes& \mathbf{r} = \mathbf{r}, \quad \mathbf{1'} \otimes \mathbf{1'} = \mathbf{1''}, \quad \mathbf{1'} \otimes \mathbf{1''} = \mathbf{1}, \quad \mathbf{1'} \otimes \mathbf{2} = \mathbf{2'}, \quad \mathbf{1''} \otimes \mathbf{2} = \mathbf{2''}, \quad \mathbf{1'} \otimes \mathbf{2'} = \mathbf{2''}, \nonumber \\
    \mathbf{1''} &\otimes& \mathbf{2} = \mathbf{2''}, \quad \mathbf{1'} \otimes \mathbf{2''} = \mathbf{2}, \quad \mathbf{1''} \otimes \mathbf{2''} = \mathbf{2'}, \quad \mathbf{1'} \otimes \mathbf{3} = \mathbf{1''} \otimes \mathbf{3} = \mathbf{3}, \nonumber \\
    \mathbf{2} &\otimes& \mathbf{2} = \mathbf{2'} \otimes \mathbf{2''} = \mathbf{1'} \oplus \mathbf{3} \quad \mathbf{2} \otimes \mathbf{2'} = \mathbf{2''} \otimes \mathbf{2''} = \mathbf{1''} \oplus \mathbf{3} \quad \mathbf{2} \otimes \mathbf{2''} = \mathbf{2'} \otimes \mathbf{2'} = \mathbf{1} \oplus \mathbf{3} \\
    \mathbf{2} &\otimes& \mathbf{3} = \mathbf{2'} \otimes \mathbf{3} = \mathbf{2''} \otimes \mathbf{3}  =\mathbf{2} \oplus \mathbf{2'} \oplus \mathbf{2''}, \quad \mathbf{3} \otimes \mathbf{3} = \mathbf{1} \oplus \mathbf{1'} \oplus \mathbf{1''} \oplus \mathbf{3_S} \oplus \mathbf{3_A}, \nonumber 
\end{eqnarray}
where $\mathbf{r}$ represents any of the $T'$ irreducible representation, while $\mathbf{3_S}$ and $\mathbf{3_A}$ denote the symmetric and antisymmetric contractions, respectively. Using $a_i$ and $b_i$ to denote the elements of the first and second representation, respectively, the decomposition of the above tensor products are as follows
\begin{eqnarray}
    \mathbf{s} &\otimes& \mathbf{s} \sim a b, \quad \mathbf{s} \otimes \mathbf{d} \sim \begin{pmatrix} a b_1 \\ a b_2  \end{pmatrix}, \quad \mathbf{1} \otimes \mathbf{3} \sim \begin{pmatrix} a b_1 \\ a b_2 \\ a b_3 \end{pmatrix}, \quad \mathbf{1'} \otimes \mathbf{3} \sim \begin{pmatrix} a b_3 \\ a b_1 \\ a b_2 \end{pmatrix}, \quad \mathbf{1''} \otimes \mathbf{3} \sim \begin{pmatrix} a b_2 \\ a b_3 \\ a b_1 \end{pmatrix}, \nonumber \\ 
    \mathbf{2} &\otimes& \mathbf{2} = \mathbf{2'} \otimes \mathbf{2''} \sim (a_1 b_2 - a_2 b_1)_\mathbf{1'} \oplus \begin{pmatrix} a_2 b_2 \\ \frac{1}{\sqrt 2} (a_1 b_2 + a_2 b_1) \\ -a_1 b_1 \end{pmatrix}, \\
    \mathbf{2} &\otimes& \mathbf{2'} = \mathbf{2''} \otimes \mathbf{2''} \sim (a_1 b_2 - a_2 b_1)_\mathbf{1''} \oplus \begin{pmatrix} -a_1 b_1 \\ \frac{1}{\sqrt 2} a_2 b_2 \\ \frac{1}{\sqrt2} (a_1 b_2 + a_2 b_1) \end{pmatrix}, \\
    \mathbf{2} &\otimes& \mathbf{2''} = \mathbf{2'} \otimes \mathbf{2'} \sim (a_1 b_2 - a_2 b_1)_\mathbf{1} \oplus \begin{pmatrix} \frac{1}{\sqrt 2} (a_1 b_2 + a_2 b_1) \\ -a_1 b_1 \\ a_2 b_2 \end{pmatrix}, \\
    \mathbf{2} &\otimes& \mathbf{3} \sim \begin{pmatrix} a_1 b_1 + \sqrt2 a_2 b_2 \\ -a_2 b_1 + \sqrt2 a_1 b_3 \end{pmatrix}_{\mathbf{2}} \oplus \begin{pmatrix} a_1 b_2 + \sqrt2 a_2 b_3 \\ -a_2 b_2 + \sqrt2 a_1 b_1 \end{pmatrix}_{\mathbf{2'}} \oplus \begin{pmatrix} a_1 b_3 + \sqrt2 a_2 b_1 \\ -a_2 b_3 + \sqrt2 a_1 b_2 \end{pmatrix}_{\mathbf{2''}}, \nonumber \\
    \mathbf{2'} &\otimes& \mathbf{3} \sim \begin{pmatrix} a_1 b_3 + \sqrt2 a_2 b_1 \\ -a_2 b_3 + \sqrt2 a_1 b_2 \end{pmatrix}_{\mathbf{2}} \oplus \begin{pmatrix} a_1 b_1 + \sqrt2 a_2 b_2 \\ -a_2 b_1 + \sqrt2 a_1 b_3 \end{pmatrix}_{\mathbf{2'}} \oplus \begin{pmatrix} a_1 b_2 + \sqrt2 a_2 b_3 \\ -a_2 b_2 + \sqrt2 a_1 b_1 \end{pmatrix}_{\mathbf{2''}}, \nonumber \\
    \mathbf{2''} &\otimes& \mathbf{3} \sim \begin{pmatrix} a_1 b_2 + \sqrt2 a_2 b_3 \\ -a_2 b_2 + \sqrt2 a_1 b_1 \end{pmatrix}_{\mathbf{2}} \oplus \begin{pmatrix} a_1 b_3 + \sqrt2 a_2 b_1 \\ -a_2 b_3 + \sqrt2 a_1 b_2 \end{pmatrix}_{\mathbf{2'}} \oplus \begin{pmatrix} a_1 b_1 + \sqrt2 a_2 b_2 \\ -a_2 b_1 + \sqrt2 a_1 b_3 \end{pmatrix}_{\mathbf{2''}},
\end{eqnarray}
where $\mathbf{s}$ stands for any $T'$ singlet, and $\mathbf{d}$ denotes any doublet representation of $T'$. The contraction rules for two generic $T'$ triplets, $a = (a_1, a_2, a_3)^T$ and $b = (b_1, b_2, b_3)^T$, is the same as the one for two $A_4$ triplet. See appendix A of Ref. \cite{Loualidi:2025tgw} for this tensor product decomposition using the same notations as above.
\section{Loop functions}
\label{app2}
The appendix gathers the loop functions entering the electroweak precision analysis of Sec. \ref{EWPO} and the cLFV amplitudes of Sec. \ref{cLFV}.
The functions $F(X,Y)$, $\xi(X,Y)$, and $G(X,Y,Z)$ are the standard scalar contributions to the oblique parameters~\cite{Lavoura:1993nq,Cheng:2022hbo}, whereas $F_2(x)$, $G_2(x)$, $D_1(x,y)$, and $D_2(x,y)$ are the one-loop kernels used in the radiative and three-body cLFV amplitudes~\cite{Toma:2013zsa}.
For the oblique parameters we use
\begin{eqnarray}
    F(X,Y) &=& \frac{X+Y}{2} - \frac{XY}{X-Y}~ \ln\frac{X}{Y} \\
    \xi(X,Y) &=& \frac{4}{9} - \frac{5}{12} (X+Y) + \frac{1}{6} (X-Y)^2 + \frac{1}{4} \left[ X^2 - Y^2 - \frac{1}{3}(X-Y)^3 - \frac{X^2+Y^2}{X-Y} \right]\ln\frac{X}{Y} - \frac{1}{12} d(X,Y) f(X,Y) \nonumber
\end{eqnarray}
where $d(X,Y)= -1 +2 (X + Y) - (X - Y)^2$ and
\begin{eqnarray}
    f(X,Y) =\left\{
	\begin{array}{cl}
		-2\sqrt{d(X,Y)}\left[\arctan \frac{X-Y+1}{\sqrt{d(X,Y)}}
		-\arctan \frac{X-Y-1}{\sqrt{d(X,Y)}}\right], &  \;\;\; d(X,Y) > 0 \\
        0  &  \;\;\; d(X,Y) = 0 \\
		\sqrt{-d(X,Y)}\ln \frac{X+Y-1+\sqrt{-d(X,Y)}}{X+Y-1-\sqrt{-d(X,Y)}}, ~~~ & ~~~ d(X,Y) < 0
	\end{array} \right.
\end{eqnarray}
In the numerical implementation we treat the equal-mass limit $X\to Y$ analytically rather than by introducing an artificial mass splitting. This is particularly important in the near-degenerate inert regions explored in the DM scans, where the exact limit of $\xi(X,Y)$ stabilizes the evaluation of the correlated oblique parameter contribution.
\begin{eqnarray}
    G(X,Y,Z) = -\frac{16}{3} + \frac{5(X+Y)}{Z} - \frac{2(X-Y)^2}{Z^2} + \frac{3}{Z} \left[ \frac{X^2+Y^2}{X-Y} - \frac{X^2-Y^2}{Z} + \frac{(X-Y)^3}{3Z^2} \right] \ln\frac{X}{Y} + \frac{r}{Z^3} f(t,r)
\end{eqnarray}
with $t = X+Y-Z$ and $r = Z^2 - 2Z(X+Y) + (X-Y)^2$ while
\begin{eqnarray}
    f(t,r) =\left\{
	\begin{array}{cl} \sqrt{r} \ln |\frac{t - \sqrt{r}}{t + \sqrt{r}}|   &  \text{for} ~~ r>0 \\
	0  &  \text{for} ~~ r =0 \\
    2\sqrt{-r} \arctan\frac{\sqrt{-r}}{t}  ~~~ & \text{for}  ~~ r<0
	\end{array} \right.
\end{eqnarray}
For the cLFV sector we use the on-shell photon-dipole loop function
\begin{equation}
F_2(x)=
\frac{1-6x+3x^2+2x^3-6x^2\ln x}{6(1-x)^4},
\end{equation}
which appears in the radiative-dipole coefficient $A_R^{\beta\alpha}$ of Eq.~(\ref{eq:ARclfv}). In the three-body notation of Sec.~III\,C, the same coefficient is denoted by $A_D^{\beta\alpha}$. The corresponding off-shell photon function is
\begin{equation}
G_2(x)=
\frac{2-9x+18x^2-11x^3+6x^3\ln x}{6(1-x)^4},
\end{equation}
which enters the non-dipole photon penguin coefficient $A_{ND}$ in Eq.~\ref{eq:3bodyfull}. The box contribution is built from
\begin{eqnarray}
D_1(x,y) &=& -\frac{1}{(1-x)(1-y)}
-\frac{x^2\ln x}{(1-x)^2(x-y)}
-\frac{y^2\ln y}{(1-y)^2(y-x)}, \\
D_2(x,y) &=& -\frac{1}{(1-x)(1-y)}
-\frac{x\ln x}{(1-x)^2(x-y)}
-\frac{y\ln y}{(1-y)^2(y-x)} .
\end{eqnarray}
In Sec.~\ref{cLFV} the arguments are the heavy-neutrino-to-scalar mass ratios $x_i^{(s)}=M_i^2/m_s^2$ for the photon penguins and the pairwise ratios $(x_i^{(s)},x_j^{(t)})$ for the box amplitudes, with $s,t=\rho,\phi$. The functions $D_1$ and $D_2$ therefore determine the pure and mixed box contributions entering the coefficient $B$ in Eq.~(\ref{eq:3bodyfull}). In the equal-mass limit the code evaluates them analytically as
\begin{eqnarray}
D_1(x,x) = -\frac{1-x^2+2x\ln x}{(1-x)^3}, \quad
D_2(x,x) = -\frac{2-2x+(1+x)\ln x}{(1-x)^3},
\end{eqnarray}
\section{Effective coupling}
\label{app3}
The effective interaction is defined as
\begin{equation}
{\cal L}_{\text{eff}} \supset -\frac12\,g_{hN_1N_1} h \overline{N_1^c}N_1 .
\end{equation}
At one loop, $g_{hN_1N_1}$ is obtained from diagrams in which the Higgs couples to a pair of inert scalars and the $N_1$ external legs are connected through a lepton-scalar loop. The generic scalar pair contribution can be written as
\begin{equation}
g_{hN_1N_1}^{(ab)} = \frac{M_{N_1}}{16\pi^2}
g_{hab} \sum_{\alpha=e,\mu,\tau} \Re\!\left(y_a^\alpha y_b^{\alpha *}\right)
{\cal I}(M_{N_1}^2;m_a^2,m_b^2,m_{\ell_\alpha}^2),
\label{eq:gEff_generic}
\end{equation}
where $a,b$ denote the inert scalar states running in the loop and $g_{hab}$ is the trilinear Higgs--scalar--scalar coupling. In our implementation those couplings are taken directly from the {\tt CalcHEP} model file and therefore remain exactly consistent with the scalar potential used by {\tt micrOMEGAs}. The loop function ${\cal I}$ is the scalar three-point function associated with the one-particle-irreducible $h\overline{N_1^c}N_1$ vertex. For a general external Higgs momentum $q^2$, the exact form factor may be expressed in terms of Passarino--Veltman functions,
\begin{equation}
{\cal I}(q^2;m_a^2,m_b^2,m_\ell^2) = C_0 + C_1 + C_2 ,
\end{equation}
evaluated at
\begin{equation}
C_i \equiv C_i(M_{N_1}^2,q^2,M_{N_1}^2;m_a^2,m_\ell^2,m_b^2).
\end{equation}
For the numerical matching used in the scan, we work in the zero-momentum limit relevant for a local effective operator inserted into CalcHEP. In that limit the triangle reduces to the finite difference of two-point functions,
\begin{equation}
{\cal I}(0;m_a^2,m_b^2,m_\ell^2) = \frac{B_0(0;m_a^2,m_\ell^2)-B_0(0;m_b^2,m_\ell^2)}{m_a^2-m_b^2},
\label{eq:b0diff}
\end{equation}
with the diagonal case obtained as the derivative limit,
\begin{equation}
{\cal I}(0;m_a^2,m_a^2,m_\ell^2) = \left.\frac{\partial B_0(0;x,m_\ell^2)}{\partial x}\right|_{x=m_a^2}.
\end{equation}
This representation is exact at zero momentum and is numerically stable both for diagonal and off-diagonal scalar insertions. In the setup used throughout the scan, $M_{A_1}=M_{S_1}$, $M_{A_2}=M_{S_2}$ and $\theta_a=\theta_n$. The total effective coupling is then the sum of the charged and neutral inert contributions,
\begin{eqnarray}
g_{hN_1N_1} &=&
\frac{M_{N_1}}{16\pi^2}\sum_{\alpha}
\Big[
+g_{h\rho^+\rho^-}\,|y_\rho^\alpha|^2\,{\cal I}(0;m_{\rho^\pm}^2,m_{\rho^\pm}^2,m_{\ell_\alpha}^2)
+g_{h\phi^+\phi^-}\,|y_\phi^\alpha|^2\,{\cal I}(0;m_{\phi^\pm}^2,m_{\phi^\pm}^2,m_{\ell_\alpha}^2)
\nonumber\\[1mm]
&&\qquad\qquad
+g_{hS_1S_1}\,|y_{S_1}^\alpha|^2\,{\cal I}(0;m_{S_1}^2,m_{S_1}^2,m_{\nu_\alpha}^2)
+g_{hS_2S_2}\,|y_{S_2}^\alpha|^2\,{\cal I}(0;m_{S_2}^2,m_{S_2}^2,m_{\nu_\alpha}^2)
\nonumber\\[1mm]
&&\qquad\qquad
+g_{hS_1S_2}\,\Re\!\left(y_{S_1}^\alpha y_{S_2}^{\alpha *}\right)\,{\cal I}(0;m_{S_1}^2,m_{S_2}^2,m_{\nu_\alpha}^2)
+g_{hA_1A_1}\,|y_{A_1}^\alpha|^2\,{\cal I}(0;m_{A_1}^2,m_{A_1}^2,m_{\nu_\alpha}^2)
\nonumber\\[1mm]
&&\qquad\qquad
+g_{hA_2A_2}\,|y_{A_2}^\alpha|^2\,{\cal I}(0;m_{A_2}^2,m_{A_2}^2,m_{\nu_\alpha}^2)
+g_{hA_1A_2}\,\Re\!\left(y_{A_1}^\alpha y_{A_2}^{\alpha *}\right)\,{\cal I}(0;m_{A_1}^2,m_{A_2}^2,m_{\nu_\alpha}^2)
\Big].
\label{gEff_final}
\end{eqnarray}
Equation~\ref{gEff_final} is the effective coupling inserted into the {\tt CalcHEP} model. The trilinear Higgs couplings are read directly from the exact scalar-sector implementation, while the Yukawa combinations are reconstructed from the mass-basis matrices used also in the neutrino-mass and cLFV calculations. In this way the direct-detection calculation is fully aligned with the treatment of neutrino masses, charged-lepton diagonalization, and flavor observables.
\section{Perturbative unitarity of the scalar sector}
\label{app4}
In the high-energy limit, the tree-level scalar $2\to2$ amplitudes are determined solely by the quartic potential $V_4$ of Eq.~\ref{potential}. For canonically normalized real scalar components ${\cal S}_a$, the amplitude matrix is
\begin{equation}
({\cal M}_0)_{ij,kl} = -\frac{\partial^4V_4/\partial{\cal S}_i\partial{\cal S}_j \partial{\cal S}_k\partial{\cal S}_l} {\sqrt{(1+\delta_{ij})(1+\delta_{kl})}}.
\label{unitarity_matrix}
\end{equation}
Normalized two-particle states are denoted by $[XY]$, with the factor $1/\sqrt2$ understood for $X=Y$. The real neutral fields are $h_d$, $\omega_0$, $H_1$, $\omega_1$, $H_2$, $\omega_2$, $h_t$, and $\omega_3$, while the charged fields are $\eta^\pm$, $\rho^\pm$, $\phi^\pm$, $\delta^\pm$, and $\delta^{\pm\pm}$. The charge basis below is related to the real-component basis by a unitary transformation. For an eigenvalue $e_i$ of ${\cal M}_0$, one has $a_{0,i}=e_i/(16\pi)$; hence $|\operatorname{Re}a_{0,i}|<1/2$ is equivalent to $|e_i|<8\pi$~\cite{Lee:1977yc,Goodsell:2018tti}. The overall sign of Eq.~\eqref{unitarity_matrix} is therefore irrelevant.
Charge conservation decomposes the neutral sector into 53 states: one $9\times9$, twelve $2\times2$, and twenty one-state blocks. The $Q=+1$ sector has 36 states: one $5\times5$, nine $2\times2$, and thirteen one-state blocks. The $Q=+2$, $+3$, and $+4$ sectors contain, respectively, four $2\times2$ plus ten one-state blocks, four one-state blocks, and one one-state block. The negative-charge sectors are their charge conjugates and have identical eigenvalues. The full calculation thus comprises one $9\times9$, two $5\times5$, thirty-eight $2\times2$, and seventy-six one-state blocks, yielding 171 eigenvalues.

The nine-state neutral basis is
\begin{align}
{\cal B}_0=\bigl(&[\eta^+\eta^-],[H^0H^{0*}];
[\delta^+\delta^-],[\delta^{++}\delta^{--}],[\delta^0\delta^{0*}]; [\rho^0\rho^{0*}],[\rho^-\rho^+]; [\phi^+\phi^-],[\phi^0\phi^{0*}]\bigr),
\end{align}
and can be written as
\begin{equation}
{\mathbb M}_0 = 
\begin{pmatrix}
A_H&X_{H\Delta}&X_{H\rho}&X_{H\phi}\\
X_{H\Delta}^{T}&A_\Delta&X_{\Delta\rho}&X_{\Delta\phi}\\
X_{H\rho}^{T}&X_{\Delta\rho}^{T}&A_\rho&X_{\rho\phi}\\
X_{H\phi}^{T}&X_{\Delta\phi}^{T}&X_{\rho\phi}^{T}&A_\phi
\end{pmatrix},
\label{eq:M0_block}
\end{equation}
where
\begin{align}
A_H=2\lambda_H\begin{pmatrix}2&1\\1&2\end{pmatrix}, \quad 
A_\rho =2\lambda_\rho\begin{pmatrix}2&1\\1&2\end{pmatrix}, \quad 
A_\phi = 2\lambda_\phi\begin{pmatrix}2&1\\1&2\end{pmatrix},\quad
A_\Delta =
\begin{pmatrix}
2\lambda_\Delta+4\lambda_\Delta'&2\lambda_\Delta+2\lambda_\Delta'&2\lambda_\Delta+2\lambda_\Delta'\\
2\lambda_\Delta+2\lambda_\Delta'&4\lambda_\Delta+4\lambda_\Delta'&2\lambda_\Delta'\\
2\lambda_\Delta+2\lambda_\Delta'&2\lambda_\Delta'&4\lambda_\Delta+4\lambda_\Delta'
\end{pmatrix},
\label{eq:self_blocks}
\end{align}
\begin{align}
X_{H\Delta}&=
\begin{pmatrix}
\lambda_4+\lambda_8/2&\lambda_4+\lambda_8&\lambda_4\\
\lambda_4+\lambda_8/2&\lambda_4&\lambda_4+\lambda_8
\end{pmatrix},&
X_{H\rho}&=
\begin{pmatrix}
\lambda_1+\lambda_1'&\lambda_1\\
\lambda_1&\lambda_1+\lambda_1'
\end{pmatrix},\nonumber\\
X_{H\phi}&=
\begin{pmatrix}
\lambda_2+\lambda_2'&\lambda_2\\
\lambda_2&\lambda_2+\lambda_2'
\end{pmatrix},&
X_{\rho\phi}&=
\begin{pmatrix}
\lambda_3+\lambda_5&\lambda_5\\
\lambda_5&\lambda_3+\lambda_5
\end{pmatrix},
\label{eq:doublet_blocks}
\end{align}
\begin{align}
X_{\Delta\rho}&=
\begin{pmatrix}
\lambda_6+\lambda_9/2&\lambda_6+\lambda_9/2\\
\lambda_6+\lambda_9&\lambda_6\\
\lambda_6&\lambda_6+\lambda_9
\end{pmatrix},&
X_{\Delta\phi}&=
\begin{pmatrix}
\lambda_7+\lambda_{10}/2&\lambda_7+\lambda_{10}/2\\
\lambda_7+\lambda_{10}&\lambda_7\\
\lambda_7&\lambda_7+\lambda_{10}
\end{pmatrix}.
\label{eq:triplet_cross_blocks}
\end{align}
The five-state $Q=+1$ basis is
\begin{equation}
{\cal B}_{1}=\bigl([\eta^+H^{0*}],[\delta^+\delta^{0*}],
-[\delta^-\delta^{++}],[\rho^0\rho^+],[\phi^+\phi^{0*}]\bigr),
\end{equation}
with the amplitude matrix given by
\begin{equation}
{\mathbb M}_{1}=
\begin{pmatrix}
2\lambda_H&\lambda_8/\sqrt2&\lambda_8/\sqrt2&\lambda_1'&\lambda_2'\\
\lambda_8/\sqrt2&2(\lambda_\Delta+\lambda_\Delta')&2\lambda_\Delta&\lambda_9/\sqrt2&\lambda_{10}/\sqrt2\\
\lambda_8/\sqrt2&2\lambda_\Delta&2(\lambda_\Delta+\lambda_\Delta')&\lambda_9/\sqrt2&\lambda_{10}/\sqrt2\\
\lambda_1'&\lambda_9/\sqrt2&\lambda_9/\sqrt2&2\lambda_\rho&\lambda_3\\
\lambda_2'&\lambda_{10}/\sqrt2&\lambda_{10}/\sqrt2&\lambda_3&2\lambda_\phi
\end{pmatrix}.
\label{eq:M1_block}
\end{equation}
The minus sign in the third basis state is only a phase convention. For later use, any real symmetric two-state block has the form
\begin{equation}
{\mathbb M}(A,B,C)=\begin{pmatrix}A&B\\B&C\end{pmatrix},
\qquad
e_\pm=\frac{A+C\pm\sqrt{(A-C)^2+4B^2}}{2}.
\label{eq:two_state_eigenvalues}
\end{equation}
Thus the eigenvalues of every $2\times2$ block below follow directly from its three entries $(A,B,C)$. The $Q=+2$ sector contains the following four two-state blocks:
\begin{align}
{\cal B}_{2}(H\Delta)&=\bigl([\eta^+\delta^+],[H^0\delta^{++}]\bigr),&
{\mathbb M}_{2}(H\Delta)&=
\begin{pmatrix}\lambda_4+\lambda_8/2&-\lambda_8/\sqrt2\\-\lambda_8/\sqrt2&\lambda_4\end{pmatrix},\nonumber\\
{\cal B}_{2}(\Delta\Delta)&=\bigl([\delta^+\delta^+],[\delta^{++}\delta^0]\bigr),&
{\mathbb M}_{2}(\Delta\Delta)&=
\begin{pmatrix}\lambda_\Delta+2\lambda_\Delta'&-\sqrt2\lambda_\Delta\\-\sqrt2\lambda_\Delta&2\lambda_\Delta'\end{pmatrix},\nonumber\\
{\cal B}_{2}(\rho\Delta)&=\bigl([\delta^+\rho^+],[\delta^{++}\rho^{0*}]\bigr),&
{\mathbb M}_{2}(\rho\Delta)&=
\begin{pmatrix}\lambda_6+\lambda_9/2&-\lambda_9/\sqrt2\\-\lambda_9/\sqrt2&\lambda_6+\lambda_9\end{pmatrix},\nonumber\\
{\cal B}_{2}(\phi\Delta)&=\bigl([\delta^+\phi^+],[\delta^{++}\phi^0]\bigr),&
{\mathbb M}_{2}(\phi\Delta)&=
\begin{pmatrix}\lambda_7+\lambda_{10}/2&-\lambda_{10}/\sqrt2\\-\lambda_{10}/\sqrt2&\lambda_7\end{pmatrix}.
\label{eq:Q2blocks}
\end{align}
The ten one-state blocks in this sector are 
\begin{eqnarray}
    [H^{0*} \delta^{++}] &:& \lambda_4, \quad [\delta^{++} \delta^{0*}] : 2\lambda_\Delta', \quad [\delta^{++} \rho^0] : \lambda_6+\lambda_9, \quad [\delta^{++} \phi^{0*}] : \lambda_7, \quad [\eta^+ \eta^+] : 2\lambda_H, \quad [\rho^+\rho^+] : 2\lambda_\rho, 
    \nonumber \\ 
    {} [\phi^+ \phi^+] &:& 2\lambda_\phi, \quad [\eta^+ \rho^+] : \lambda_1, \quad [\eta^+ \phi^+] : \lambda_2+\lambda_2', \quad [\rho^+\phi^+] : \lambda_5,
    \label{eq:Q2blocks}
\end{eqnarray}
where each expression after the colon is the corresponding eigenvalue. For $Q=+3$ and $Q=+4$ one obtains
\begin{eqnarray}
{\cal B}_3 &= \bigl([\eta^+\delta^{++}],[\delta^+\delta^{++}],
[\delta^{++}\rho^+],[\delta^{++}\phi^+]\bigr), \qquad {\mathbb M}_3 &= \operatorname{diag}\!\left( \lambda_4+\lambda_8,2(\lambda_\Delta+\lambda_\Delta'),
\lambda_6,\lambda_7+\lambda_{10}\right), \nonumber \\
{\cal B}_4 &= \bigl([\delta^{++}\delta^{++}]\bigr),
\qquad {\mathbb M}_4 =\bigl(2(\lambda_\Delta+\lambda_\Delta')\bigr).
\label{eq:Q34blocks}
\end{eqnarray}
The three remaining $Q=+1$ two-state blocks that contain $\delta^{++}$ are
\begin{align}
{\cal B}_{1}(H\Delta)=\bigl([\eta^-\delta^{++}],[H^{0*}\delta^+]\bigr):&\quad
\begin{pmatrix}\lambda_4+\lambda_8&-\lambda_8/\sqrt2\\-\lambda_8/\sqrt2&\lambda_4+\lambda_8/2\end{pmatrix},\nonumber\\
{\cal B}_{1}(\rho\Delta)=\bigl([\delta^+\rho^0],[\delta^{++}\rho^-]\bigr):&\quad
\begin{pmatrix}\lambda_6+\lambda_9/2&-\lambda_9/\sqrt2\\-\lambda_9/\sqrt2&\lambda_6\end{pmatrix},\nonumber\\
{\cal B}_{1}(\phi\Delta)=\bigl([\delta^+\phi^{0*}],[\delta^{++}\phi^-]\bigr):&\quad
\begin{pmatrix}\lambda_7+\lambda_{10}/2&-\lambda_{10}/\sqrt2\\-\lambda_{10}/\sqrt2&\lambda_7+\lambda_{10}\end{pmatrix}.
\label{eq:Q1doublycharged}
\end{align}
For each doublet pair, the remaining two-state blocks occur in two simple forms,
\begin{equation}
\begin{pmatrix}a&b\\b&a\end{pmatrix}: e=\{a-b,\,a+b\},
\qquad
\begin{pmatrix}a+b&b\\b&a+b\end{pmatrix}: e=\{a,\,a+2b\}.
\label{eq:D_blocks}
\end{equation}
The assignments $(a,b)=(\lambda_1,\lambda_1')$, $(\lambda_2,\lambda_2')$, and $(\lambda_5,\lambda_3)$ apply to the $H\rho$, $H\phi$, and $\rho\phi$ sectors, respectively. The doublet--triplet blocks have the analogous forms
\begin{equation}
\begin{pmatrix}a+b/2&b/\sqrt2\\b/\sqrt2&a\end{pmatrix}: e=\{a-b/2,\,a+b\},
\qquad
\begin{pmatrix}a+b/2&b/\sqrt2\\b/\sqrt2&a+b\end{pmatrix}: e=\{a,\,a+3b/2\}.
\label{eq:T_blocks}
\end{equation}
Here $(a,b)=(\lambda_4,\lambda_8)$, $(\lambda_6,\lambda_9)$, and $(\lambda_7,\lambda_{10})$ for $H\Delta$, $\rho\Delta$, and $\phi\Delta$, respectively. Other charge assignments are related by conjugation, state permutations, or rephasing and therefore have the same eigenvalues.

The remaining one-state blocks are already diagonal, so each entry is itself an eigenvalue. For every parameter point in the scan, all 117 charge-sector matrices were constructed from the corresponding quartic couplings and diagonalized numerically. We find that all eigenvalues satisfy the perturbative-unitarity condition $|e_i|<8\pi$.
\section{Renormalization of electroweak precision observables}
\label{app5}
The triplet VEV modifies the tree-level gauge-boson mass relations
\begin{equation}
m_W^2=\frac{g_2^2}{4}(v_H^2+2v_\Delta^2),\qquad
m_Z^2=\frac{g_2^2}{4c_W^2}(v_H^2+4v_\Delta^2),\qquad
\rho_{\rm EW}^{\rm tree}=\frac{v_H^2+2v_\Delta^2}{v_H^2+4v_\Delta^2}.
\label{rho-tree-exact}
\end{equation}
Because $\rho_{\rm EW}^{\rm tree}\neq1$, the usual three electroweak inputs do not determine all parameters of the gauge sector. An on-shell realization may use $\{\alpha_{\rm em},G_F,m_Z,\hat s_W^2\}$, with $\hat s_W^2$ defined by the effective leptonic weak mixing angle at the $Z$ pole~\cite{Blank:1997qa,Kanemura:2012rs}. Its renormalization condition fixes the additional counterterm $\delta\hat s_W^2$, equivalently $\delta v_\Delta$. The complete one-loop predictions for $m_W$, $\rho_{\rm EW}$, and related observables then require the renormalized gauge-boson self-energies and the accompanying vertex and box corrections. The finite renormalization condition is chosen such that the active $H$--$\Delta$ contribution to $\Delta T$ vanishes at the electroweak input point. The electroweak validation therefore separates the exact tree-level prediction in Eq.~\ref{rho-tree-exact} from the correlated test of the finite loop contributions. Our numerical scan employs the reduced quantities
\begin{equation}
\Delta S=\Delta S_{H\Delta}^{(0)}+\Delta S_{\rm inert},\qquad
\Delta T=\Delta T_{\rm inert},\qquad
\chi^2_{ST}=\Delta X^{T}C_{ST}^{-1}\Delta X,
\label{ST-scan}
\end{equation}
These definitions contain the finite inert-doublet contributions and the leading active-sector contribution to $\Delta S$ in the custodial reference limit. The comparison is evaluated with $\Delta U = 0$.

Evaluation on the reported viable points gives
$\rho_{\rm EW}^{\rm tree}=0.99999475$--$0.99999795$ for NO and
$0.99999576$--$0.99999810$ for IO. The largest displacement from unity is $5.25\times10^{-6}$, corresponding to $|\delta\rho|/\alpha\simeq6.7\times10^{-4}$ for $\alpha^{-1}=128$. The oblique shifts are
\begin{table}[H]
\centering
\renewcommand{\arraystretch}{1.15}
\begin{tabular}{lccc}
\hline
Ordering & $\Delta S$ & $\Delta T$ & $\max\chi^2_{ST}$\\
\hline
NO & $[-0.00359,\,0.01730]$ & $[0.00403,\,0.10522]$ & $9.84$\\
IO & $[-0.00185,\,0.01907]$ & $[0.01298,\,0.09960]$ & $8.18$\\
\hline
\end{tabular}
\caption{Oblique parameters evaluated on the accepted samples.}
\label{tab:oblique-results}
\end{table}
All $62$ NO points and all $56$ IO points lie inside the adopted tree-level $\rho_{\rm EW}$ interval. Their largest $\chi^2_{ST}$ values are $9.84$ and $8.18$, respectively, both below the two-degree-of-freedom $3\sigma$ value $11.83$. The accepted samples therefore satisfy the tree-level $\rho_{\rm EW}$ constraint and the correlated $(\Delta S,\Delta T)$ condition defined in Eq.~\eqref{ST-scan}. A complete one-loop four-input electroweak fit remains beyond the scope of the present numerical analysis.
\bibliographystyle{JHEP}
\bibliography{bibliography.bib}
\end{document}